\documentclass[preprint, showpacs, amsmath, amssymb, prb, 12pt]{revtex4}
\usepackage{graphicx}
\usepackage{dcolumn}
\usepackage{bm}
\begin{document}
\title{Strongly-correlated multi-particle transport in one-dimension through a quantum impurity: an outline of exact and complete solutions}
\author{Jung-Tsung Shen}
\author{Shanhui Fan}
\email{shanhui@stanford.edu}
\affiliation{Department of Electrical Engineering and Ginzton Laboratory, Stanford University, Stanford, California
 94305, USA}
\date{\today}
\begin{abstract}
We consider the transport properties of multiple-particle quantum states in a class of one-dimensional systems with a single quantum impurity. In these systems, the local interaction at the quantum impurity induces strong and non-trivial correlations between the multi-particles. We outline an exact theoretical approach, based upon real-space equations of motion and the Bethe ansatz, that allows one to construct the full scattering matrix (S-matrix) for these systems.  In particular, we emphasize the need for completeness check upon the eigenstates of the S-matrix, when these states obtained from Bethe Ansatz are used for describing the scattering properties. As a detailed example of our approach, we solve the transport properties of two photons incident on a single two-level atom, when the photons are restricted to a one-dimensional system such as a photonic crystal waveguide. Our approach predicts a number of novel nonlinear effects involving only two photons, including background fluorescence, spatial attraction and repulsion between the photons, as well as the emergence of a two-photon bound state.
\end{abstract}
 \maketitle

\section{Introduction}\label{P:Introduction}

Understanding the properties of a single quantum impurity embedded in a continuum of extended states is of central importance in both condensed matter physics and quantum optics. In general, a quantum impurity problem is defined by a Hamiltonian of the following form:
\begin{equation}\label{E:TotalHamiltonian}
H = H_{p} + H_{a} + H_{\mbox{\scriptsize int}},
\end{equation}where $H_{p}$ describes free propagating quantum particles, $H_{a}$ describes the internal dynamics of the impurity (henceforth we will also use ``atom'' interchangeably), and $H_{\mbox{\scriptsize int}}$ describes the tunneling processes between the impurity and the free propagating states. In condensed matter physics, a notable example of a quantum impurity is described by the Anderson Hamiltonian~\cite{Anderson:1961}, where the continuum is the band formed by a free-electron gas, the impurity is a local site with a single $d$-orbital, and the electrons can tunnel between the impurity and the Fermi sea. In quantum optics, the Dicke Hamitonian~\cite{Dicke:1954}, which describes in a full quantized fashion the interactions of a two-level atom with photons, also falls into this category. Here the extended states are free-propagating photon states, the impurity is the two-level atom, and the tunneling term $H_{\mbox{\scriptsize int}}$ describes the emission and absorption processes. In each case, due to the interactions at the localized impurity site, the overall system possesses highly-nontrivial and strongly correlated behaviors.  


In this article we focus on the scattering properties of such a quantum impurity, when one or more quantum particles are incident upon it. The quantum particles are restricted to propagate in a one-dimensional continuum. Such a one-dimensional model is relevant to recent experiments on the transport properties of electrons through quantum dots~\cite{Cronenwett:1998, Nygard:2000, Wiel:2000}, and photons through quantum dots~\cite{Badolato:2005} or trapped atoms~\cite{Birnbaum:2005}. Moreover, such a one-dimensional model can also be relevant for three-dimensional scattering problems. Since the impurity is typically far smaller in its spatial extent compared with the wavelengths of incident particles, most of the three-dimensional problems involving a single impurity can be mapped into a one-dimensional problem, because only $S$ waves need to be taken into account.

It is known that many quantum impurity problems can not be solved using perturbation theory. Instead, since the 1980's, significant efforts have been devoted to non-perturbative approaches, such as Bethe ansatz that directly diagonalizes the Hamiltonian~\cite{Andrei:1980, Wiegmann:1980, Wiegmann:1981, Wiegmann:1983a, Rupasov:1984, John:1997}.  However, most of these papers assume a periodic boundary condition in order to obtain thermodynamic information of the overall system. Only until very recently were the Bethe ansatz approach employed to solve for the scattering properties~\cite{Mehta:2006, Shen:2007a}. The scattering problems involve open boundary conditions and are in fact subtle and require very careful treatments. 

Here we develop a full quantum-mechanically theoretical framework that allows one to extract scattering information from the eigenstates of the full interacting Hamiltonian. In particular, we emphasize the necessity of the completeness check of the computational scheme, in order to obtain the correct description of scattering properties. As an illustration of our formalism, the multi-photon problem is completely solved using this approach. The formalism, however, is general and can be readily applied to electrons as well.  

The paper is organized as follows: in Sec.~\ref{P:GeneralProperties} we summarize some of basic results of quantum scattering theory. In particular, we discuss the Lippman-Schwinger formalism, with emphasis on those aspects that are relevant for our purpose. In Sec.~\ref{P:System} we then discuss in details the photon Hamiltonian, and its connections to the Anderson Hamiltonian. Sec.~\ref{P:SRelation} discusses the decomposition of the scattering matrix (S-matrix), which enables and greatly simplifies the calculations. Finally, in Sec.~\ref{P:OnePhoton},  and Sec.~\ref{P:TwoPhotonI} and \ref{P:TwoPhotonII}, respectively, we  present a detailed discussion of solving the photon Hamiltonian for its one and two photon transport properties. In Sec.~\ref{P:ThreePhoton} we briefly discuss the three-particle case.

\section{General Aspects of the Many-Body Quantum Impurity Scattering Problem}\label{P:GeneralProperties}

Before we begin the mathematical adventures of solving for the quantum impurity scattering problem, we first provide a brief review of relevant theoretical background.

\subsection{Revisit of the Lippmann-Schwinger Formalism}\label{A:Lippmann-Schwinger}

In general, quantum scattering theory deals with a Hamiltonian of the form $H = H_0 + H_{\mbox{\scriptsize int}}$, where $H_0$ defines the free constituents. $H_0$, for example, can be $H_p + H_a$ in Sec.~\ref{P:Introduction}, describing particles and the single impurity. $H_{\mbox{\scriptsize int}}$ defines the interactions between these contituents. We will restrict to the case where the interaction range of the quantum impurity is finite in space. 

The scattering theory aims to answer the following question: for a given incident multi-particle state, after scattering off the quantum impurity, what is the outgoing multi-particle state? Long before ($t\rightarrow -\infty$) and after ($t\rightarrow +\infty$) the scattering, the incoming and outgoing states are away from the quantum impurity and accordingly are outside of the interaction range. As a result, both the initial and the final states are \emph{free} particle states, governed by $H_0$. The quantum impurity therefore can be viewed as an intermediary inducing a mapping from one free state to another. The mapping is represented by the S-matrix, which encodes all scattering properties. 

In the Lippmann-Schwinger formalism, in order to define the S-matrix, one assumes that the interaction Hamiltonian $H_{\mbox{\scriptsize int}}$ was adiabatically ``switched on'' very slowly from the distant past ($t\rightarrow -\infty$), to its full strength at $t=0$, and will be adiabatically ``switched off'' very slowly in the distant future ($t\rightarrow +\infty$)~\cite{Taylor:1972, Greiner:1996,Huang:1998}. That is, the interaction $H_{\mbox{\scriptsize int}}$ is replaced by
\begin{equation}
H_{\mbox{\scriptsize int}}'(t) \equiv e^{-\epsilon |t|} H_{\mbox{\scriptsize int}}, \quad \epsilon \rightarrow 0^+.
\end{equation}
The limit $\epsilon \rightarrow 0^+$ is to be taken last, after all calculations. The adiabatic switching is designed to represent the situation that the incoming particles only interact with the target for a short period of time and then fly away~\cite{Huang:1998}.

Let the interacting state of the overall system at $t=0$ be $|i^+\rangle$. The time-evolution of the interacting state is described by $\mathbf{U}(t)|i^+\rangle$, where $\mathbf{U}(t)$ is the evolution operator related to the time-dependent Hamiltonian $H_0 + H_{\mbox{\scriptsize int}}'(t)$.  Following the adiabatic switching on of the interaction aforementioned, in the remote past ($t\rightarrow -\infty$), $\mathbf{U}(t)|i^+\rangle$ asymptotically approaches $e^{-iH_0 t}|i\rangle \equiv \mathbf{U}^{0}(t)|i\rangle$, where $|i\rangle$ is a \emph{free} state. Similarly, with the adiabatic switching off of the interaction, in the remote future ($t\rightarrow +\infty$), $\mathbf{U}(t)|i^+\rangle$ asymptotically approaches $ \mathbf{U}^{0}(t)|f_i\rangle$, where $|f_i\rangle$ is a \emph{free} state. Both $|i\rangle$ and $|f_i\rangle$ are governed by the free Hamiltonian $H_0$. The subscript $i$ in the state $|f_i\rangle$ indicates its dependence on $|i\rangle$. $|i\rangle$ and $|f_i\rangle$ directly correspond to the incoming free states prepared, and outgoing free states detected in the experiments. Hence they are referred to as ``in-'' and ``out-''state, respectively.

\begin{figure}[thb]
\scalebox{1}{\includegraphics[width=\columnwidth]{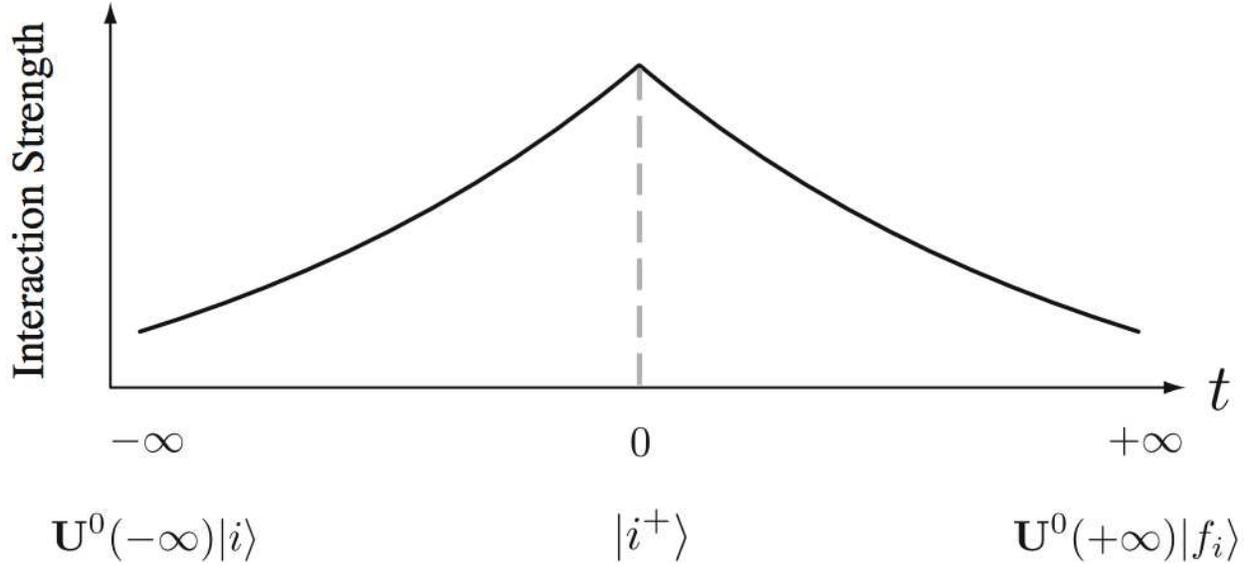}}
\caption{The evolution of the state of the system with adiabatic switching. $|i\rangle$ and $|f_i\rangle$ are free states governed by the free Hamiltonian, $H_0$. When the interaction is adiabatically switched on from the distant past ($t\rightarrow -\infty$) to its full strength $H_{\mbox{\scriptsize int}}$ at $t=0$, the state evolves from asymptotic state $\mathbf{U}^{0}(t\rightarrow-\infty)|i\rangle$ to $|i^+\rangle$, which is governed by the full interacting Hamiltonian $H = H_0 + H_{\mbox{\scriptsize int}}$. When the interaction is adiabatically switched off in the remote future ($t\rightarrow +\infty$), the state asymptotically approaches $\mathbf{U}^{0}(t\rightarrow+\infty)|f_i\rangle$.}\label{Fi:Adiabatic}
\end{figure}

The three states $|i\rangle$, $|i^+\rangle$ and $|f_i\rangle$, as defined above (Fig.~\ref{Fi:Adiabatic}), satisfy the Lippmann-Schwinger equations~\cite{Taylor:1972, Sakurai:1994}:
\begin{align}
|i ^+\rangle &= |i\rangle +  \frac{1}{E-H_0 + i\epsilon}H_{\mbox{\scriptsize int}} |i^+\rangle,\label{E:LippmannR}\\
|i ^+\rangle &= |f_i\rangle +  \frac{1}{E-H_0 - i\epsilon}H_{\mbox{\scriptsize int}} |i^+\rangle,\label{E:LippmannA}
\end{align}where 
$\frac{1}{E-H_0 + i\epsilon}\equiv G^{R}_0$ is the \emph{free} retarded Green's function; while $\frac{1}{E-H_0 - i\epsilon}\equiv G^{A}_0$ is the \emph{free} advanced Green's function. Eqs.~(\ref{E:LippmannR}) and (\ref{E:LippmannA}) are applicable for energy eigenstates, \emph{i.e.}, $H_0 |i\rangle = E|i\rangle$, $H |i^+\rangle = E|i^+\rangle$, and $H_0 |f_i\rangle = E|f_i\rangle$. It can be proved that the energies of $|i\rangle$ and $|f_i\rangle$ are the same as that of $|i^+\rangle$~\cite{Taylor:1972, Greiner:1996, Sakurai:1994, Huang:1998}. 

The S-matrix, in general, is defined as
\begin{equation}\label{E:SMatrixConstructed}
\mathbf{S}\equiv \sum_{i}|f_i\rangle \langle i|,
\end{equation}where the summation is taken over a complete basis $\{|i\rangle\}$ of the Hilbert space defined by $H_0$. Once the S-matrix is determined, one can then calculate the scattering properties for an arbitrary incident state. For a given in-state $|\mbox{in}\rangle$, the out-state is
\begin{equation}
|\mbox{out}\rangle = \mathbf{S} |\mbox{in}\rangle = \sum_{i} |f_i\rangle \langle i|\mbox{in}\rangle,
\end{equation}and thereby the probability amplitude of finding the out-going particles to be in a state $|\chi\rangle$ is
\begin{equation}
\langle \chi|\mbox{out}\rangle = \langle \chi|\mathbf{S} |\mbox{in}\rangle = \sum_{i} \langle \chi|f_i\rangle \langle i|\mbox{in}\rangle.
\end{equation}

In most practical scattering calculations, one starts with a given $|i\rangle$ and computes $|i^+\rangle$ from Eq.~(\ref{E:LippmannR}), and then obtains $|f_i\rangle$ from Eq.~(\ref{E:LippmannA}). By repeating this process for a complete set of eigenstates $\{|i\rangle\}$ of $H_0$, the S-matrix is constructed. In this route of constructing the S-matrix, the unitarity of the S-matrix, \emph{i.e.}, $\mathbf{S}^{\dagger}\mathbf{S}= \mathbf{S}\mathbf{S}^{\dagger}=\mathbf{1}$, is automatically guaranteed by starting with a complete basis set $\{|i\rangle\}$ for the free Hamiltonian $H_0$.

For the impurity scattering problem that we deal with here, however, as it turns out, we will be in an unusual situation wherein the interacting state $|i^+\rangle$ is first obtained through a non-perturbative Bethe-ansatz technique. In this situation, to construct the S-matrix, one has to turn around Eq.~(\ref{E:LippmannR}) and (\ref{E:LippmannA}) to compute $|i\rangle$ and $|f_i\rangle$ from $|i^+\rangle$. In doing so, the completeness of the set $\{|i\rangle\}$ thus obtained needs to be explicitly checked, especially since the state $|i\rangle$ thus obtained can itself possess rich and entangled structures. (A completeness check for \{$|i^+\rangle\}$ typically is far more involved.) This route of constructing the S-matrix, and the completeness check, will be explicitly carried out for photon-impurity scattering problem in Sec.~\ref{P:OnePhoton} for one-photon case,  and in Sec.~\ref{P:TwoPhotonI} and \ref{P:TwoPhotonII} for two-photon case.

As a remark, we note that Eq.~(\ref{E:LippmannR}) and (\ref{E:LippmannA}) can also be expressed using the \emph{exact}  Green's functions\cite{Taylor:1972, Greiner:1996,Huang:1998}:
\begin{align}
|i ^+\rangle &= |i\rangle +  \frac{1}{E-H + i\epsilon}H_{\mbox{\scriptsize int}} |i\rangle \equiv |i\rangle +  G^R H_{\mbox{\scriptsize int}} |i\rangle,\label{E:LippmannRExact}\\
|i ^+\rangle &= |f_i\rangle +  \frac{1}{E-H - i\epsilon}H_{\mbox{\scriptsize int}} |f_i\rangle \equiv |f_i\rangle +  G^A H_{\mbox{\scriptsize int}} |f_i\rangle.\label{E:LippmannAExact}
\end{align}
Our approach therefore also provides a way to compute the exact Green's functions.

\subsection{A Simple Example}

\begin{figure}[thb]
\scalebox{1}{\includegraphics{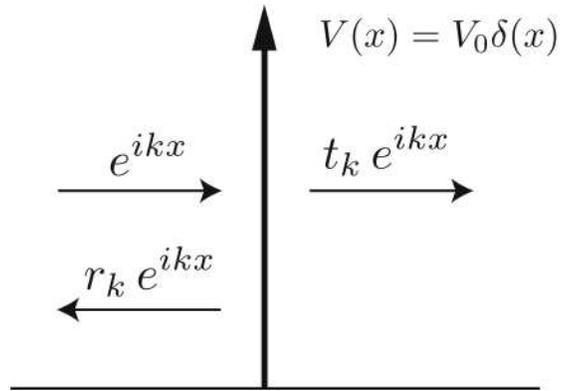}}
\caption{A quantum particle incident upon a delta potential barrier characterized by $V_0 \delta(x)$.}\label{Fi:Example}
\end{figure}

As an illustration of the approach of constructing the full S-matrix starting from $|i^+\rangle$, here we give one simple example of a one-dimensional scattering problem wherein a quantum particle (or a wave) scatters off a delta potential characterized by $V(x) = V_0 \delta(x)$, as shown in Fig.~\ref{Fi:Example}. The quantum particle is described by the free Hamiltonian $H_0 = -d^2/d x^2$. 


An eigenstate $|i^+\rangle$ of the full Hamiltonian $H = H_0 + V(x)$ can be computed  straightforwardly as:
\begin{equation}\label{E:ExampleEigenstate} 
\langle x|i^+\rangle = \left(\frac{e ^{i k x}}{\sqrt{2\pi}} + r_k \frac{e^{- i k x}}{\sqrt{2\pi}}\right)\theta(-x) + t_k \frac{e^{i k x}}{\sqrt{2\pi}}\theta(x),
\end{equation}where $r_k = \frac{-i V_0}{2 k + i V_0}$, $t_k = \frac{2 k}{2 k + i V_0}$, and $1 + r_k = t_k$. To compute $|f_i\rangle$, we write Eq.~(\ref{E:LippmannA}) in the real-space representation:
\begin{align}\label{E:Example1}
\langle x|i^+\rangle &= \langle x|f_i\rangle +  \int dx' \langle x|\frac{1}{E_k - H_0 - i\epsilon}|x'\rangle V(x')\langle x' |i^+\rangle\notag\\
&=\langle x|f_i\rangle +  \langle x|\frac{1}{E_k - H_0 - i\epsilon}|0\rangle V_0 \langle 0 |i^+\rangle,
\end{align}where $\langle 0 |i^+\rangle \equiv \left(\langle 0^- |i^+\rangle + \langle 0^+ |i^+\rangle\right)/2 = t_k/\sqrt{2\pi}$, and $E_k \equiv k^2$.

Using the fact that the advanced Green's function for $H_0 = -d^2/d x^2$ is
\begin{equation}
\langle x|\frac{1}{E_k - H_0 - i\epsilon}|x'\rangle = +\frac{i}{2 k} e^{-i k |x-x'|},
\end{equation}one can easily verify using Eq.~(\ref{E:Example1}) that the out-state $|f_i\rangle$ is
\begin{equation}
\langle x|f_i\rangle = r_k \frac{e^{- i k x}}{\sqrt{2\pi}} + t_k \frac{e^{i k x}}{\sqrt{2\pi}}, \quad \mbox{for all} \,\,x.
\end{equation}Similarly, by using the retarded Greens' function for $H_0$:
\begin{equation}
\langle x|\frac{1}{E_k - H_0 + i\epsilon}|x'\rangle = -\frac{i}{2 k} e^{+i k |x-x'|},
\end{equation}and Eq.~(\ref{E:LippmannR}), the in-state $|i\rangle$ can be obtained as
\begin{equation}
\langle x|i\rangle = \frac{e^{i k x}}{\sqrt{2\pi}}, \quad \mbox{for all} \,\,x.
\end{equation}Hence, $|i\rangle$ is a plane wave state $|k\rangle$. Since the set $\{|i\rangle\}$ forms a complete basis set of eigenstates of $H_0$:
\begin{align}
H_0 |k\rangle &= k^2 |k\rangle,\notag\\
\sum_{k} |k\rangle \langle k| &= \mathbf{1},
\end{align} 
the S-matrix is
\begin{equation}
\mathbf{S} = \sum_{i} |f_i\rangle\langle i| = \sum_{k} r_k |-k\rangle \langle k|+ t_k |k\rangle \langle k|,
\end{equation}with the relations $\langle -k|\mathbf{S}|k\rangle = r_k$, and  $\langle k|\mathbf{S}|k\rangle = t_k$. 

These relations are consistent with the usual reading of $|i^+\rangle$ in Eq.~(\ref{E:ExampleEigenstate}), where $r_k$ and $t_k$ are interpreted as the reflection and transmission amplitude, respectively. The derivations here put such an intuitive reading of the interacting eigenstates on a firm theoretical foundation.



\subsection{One- and Two-Particle States Expressed in Second-Quantized Form}\label{P:SecondQuantization}

When describing scattering processes involving multiple particles, it is advantageous to express the states in second-quantized form. Moreover, similar to the example above, the eigenstate $|i^+\rangle$ in the Bethe ansatz calculations is best expressed in real space representation. Here, for convenience, we list the expressions of one- and two-particle states in the second quantized form using a real-space representation.

The basis for real-space representation of one- and two-particle states are
\begin{align}
|x\rangle &\equiv c^{\dagger}(x)|\emptyset\rangle,\notag\\
|x_1, x_2\rangle &\equiv \frac{1}{\sqrt{2}}c^{\dagger}(x_1)c^{\dagger}(x_2)|\emptyset\rangle,
\end{align}where $|\emptyset\rangle$ is vacuum. These states are normalized as
\begin{align}
\langle x|x'\rangle &= \delta(x-x'),\notag\\
\langle x_1, x_2|x_1', x_2'\rangle &= \frac{1}{2}[\delta(x_1-x_1')\delta(x_2-x_2')\pm \delta(x_1-x_2')\delta(x_2-x_1')],
\end{align}
with $+$ sign for bosons, and $-$ sign for fermions. ($\langle x_1, x_2|\equiv \langle \emptyset| \frac{1}{\sqrt{2}}c(x_2)c(x_1)$.) Using this basis, any two-particle states $|f\rangle$ is defined as
\begin{equation}
|f\rangle \equiv \int dx_1' dx_2'\, f(x_1', x_2') \frac{1}{\sqrt{2}}c^{\dagger}(x_1')c^{\dagger}(x_2')|\emptyset\rangle,
\end{equation}with $f(x_1, x_2) = +f(x_2, x_1)$ for bosons, $f(x_1, x_2) = -f(x_2, x_1)$ for fermions. 

$f(x_1, x_2)$ is in fact the two-particle wavefunction. Let the free Hamiltonian $H_0$ take the following form in the second quantization:
\begin{equation}\label{E:SchrodingerSecond}
H_0 = \int dx \,c^{\dagger}(x)\hat{H}_0(x)c(x),
\end{equation}which is relevant to our purpose. One can show that the second-quantized Schr\"{o}dinger equation $H_0|f\rangle = E|f\rangle$ leads to
\begin{equation}\label{E:SchrodingerFirst}
\left[\hat{H}_0(x_1)+ \hat{H}_0(x_2)\right] f(x_1, x_2) = E f(x_1, x_2),
\end{equation}the Schr\"{o}dinger equation in the first quantization form~\cite{Greiner:1996, Huang:1998}. Thus $f(x_1, x_2)$ has the same properties of the two-particle wavefunction when expressed in the first quantization form. (Note that similar relations between Eq.~(\ref{E:SchrodingerSecond}) and (\ref{E:SchrodingerFirst}) hold true, for one-particle wavefunction, as well as when one-particle external potential, and two-particle interaction are included~\cite{Greiner:1996, Huang:1998}.) 

Below we provide further evidence that $f(x_1, x_2)$ is indeed a ``two-particle'' wavefunction. For example,
\begin{widetext}  
\begin{align}
\langle x_1, x_2|f\rangle &=  \int dx_1' dx_2' \frac{1}{2}[\delta(x_1-x_1')\delta(x_2-x_2')\pm \delta(x_1-x_2')\delta(x_2-x_1')] f(x_1', x_2')\notag\\
&=\frac{1}{2}\left(f(x_1, x_2) \pm f(x_2, x_1)\right)\notag\\
&= f(x_1, x_2).
\end{align}
\end{widetext}
Moreover, for any two-particle states $|f\rangle$ and $|g\rangle$, where
\begin{equation}
|g\rangle \equiv \int dx_1' dx_2'\, g(x_1', x_2') \frac{1}{\sqrt{2}}c^{\dagger}(x_1')c^{\dagger}(x_2')|\emptyset\rangle,
\end{equation}with $g(x_1, x_2) = \pm g(x_2, x_1)$ ($+$ sign for bosons, and $-$ sign for fermions), one has
\begin{widetext}
\begin{align}
\langle g|f\rangle &= \int dx_1 dx_2 dx_1' dx_2'\, f^{*}(x_1, x_2) g(x_1', x_2')\frac{1}{2}[\delta(x_1-x_1')\delta(x_2-x_2')\pm \delta(x_1-x_2')\delta(x_2-x_1')]\notag\\
&= \int dx_1 dx_2 \frac{1}{2}\left(f^{*}(x_1, x_2) g(x_1, x_2)\pm f^{*}(x_1, x_2) g(x_2, x_1)\right)\notag\\
&=\int dx_1 dx_2 f^{*}(x_1, x_2) g(x_1, x_2)\notag\\
&=\int dx_1 dx_2 \langle f|x_1, x_2\rangle \langle x_1, x_2|g\rangle.
\end{align}
\end{widetext}

\section{The System and the Hamiltonian}\label{P:System}

\begin{figure}[thb]
\scalebox{1}{\includegraphics{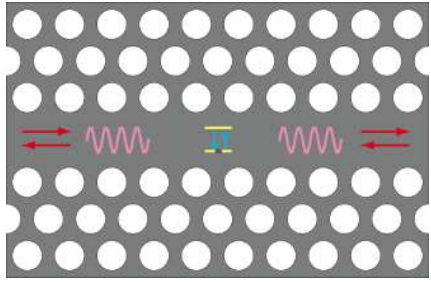}}
\caption{(Color online) Schematics of the system. A two-level system is coupled to a one-dimensional continuum in which the photons, shown as wiggly waves, propagate in each direction.}\label{Fi:Geometry}
\end{figure}

We now apply the general ideas developed above to the problem of photons scattering off a two-level system. Fig.~\ref{Fi:Geometry} shows the schematics of the overall system of interest. The two-level system is embedded in a one-dimensional waveguide in which the photons propagate. The one-dimensional waveguide can be, for example, a line-defect waveguide in a photonic crystal with a complete photonic band gap. A discussion of such a problem for photonic crystal experiments is provided in Ref.~[\onlinecite{Shen:2007a}]. The focus here is on the formalism itself.

The system 
is modeled by the Hamiltonian~\cite{Shen:2005,Shen:2005a}:
\begin{align}\label{E:hamiltonian}
H&=\int dx \left\{-i v_g c_{R}^{\dagger}(x)\frac{\partial}{\partial
x} c_{R}(x)
+ i v_g c_{L}^{\dagger}(x)\frac{\partial}{\partial x} c_{L}(x)\right. \notag\\
&\quad \left.+ \bar{V} \delta(x) \left(c^{\dagger}_{R}(x) \sigma_{-} +
c_{R}(x) \sigma_{+}
+ c^{\dagger}_{L}(x) \sigma_{-} + c_{L}(x) \sigma_{+}\right)\right\} \notag\\
&\quad +E_{e} a^{\dagger}_{e} a_{e}+ E_{g} a^{\dagger}_{g} a_{g}
\end{align}
where $v_g$ is the group velocity of the photons, and
$c_{R}^{\dagger}(x)$($c_{L}^{\dagger}(x)$) is a bosonic operator
creating a right-going(left-going) photon at $x$. $\bar{V}$ is the coupling constant,
$a^{\dagger}_{g}$($a^{\dagger}_{e}$) is the creation operator of the
ground (excited) state of the atom, $\sigma_{+}=a^{\dagger}_{e}
a_{g}$($\sigma_{-}=a^{\dagger}_{g} a_{e}$) is the atomic raising (lowering) ladder
operator satisfying $\sigma_{+}|n,-\rangle =|n,+\rangle$ and $\sigma_{+}|n,+\rangle =0$, where $|n, \pm\rangle \equiv |n\rangle\otimes|\pm\rangle$ describes the state of the system with $n$ photons and the atom in the excited ($+$) or ground ($-$) state. $E_{e}-E_{g}
(\equiv\Omega)$ is the transition energy. $\hbar$ is set to 1. This Hamiltonian describes the situation where the propagating photons can run in both directions, and is referred to as ``two-mode'' model. 

The aim of this paper is to solve the two-photon transport properties of this Hamiltonian. Specifically, we imagine a physical scattering experiment, where two photons incident upon a two-level quantum impurity embedded in a one-dimensional waveguide. When
 the two photons arrive at the impurity within a time interval comparable to the spontaneous emission lifetime of the impurity, one should expect that the transport properties of the photons are strongly correlated, as mediated by the quantum impurity. Here, we develop the theoretical formalism to describe such a correlation.

As a first step, we note that by employing the following transformation 
\begin{align}\label{E:Transformation}
c^{\dagger}_e (x)&\equiv \frac{1}{\sqrt{2}}(c^{\dagger}_{R}(x)+ c^{\dagger}_{L}(-x)),\notag\\ 
c^{\dagger}_o (x)&\equiv
\frac{1}{\sqrt{2}}(c^{\dagger}_{R}(x)- c^{\dagger}_{L}(-x)),
\end{align} the original Hamiltonian is transformed into two decoupled ``one-mode''
Hamiltonians, \emph{i.e.,} $H=H_e + H_o$, where
\begin{widetext}
\begin{subequations}\label{E:PhotonHamiltonianDecomposed}
\begin{align}
H_e &= \int dx (-i) v_g c_{e}^{\dagger}(x)\frac{\partial}{\partial x}
c_{e}(x) + \int dx V \delta(x)\left(c^{\dagger}_{e}(x)
\sigma_{-} + c_{e}(x) \sigma_{+}\right)+E_{e} a^{\dagger}_{e} a_{e}+
E_{g} a^{\dagger}_{g}
a_{g}\label{E:He},
\end{align}
\begin{align}
H_o &= \int dx (-i) v_g c_{o}^{\dagger}(x)\frac{\partial}{\partial x}
c_{o}(x)\label{E:Ho},
\end{align}
\end{subequations}
\end{widetext} with $[H_e, H_o]=0$. $H_o$ is an interaction-free one-mode Hamiltonian, while
$H_e$ describes a non-trivial one-mode interacting model with coupling strength $V\equiv\sqrt{2} \bar{V}$. For notational simplicity, $v_g$ is set to 1 hereafter.

As a side note, the Hamiltonian $H_e$ is closely related to the extensively studied Anderson model in condensed matter physics. The Anderson model describes the interaction of the conduction electrons with a single quantum impurity~\cite{Anderson:1961}. In real-space, the one-mode Anderson Hamiltonian takes the following form~\cite{Wiegmann:1983a, Hewson:1997}:
\begin{widetext}
\begin{equation}\label{E:AndersonOneMode}
H_A = \int -i\sum_{\sigma} c_{\sigma}^{\dagger}(x)\frac{\partial c_{\sigma}(x)}{\partial x}dx + \int V\delta(x)\left(c^{\dagger}_{\sigma}(x) c_{d,\sigma} + c^{\dagger}_{d, \sigma}c_{\sigma}(x)\right)dx +\sum_{\sigma}\epsilon_d n_{d,\sigma}+ U n_{d, \uparrow}n_{d, \downarrow},
\end{equation}
\end{widetext}where $c^{\dagger}_{\sigma}(x)$ ($c_{\sigma}(x)$) is the creation (annihilation) operator of the conduction electron with spin $\sigma$, while the operator $c^{\dagger}_{d, \sigma}$ ($c_{d,\sigma}$) creates an electron of spin $\sigma$ on the local impurity at $x=0$. $n_{d,\sigma}$ is the number operator of electrons on the impurity. $V$ is the coupling strength. $\epsilon_d$ is the energy of the electron on the impurity, and is degenerate for both spins. 

The first term in Eq.~(\ref{E:AndersonOneMode}) describes the kinetic energy of the conduction electrons, and the second term describes the interaction (``hybridization'') between the conduction electrons and the impurity. These two terms closely resemble the first two terms of $H_e$ in Eq.~(\ref{E:He}). The last term, $U n_{d, \uparrow}n_{d, \downarrow}$, describes the on-site Coulomb interaction between the electrons on the impurity. For an isolated impurity, this term has three energy configurations: (i) zero occupation with energy $E_0= 0$; (ii) single occupation by an electron of spin $\sigma$. The energy is $E_{1,\sigma}=\epsilon_d$ where $\sigma = \uparrow$ or $\downarrow$; (iii) double occupation with a spin $\uparrow$ and a spin $\downarrow$ electron with an energy $E_2 = 2\epsilon_d + U$. When $U$ is large, double occupation is energetically unfavorable. In the limit where $U \rightarrow +\infty$, double occupation becomes prohibited, and the quantum impurity can only accommodate at most one electron, a situation similar to the photon Hamiltonian wherein the two-level system can at most absorb one photon at a time. In this infinite $U$ limit, since there is no double occupation, the term $U n_{d, \uparrow}n_{d, \downarrow}$ effectively drops out, and the Anderson Hamiltonian $H_A$ is exactly the same as the photon Hamiltonian, $H_e$ (except the spin degeneracy). In fact, the general procedures and formalism detailed in this article for two photons can be directly applied to the Anderson model for two electrons in the spin-singlet state for \emph{arbitrary} $U$~\cite{Shen:2007b}. Our procedures thus provide a unified computation schemes for the transport properties of strongly correlated photons as well as electrons. 

Furthermore, when $\langle \sum_{\sigma} n_{d, \sigma}\rangle \simeq 1$, \emph{i.e.},  in the so-called local moment phase, the Anderson model and the Kondo Hamiltonian (s-d model) are equivalent~\cite{Schrieffer:1966}. Both the Anderson model and the Kondo model have recently been  applied to nano-structures such as quantum dots and single electron transistor~\cite{Ng:1988, Meir:1993,Goldhaber-Gordon:1998a,Wiel:2000, Konik:2002}. This connection hints the rich structures of the problem of strongly correlated photon transport. On the other hand, unlike the Anderson model, where fermionic operators describe electrons, here we have bosonic operators describing photons. Consequently, the physics that arises from the existence of a Fermi surface does not occur in our system. The transport properties for multi-electrons and for multi-photons will be correspondingly different. 

In the following three sections, we will provided a detailed account of our solutions to the two-photon transport properties for the Hamiltonian in Eq.~(\ref{E:hamiltonian}).

\section{Relations between the Two-Mode and One-Mode S-matrix}\label{P:SRelation}

The decomposition of the two-mode Hamiltonian $H$ into two decoupled one-mode Hamiltonians [Eq.~(\ref{E:He}) and  (\ref{E:Ho})] greatly simplifies the calculations. In this section, we present a strategy, followed by explicitly detailed calculations, to construct the exact S-matrix of $H$ from the scattering properties of $H_e$ and $H_o$.

Since both $c^{\dagger}_{R}(x)$ and $c^{\dagger}_{L}(x)$ can be decomposed into a linear combination of $c^{\dagger}_{e}(x)$ and $c^{\dagger}_{o}(x)$ via Eq.~(\ref{E:Transformation}), any free one-photon state $|\Psi_1\rangle$ can be written as
\begin{equation}
|\Psi_1\rangle = |\Psi\rangle_{e} + |\Psi\rangle_{o},
\end{equation}where the subscripts label the subspace spanned by $c^{\dagger}_{e}(x)|\emptyset, -\rangle$ or $c^{\dagger}_{o}(x)|\emptyset, -\rangle$, respectively. Similarly, since $c_R^{\dagger}(x_1) c_R^{\dagger}(x_2)$, $c_R^{\dagger}(x_1) c_L^{\dagger}(x_2)$, and $c_L^{\dagger}(x_1) c_L^{\dagger}(x_2)$ can all be decomposed into a linear combination of $c_e^{\dagger}(x_1) c_e^{\dagger}(x_2)$, $c_e^{\dagger}(x_1) c_o^{\dagger}(x_2)$, and $c_o^{\dagger}(x_1) c_o^{\dagger}(x_2)$, $|\Psi_2\rangle$ can also be written as 
\begin{equation}
|\Psi_2\rangle = |\Psi\rangle_{ee} + |\Psi\rangle_{eo} + |\Psi\rangle_{oo},
\end{equation} where the subscripts label the subspace spanned by $\frac{1}{\sqrt{2}}c^{\dagger}_{e}(x_1)c^{\dagger}_{e}(x_2)|\emptyset, -\rangle$, $c^{\dagger}_{e}(x_1)c^{\dagger}_{o}(x_2)|\emptyset, -\rangle$, or $\frac{1}{\sqrt{2}}c^{\dagger}_{o}(x_1)c^{\dagger}_{o}(x_2)|\emptyset, -\rangle$,  respectively. 

Let the two-mode S-matrix be $\mathbf{S}$, we assert the following \emph{decomposition relation}: 
\begin{align}\label{E:Decomposition}
\mathbf{S}|\Psi_{1}\rangle &= \mathbf{S}_{e}|\Psi\rangle_{e} + \mathbf{S}_{o}|\Psi\rangle_{o},\notag\\
\mathbf{S} |\Psi_{2}\rangle &= \mathbf{S}_{ee} |\Psi\rangle_{ee}+ \mathbf{S}_{eo} |\Psi\rangle_{eo}+ \mathbf{S}_{oo} |\Psi\rangle_{oo},
\end{align} where $\mathbf{S}_{e}$ is the one-photon S-matrix in the $e$ subspace governed by $H_e$,  $\mathbf{S}_{o} = \mathbf{1}$, the identity operator, is the one-photon S-matrix in the $o$ subspace governed by $H_o$. $\mathbf{S}_{ee}$ is the two-photon S-matrix in the $ee$ subspace governed by $H_e$, $\mathbf{S}_{eo} = \mathbf{S}_{e} \mathbf{S}_{o}$, and $\mathbf{S}_{oo} = \mathbf{1}$. Once the terms on the right hand side of Eq.~(\ref{E:Decomposition}) are computed, $\mathbf{S}$ can be constructed correspondingly. We will calculate $\mathbf{S}_{e}$ in the next section. Obtaining $\mathbf{S}_{ee} |\Psi\rangle_{ee}$ involves non-trivial calculations, and will be done via the Bethe-ansatz approach in Sec.~\ref{P:TwoPhotonI}. 

To prove the decomposition relation of the two-mode S-matrix [Eq.~(\ref{E:Decomposition})], we start from the asymptotic conditions stated in Sec.~\ref{A:Lippmann-Schwinger} which relates the in-state $|i\rangle$, out-state $|f_i\rangle$, and the interacting eigenstate $|i^+\rangle$:
\begin{subequations}
\begin{equation}\label{E:AsymptoticInT}
\mathbf{U}(t) |i^+\rangle\stackrel{t\to -\infty}{\overrightarrow{\qquad\qquad}} \mathbf{U}^0(t)|i\rangle,
\end{equation}
\begin{equation}\label{E:AsymptoticOutT}
\mathbf{U}(t) |i^+\rangle \stackrel{t\to +\infty}{\overrightarrow{\qquad\qquad}} \mathbf{U}^0(t)|f_i\rangle,
\end{equation}
\end{subequations}where $\mathbf{U}^{0}(t)\equiv e^{- i H_0 t}$ is the unitary evolution operators for $H_0$, while $\mathbf{U}(t)\equiv e^{-i H t}$ is the unitary evolution operator for $H$. Combining Eqs.~(\ref{E:AsymptoticInT}) and (\ref{E:AsymptoticOutT}), one then has
\begin{align}\label{E:SDefAlternative}
|f_i\rangle &= \lim_{t_f \to+\infty}\lim_{t_i \to -\infty} \left({\mathbf{U}^{0}}^{\dagger}(t_f)\mathbf{U}(t_f)\right)\left(\mathbf{U}^{\dagger}(t_i){\mathbf{U}^{0}}^{\phantom{\dagger}}\!\!(t_i)\right)|i\rangle\notag\\
&\equiv \mathbf{S}|i\rangle
\end{align}We note this form of the S-matrix is equivalent to that of Eq.~(\ref{E:SMatrixConstructed}) and both have exactly the same matrix elements. Also, when $|i\rangle$ is an eigenstate of $H_0$, Eq.~(\ref{E:SDefAlternative}) directly gives rise to the Lippmann-Schwinger formalism~\cite{Sakurai:1994}.

To proceed, one recognizes that the photon Hamiltonian $H$, Eq.~(\ref{E:hamiltonian}), can be separated as $H = H_e + H_o$ [Eq.~(\ref{E:PhotonHamiltonianDecomposed})], and so is the free Hamiltonian $H_0 = H_0^o + H_0^e$, where
\begin{subequations}
\begin{equation}\label{E:HFreee}
H_0^e = \int dx (-i) v_g c_{e}^{\dagger}(x)\frac{\partial}{\partial x}
c_{e}(x) +E_{e} a^{\dagger}_{e} a_{e}+
E_{g} a^{\dagger}_{g}
a_{g},
\end{equation}
\begin{equation}\label{E:HFreeo}
H_0^o = \int dx (-i) v_g c_{o}^{\dagger}(x)\frac{\partial}{\partial x}
c_{o}(x),
\end{equation}
\end{subequations}with $[H_0^e, H_0^o]=0$. It thus is easily seen that the S-matrix in Eq.~(\ref{E:SDefAlternative}) can be factored as
\begin{align}\label{E:SDefAlternativeeo}
\mathbf{S} &= \lim_{t_f \to+\infty}\lim_{t_i \to -\infty} \left({\mathbf{U}_e^{0}}^{\dagger}(t_f)\mathbf{U}_e(t_f)\right)\left(\mathbf{U}_e^{\dagger}(t_i){\mathbf{U}_e^{0}}^{\phantom{\dagger}}\!\!(t_i)\right) \left({\mathbf{U}_o^{0}}^{\dagger}(t_f)\mathbf{U}_o(t_f)\right)\left(\mathbf{U}_o^{\dagger}(t_i){\mathbf{U}_o^{0}}^{\phantom{\dagger}}\!\!(t_i)\right)\notag\\
&\equiv \mathbb{S}_e  \mathbb{S}_o,
\end{align}where
\begin{align}
\mathbf{U}_{e,o}(t)&\equiv e^{- i H_{e,o} t},\notag\\
\mathbf{U}_{e,o}^0 (t)&\equiv e^{- i H_0^{e,o} t}.
\end{align}The factoring of the S-matrix also occurs in situations such as when the Hamiltonian can be separated into degrees of freedom of center of mass and relative variables, or the spin and spatial coordinates~\cite{Taylor:1972}.
 
The decomposition relation, Eq.~(\ref{E:Decomposition}), follows naturally as a consequence of the factoring of the S-matrix. For the one-photon case, we have
\begin{align}
\mathbf{S}|i\rangle &= \mathbb{S}_e  \mathbb{S}_o |i\rangle_e + \mathbb{S}_e  \mathbb{S}_o |i\rangle_o\notag\\
&\equiv \mathbf{S}_e |i\rangle_e + \mathbf{S}_o |i\rangle_o,
\end{align}where $\mathbf{S}_e$ and $\mathbf{S}_o$ take the form as in Eq.~(\ref{E:SMatrixConstructed}), with $|i\rangle$ restricted to one-photon ``$e$'' and ``$o$'' subspaces. In this derivation, we have used $\mathbb{S}_e |i\rangle_o = |i\rangle_o$, and $\mathbb{S}_o |i\rangle_e = |i\rangle_e$. Also, for our case, $\mathbb{S}_o = \mathbf{S}_o =\mathbf{1}$, as can be seen from Eq.~(\ref{E:SDefAlternativeeo}), since $H_o=H_0^o$.

For the two-photon case, we have
\begin{align}
\mathbf{S}|i\rangle &=  \mathbb{S}_e  \mathbb{S}_o |i\rangle_{ee} +  \mathbb{S}_e  \mathbb{S}_o|i\rangle_{eo} +  \mathbb{S}_e  \mathbb{S}_o|i\rangle_{oo}\notag\\
&= \mathbb{S}_e  |i\rangle_{ee} +\mathbb{S}_e \mathbb{S}_o |i\rangle_{eo} +  \mathbb{S}_o |i\rangle_{oo}\notag\\
&\equiv \mathbf{S}_{ee} |i\rangle_{ee} + \mathbf{S}_{eo} |i\rangle_{eo}+ \mathbf{S}_{oo}|i\rangle_{oo},
\end{align}where, again, both $\mathbf{S}_{ee}$, $\mathbf{S}_{eo} = \mathbf{S}_e \mathbf{S}_o$, can be calculated using Eqs.~(\ref{E:LippmannR}) -- (\ref{E:SMatrixConstructed}), and $\mathbf{S}_{oo}=\mathbf{1}$. 

With the decomposition relation established, we now concentrate on constructing 
$\mathbf{S}_e$ and $\mathbf{S}_{ee}$ for  the non-tivial $H_e$ in next two sections. And finally, we will use the solutions of $H_e$ to construct the two-mode two-photon scattering solutions of $H = H_e + H_o$ in Sec.~\ref{P:TwoPhotonII}.

\section{One photon S-matrix: $S_e$}\label{P:OnePhoton}

As a preparation of the two-photon solution, we first briefly summarize the one-photon solution for the Hamiltonian $H_e$, which is needed for constructing the two-photon S-matrix later. 

One could readily check that the full interacting one-photon eigenstate for $H_e$ takes the form~\cite{Shen:2005,Shen:2005a} 
\begin{align}\label{E:kPlus}
|k^+\rangle_{e} &\equiv \left\{\int dx \left[\frac{e^{i k x}}{\sqrt{2\pi}} \left(\theta(-x)+ t_k \theta(x)\right)c_{e}^{\dagger}(x)\right] + e_k \sigma_{+}\right\} |\emptyset, -\rangle\notag\\
&\equiv \int dx \,\phi(x) c_{e}^{\dagger}(x) |\emptyset, -\rangle +  e_k \sigma_{+} |\emptyset, -\rangle,
\end{align}where 
\begin{equation}
t_k \equiv\frac{k - \Omega - i \Gamma/2}{k -\Omega + i \Gamma/2}.
\end{equation}
The single photon thus experiences resonance when its energy $k$ is close to the transition energy $\Omega$ of the atom. $\Gamma\equiv V^2$ characterizes the width of the resonance and is related to the spontaneous emission lifetime of the atom, and $e_k =\frac{1}{\sqrt{2\pi}} \frac{V}{k-\Omega+i \Gamma/2}$ is the excitation amplitude. $|\emptyset, -\rangle$ is the state where there is no photon, and the atom is in the ground state.  

The normalized in-state $|k\rangle_e$ and the out-state $|f_k\rangle_e$, constructed from $|k^+\rangle_{e}$, as shown in Appendix~\ref{A:ReadOffOnePhoton}, are
\begin{align}\label{E:OnePhotonInOut}
|k\rangle_e &\equiv \int dx \,\phi^k_i(x) c_e^{\dagger}(x)|\emptyset, -\rangle,\notag\\
|f_k\rangle_e &\equiv \int dx \,\phi^k_f(x) c_e^{\dagger}(x)|\emptyset, -\rangle,
\end{align}with
\begin{align}
\phi^k_i(x) &= \phantom{}_e\langle x, -|f_k\rangle_e = \frac{1}{\sqrt{2\pi}}e^{i k x},\notag\\
\phi^k_f(x) &= \phantom{}_e\langle x, -|f_k\rangle_e = t_k  \left(\frac{1}{\sqrt{2\pi}}e^{i k x}\right) = t_k \phi_i (x),
\end{align}for \emph{all} $x$. The normalization condition is
\begin{equation}
\phantom{}_e\langle k|k'\rangle_{e} = \delta(k-k').
\end{equation}
This demonstrates that one can ``read off'' the in-state one-photon wavefunction $\phi_i(x)$, and the out-state one-photon wavefunctions, $\phi_f(x)$, respectively, from the ``incoming'' ($x<0$) and the ``outgoing'' ($x>0$) part of the photon wavefunction, $\phi(x)$. 

Since the set $\{|k\rangle_e\}$ forms a complete set in the ``$e$'' subspace, the one-photon S-matrix in the ``$e$'' subspace ($\equiv \mathbf{S}_e$) therefore is
\begin{align}\label{E:Se}
\mathbf{S}_e&\equiv \sum_{k} |f_k\rangle_{e}\,\phantom{}_{e}\langle k|\notag\\
&= \sum_{k} t_k |k\rangle_{e}\,\phantom{}_{e}\langle k|\notag\\
&= \sum_{k} \iint dx dx' \, \left(t_k \phi^k_i(x) {\phi^k_{i}}^{*}(x')\right) \left[c_{e}^{\dagger}(x)|\emptyset, -\rangle\langle \emptyset, -| c_{e}(x')\right].
\end{align}

For any two one-particle states
\begin{align}
|\varphi\rangle_{e} \equiv \int dx\, \varphi (x) c_{e}^{\dagger}(x)|\emptyset, -\rangle,\notag\\
|\chi\rangle_{e} \equiv \int dx\, \chi(x) c_{e}^{\dagger}(x)|\emptyset, -\rangle,
\end{align}the transition amplitude is thereby given by
\begin{align}
\phantom{}_{e}\langle \chi|\mathbf{S}_e|\varphi\rangle_{e} &= \sum_{k} t_k\, \phantom{}_{e}\langle \chi|k\rangle_{e}\phantom{}_{e}\langle k|\varphi\rangle_{e}\notag\\
&=\sum_{k} \iint dx dx' \, t_k \chi^*(x)\phi^k_i(x){\phi^k_{i}}^*(x')\varphi(x').
\end{align} 

We now proceed to solve for the one-photon properties for the two-mode Hamiltonian $H$. Since $H_o$ describes free propagating photons, the one-mode one-photon S-matrix in the ``$o$'' sector is simply an identity operator in the $o$ subspace, \emph{i.e.},
\begin{equation}\label{E:So}
\mathbf{S}_o = \mathbf{1} =\sum_{k}|k\rangle_o\,\phantom{}_o\langle k|.
\end{equation} According to the decomposition relation, Eq.~(\ref{E:Decomposition}), the two-mode one-photon S-matrix is
\begin{equation}
\mathbf{S}=\sum_{k}t_k |k\rangle_e\,\phantom{}_e\langle k| + \sum_{k}|k\rangle_o\,\phantom{}_o\langle k|.
\end{equation}

For a normalized in-state
\begin{align}
|k'\rangle_{R} &\equiv \int dx\, \frac{e^{i k' x}}{\sqrt{2\pi}}\, c^{\dagger}_{R}(x)|0\rangle\notag\\
&=\int dx\, \frac{e^{i k' x}}{\sqrt{2\pi}}\,\frac{1}{\sqrt{2}}c^{\dagger}_{e}(x)|0\rangle+\int dx\,\frac{e^{i k' x}}{\sqrt{2\pi}}\,\frac{1}{\sqrt{2}}c^{\dagger}_{o}(x)|0\rangle\notag\\
&= \frac{1}{\sqrt{2}}|k'\rangle_e + \frac{1}{\sqrt{2}}|k'\rangle_o,
\end{align}the out-state is
\begin{align}
\mathbf{S}|k'\rangle_{R} &=  \frac{1}{\sqrt{2}}\left(\mathbf{S}_e |k'\rangle_e + \mathbf{S}_o |k'\rangle_o\right)\notag\\
&=\frac{1}{\sqrt{2}}\left(\sum_{k}t_k |k\rangle_e\,\phantom{}_e\langle k|k'\rangle_e + \sum_{k}|k\rangle_o\,\phantom{}_o\langle k|k'\rangle_o\right)\notag\\
&=\frac{1}{\sqrt{2}}\left( t_{k'} |k'\rangle_e + |k'\rangle_o\right)\notag\\
&=\frac{1}{2}(t_{k'} +1) | k'\rangle_R + \frac{1}{2}(t_{k'} -1) |-k'\rangle_L\notag\\
&\equiv \bar{t}_{k'} | k'\rangle_R + \bar{r}_{k'} | -k'\rangle_L,
\end{align}
where the two-mode transmission amplitude $\bar{t}_{k'}$ and reflection amplitude $\bar{r}_{k'}$ are
\begin{align}\label{E:TwoModetr}
\bar{t}_{k'} &= \frac{1}{2}(t_{k'} +1) = \frac{k'-\Omega}{k'-\Omega + i \Gamma/2},\notag\\
\bar{r}_{k'} &= \frac{1}{2}(t_{k'} -1) = \frac{-i\Gamma/2}{k'-\Omega + i \Gamma/2},
\end{align}respectively, in agreement with previous calculations~\cite{Shen:2005, Shen:2005a}. In the derivations, we have used 
\begin{align}
|k'\rangle_e &= \int dx\, \frac{e^{i k' x}}{\sqrt{2\pi}}\,c^{\dagger}_{e}(x)|\emptyset, -\rangle\notag\\
 &=  \int dx\, \frac{e^{i k' x}}{\sqrt{2\pi}}\, \frac{1}{\sqrt{2}}\left(c^{\dagger}_{R}(x)+ c^{\dagger}_{L}(-x)\right)|\emptyset, -\rangle\notag\\
 & = \frac{1}{\sqrt{2}}\left(|k'\rangle_R + |-k'\rangle_L\right)\notag\\
 |k'\rangle_o &=\frac{1}{\sqrt{2}}\left(|k'\rangle_R - |-k'\rangle_L\right),
\end{align} and $\Gamma = V^2 = 2\bar{V}^2$.

\begin{figure}[thb]
\scalebox{1}{\includegraphics{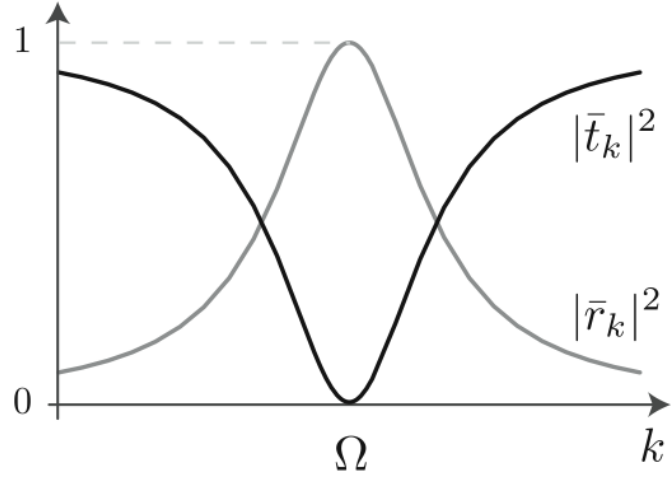}}
\caption{The single-photon transmission and the reflection spectrum. $|\bar{t}_k|^2$ is indicated by the black curve, and $|\bar{r}_k|^2$ is denoted by the gray curve. The full width at half maximum for $|\bar{r}_k|^2$ is $\Gamma$.}\label{Fi:TwoModeTR}
\end{figure}

Fig.~\ref{Fi:TwoModeTR} plots the transmission and reflection spectrum. On resonance ($k=\Omega$), $|\bar{t}_k|^2$ is 0, while $|\bar{r}_k|^2$ is 1, and the particle is $100\%$ reflected. Note that the effect of the spontaneous emission of the two-level system is explicitly included. In the one-dimensional geometry, the spontaneous emission therefore does not represent a loss mechanism, and is used here to control the coherent transport property of a single photon.


\section{Two-photon case: Constructing $\mathbf{S}_{ee}$}\label{P:TwoPhotonI}

Having solved the one-photon case, we now proceed to construct the two-photon S-matrix of the two-mode Hamiltonian $H$. Following the discussions in Eq.~(\ref{E:Decomposition}) in Sec.~\ref{P:SRelation}, the key to this is to construct the two-photon S-matrix, $\mathbf{S}_{ee}$, for $H_e$, which we will undertake in this section. 

Before we set out to construct the S-matrix, $\mathbf{S}_{ee}$, we comment on some of the general aspects of two-photon problem that would be useful for this effort. As emphasized before in Sec.~\ref{A:Lippmann-Schwinger}, the $\mathbf{S}_{ee}$ matrix is a mapping in the free two-photon Hilbert space. This Hilbert space, in its real space representation, consists of all symmetric functions of the coordinates of the photons $x_1$, $x_2$, and is spanned by a complete basis $\{|S_{k,p}\rangle_{ee}: k\leq p\}$ defined as
\begin{equation}
|S_{k,p}\rangle_{ee}  \equiv \iint dx_1 dx_2\, S_{k, p}(x_1, x_2) \frac{1}{\sqrt{2}}c_{e}^{\dagger}(x_1) c_{e}^{\dagger}(x_2)|\emptyset,-\rangle,
\end{equation}with
\begin{align}\label{E:DefSkp}
\phantom{}_{ee}\langle x_1, x_2|S_{k, p}\rangle_{ee} &= S_{k, p}(x_1, x_2)\notag\\
&\equiv \frac{1}{2\pi}\frac{1}{\sqrt{2}}\left(e^{i k x_1} e^{i p x_2} +e^{i k x_2} e^{i p x_1}\right)\notag\\
&= \frac{\sqrt{2}}{2\pi} e^{i E x_c}\cos\left(\Delta x\right), 
\end{align}where 
\begin{equation}
E=k+p
\end{equation} is the total energy of the photon pair, 
\begin{align}
x_c &\equiv (x_1 + x_2)/2,\notag\\
x &\equiv x_1 - x_2,
\end{align}are the center of mass coordinate and the relative coordinate, respectively. 
\begin{equation}
\Delta \equiv (k - E/2) = (k-p)/2,
\end{equation}measures the energy difference between two photons. The completeness of $\{|S_{k,p}\rangle_{ee}: k\leq p\}$ is expressed by 
\begin{align}
\mathbf{1} &=\int_{-\infty}^{\infty}dp\int_{-\infty}^{p}dk |S_{k, p}\rangle_{ee}\, \phantom{}_{ee}\langle S_{k, p}|\notag\\
& = \frac{1}{2}\int_{-\infty}^{\infty}dp\int_{-\infty}^{+\infty}dk |S_{k, p}\rangle_{ee}\, \phantom{}_{ee}\ \langle S_{k, p}|,
\end{align}using the fact that $|S_{p, k}\rangle = |S_{k, p}\rangle$. As a side note, in computations related to the two-photon Hilbert space below, we will adopt two equivalent set of variables: $(k, p)$, with $k \leq p$,  and $(\Delta, E)$, with $\Delta \equiv (k-p)/2$ and $E \equiv k+p$, and use the  two sets of variables interchangeably. The Jacobian between the two sets is 1. Thus,
\begin{equation}\label{E:VariablesChange}
\int_{-\infty}^{\infty}dp\int_{-\infty}^{p}dk=\int_{-\infty}^{\infty}dE\int_{-\infty}^{\frac{E}{2}}dk = \int_{-\infty}^{\infty}dE\int_{-\infty}^{0}d\Delta.
\end{equation}

Alternatively, the \emph{same} Hilbert space can instead be spanned by another basis $\{|A_{k,p}\rangle_{ee}: k\leq p\}$ defined as
\begin{equation}
|A_{k,p}\rangle_{ee} \equiv \iint dx_1 dx_2\, A_{k, p}(x_1, x_2)\frac{1}{\sqrt{2}}c_{e}^{\dagger}(x_1) c_{e}^{\dagger}(x_2)|\emptyset,-\rangle,
\end{equation}with
\begin{align}\label{E:DefAkp}
\phantom{}_{ee}\langle x_1, x_2|A_{k,p}\rangle_{ee} &= A_{k, p}(x_1, x_2)\notag\\
&\equiv \frac{1}{2\pi}\frac{1}{\sqrt{2}}\,\mbox{sgn}(x)\left(e^{i k x_1} e^{i p x_2} -e^{i k x_2} e^{i p x_1}\right)\notag\\
&= \frac{\sqrt{2}i}{2\pi}\,\mbox{sgn}(x) \, e^{i E x_c}\sin\left(\Delta x\right),
\end{align} where $\mbox{sgn}(x)\equiv \theta(x)-\theta(-x)$ is the sign function.  By definition, one has $|A_{p, k}\rangle_{ee} = -|A_{k, p}\rangle_{ee}$. We emphasize that, while both $\{|S_{k,p}\rangle_{ee}: k\leq p\}$ and $\{|A_{k,p}\rangle_{ee}: k\leq p\}$ are complete~\cite{Schulz:1982}, arbitrary linear combination $\{ a_{k, p} |S_{k,p}\rangle_{ee} + b_{k, p} |A_{k,p}\rangle_{ee}: k\leq p\}$ may not be. The properties of the two complete sets $\{|S_{k,p}\rangle_{ee}: k\leq p\}$ and $\{|A_{k,p}\rangle_{ee}: k\leq p\}$ are summarized in Appendix~\ref{A:Overlap}. Here we only emphasize that the two states $|S_{k_1,p_1}\rangle _{ee} $ and $|A_{k_2,p_2}\rangle_{ee}$ are \emph{not} orthogonal to each other.

We now proceed to construct $\mathbf{S}_{ee}$ as follows: we start in Sec.~\ref{P:EquationOfMotion} by deriving the real-space equations of motion from the Schr\"{o}dinger equation $H_e |\Phi\rangle = E |\Phi\rangle$. In Sec.~\ref{P:HeEigenstates}, we then solve the real-space equations of motion using the standard Bethe-ansatz approach to obtain a class of eigenstates of the interacting Hamiltonian $H_e$. In Sec.~\ref{P:WiegmannState} we obtain the corresponding ``in-'' and ``out-''states from the interacting eigenstates using the Lippmann-Schwinger formalism discussed in Sec.~\ref{A:Lippmann-Schwinger}. Through a completeness check,  we show that the in-states thus obtained are in fact not complete. Instead, in order to span the two-photon free Hilbert space, one must supplement it with another class of states: a two-photon bound state. In Sec.~\ref{P:BoundState} we show that the two-photon bound state is also an eigenstate of $\mathbf{S}_{ee}$. This, together with the completeness check upon the two classes of solutions, allow us to determine the exact form of $\mathbf{S}_{ee}$. Finally, in Sec.~\ref{P:S}, we discuss some of the properties of $\mathbf{S}_{ee}$.

\subsection{Equations of motion and boundary conditions in real space}\label{P:EquationOfMotion}

An  eigenstate for $H_e$ has the general form:
\begin{equation}\label{E:TwoPhotonGeneral}
|i^+\rangle \equiv |\Phi\rangle \equiv\left(\int dx_1 dx_2 \, g(x_1, x_2) \frac{1}{\sqrt{2}}c_{e}^{\dagger}(x_1) c_{e}^{\dagger}(x_2) + \int dx \, e(x) c_{e}^{\dagger}(x) \sigma_{+}\right)|\emptyset, -\rangle,
\end{equation}
where $e(x)$ is the probability amplitude distribution of one-photon while the atom in the excited state.
Due to the boson statistics, the wavefunction satisfies $g(x_1, x_2) = +g(x_2, x_1)$, and is continuous on the line $x_1 = x_2$.


From $H_e |\Phi\rangle = E |\Phi\rangle$, by equating the coefficients of $c_{e}^{\dagger}c_{e}^{\dagger}|\emptyset, -\rangle$ and $c_{e}^{\dagger}\sigma_{+}|\emptyset, -\rangle$, respectively, we obtain the equations of motion:
\begin{subequations}
\begin{equation}
\left(-i \frac{\partial}{\partial x_1}  -i \frac{\partial}{\partial x_2} - E\right) g(x_1, x_2) + \frac{V}{\sqrt{2}} \left(e(x_1) \delta(x_2) + e(x_2) \delta(x_1)\right) =0,
\end{equation}
\begin{equation}\label{E:eEnsure}
\left(-i \frac{\partial}{\partial x} - (E - \Omega)\right) e(x) + \frac{V}{\sqrt{2}} \left(g(0, x) + g(x,0)\right) = 0,
\end{equation}    
\end{subequations}
where $g(0, x) \equiv 1/2 \times(g(0^-, x) + g(0^+, x)) = g(x, 0)  \equiv 1/2 \times(g(x, 0^-) + g(x, 0^+))$. The functions 
$g(x_1, x_2)$ and $e(x)$ are piecewise continuous. 
For any such piecewise continuous function $f(x)$, the derivative of $f(x)$ is the ordinary derivative plus contributions from the jump discontinuities. If $x_0$ is such a discontinuity, we add a term~\cite{Kaplan:1984} $\left[\lim_{x\rightarrow x_0^{+}} f(x) -\lim_{x\rightarrow x_0^{-}} f(x)\right]\delta(x-x_0)$.

The interactions occur on the coordinate axes: $x_1=0$, and $x_2=0$. Applying the equations of motions on the boundaries between adjacent quadrants gives the following boundary conditions on the boundary of quadrants II and III ($x_1 < 0$):
\begin{subequations}\label{E:BC1}
\begin{equation}\label{E:BC1One}
-i \left(g(x_1, 0^+) - g(x_1, 0^-)\right)  +\frac{V}{\sqrt{2}} e(x_1)=0,
\end{equation}
\begin{equation}\label{E:BC1Two}
\left(-i\frac{\partial}{\partial x_1}-(E-\Omega)\right) e(x_1) +  \frac{V}{\sqrt{2}}(g(x_1, 0^+) + g(x_1, 0^-))=0,
\end{equation}
\end{subequations}and on the boundary of quadrants II and I ($x_2 > 0$):
\begin{subequations}\label{E:BC2}
\begin{equation}
-i \left(g(0^+, x_2) - g(0^-, x_2)\right)  +\frac{V}{\sqrt{2}} e(x_2)=0,
\end{equation}
\begin{equation}
\left(-i\frac{\partial}{\partial x_2}-(E-\Omega)\right) e(x_2) +  \frac{V}{\sqrt{2}}(g(0^+, x_2) + g(0^-, x_2))=0.
\end{equation}
\end{subequations}These boundary conditions must be supplemented by a further condition\begin{equation}\label{E:YangBaxter}
e(0^-) = e(0^+),
\end{equation}which arises directly from Eq.~({\ref{E:eEnsure}}) and ensures the self-consistency. When there are more than two photons, this condition gives rise to the Yang-Baxter relation~\cite{Yang:1967, Wiegmann:1983a}. 

\begin{figure}[thb]
\scalebox{1}{\includegraphics{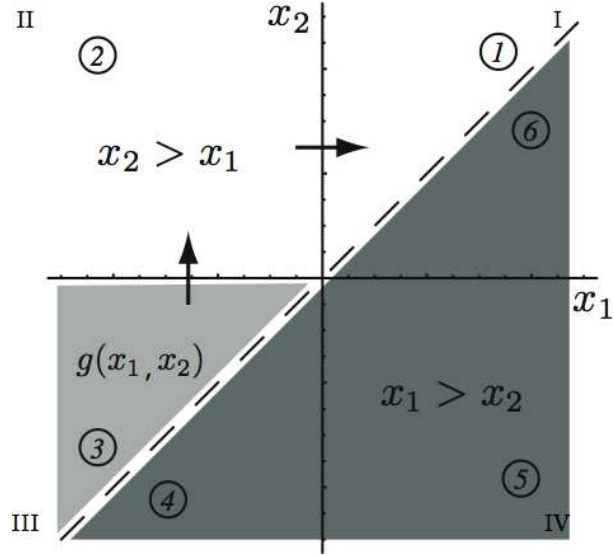}}
\caption{The $x_1$-axis, $x_2$-axis, and $x_1 = x_2$ dissect the $x_1$-$x_2$ coordinate plane into six regions, labeled by the numbers in circle. When given $g(x_1, x_2)$ in region 3 (lightly-shaded area), the boundary condition is imposed to obtain $g(x_1, x_2)$ in other regions, as denoted by the arrows.  $g(x_1, x_2)$ in $x_1 \geq x_2$ region (darkly-shaded area) is obtained from $g(x_1, x_2)$ in $x_2 \geq x_1$ region by the boson statistics. 
}\label{Fi:Coord_Plane}
\end{figure}

The $x_1$-axis, $x_2$-axis and the line $x_1=x_2$ dissect the $x_1$-$x_2$ plane into six regions (Fig.~\ref{Fi:Coord_Plane}). When $g(x_1, x_2)$ is given in either one of the six regions, one could use the boundary conditions to obtain $g(x_1, x_2)$ in all other regions.  

For example, at the boundary between quadrant II and III, using Eq.~(\ref{E:BC1One}), we have 
\begin{equation}\label{E:Eq1}
g(x_1, 0^+) = g(x_1, 0^-) - i \frac{V}{\sqrt{2}} e(x_1).
\end{equation}
Substitute this into Eq.~(\ref{E:BC1Two}), one obtains
\begin{equation}\label{E:ODE}
-i \frac{\partial}{\partial x_1} e(x_1) = \left((E-\Omega) + i \frac{\Gamma}{2}\right) e(x_1)- \sqrt{2} V g(x_1, 0^-),
\end{equation}which has the solution
\begin{equation}\label{E:ODESolution}
e(x_1)  = c \,e^{+i \left((E-\Omega) + i \frac{\Gamma}{2}\right) x_1} + i e^{+i \left((E-\Omega) + i \frac{\Gamma}{2}\right) x_1}\int_{-\infty}^{x_1} e^{-i \left((E-\Omega) + i \frac{\Gamma}{2}\right) x} \left(-\sqrt{2}V g(x, 0^-)\right) dx,
\end{equation}where $c$ is an integration constant. By requiring $e(x)$ to be zero when the coupling strength $V$ is zero, we have $c=0$. Hence
\begin{equation}\label{E:ODESolution2}
e(x_1)  =  i e^{+i \left((E-\Omega) + i \frac{\Gamma}{2}\right) x_1}\int_{-\infty}^{x_1} e^{-i \left((E-\Omega) + i \frac{\Gamma}{2}\right) x} \left(-\sqrt{2}V g(x, 0^-)\right) dx.
\end{equation}



\subsection{Constructing eigenstates of $H_e$ using Bethe ansatz}\label{P:HeEigenstates}

We now solve the two-photon equations of motions of $H_e$ in Sec.~\ref{P:EquationOfMotion} using the Bethe ansatz. The Bethe ansatz usually postulates that the eigenstates are superpositions of a few \emph{extended plane waves} when all particles are away from the impurity~\cite{Bethe:1931, Batchelor:2007}.  We shall also call these solutions of $H_e$ the Wiegmann-Andrei states, after the two authors who worked out similar solutions for the Kondo model~\cite{Andrei:1980, Wiegmann:1980} and for the Anderson model~\cite{Wiegmann:1983a}. 

For the two-particle case, the Bethe ansatz postulates that in regions 1, 2, and 3 (Fig.~\ref{Fi:Coord_Plane}), the two-photon wavefunction has the form
\begin{equation}\label{E:g}
g(x_1, x_2)=
	\begin{cases}
	B_3 e^{i k x_1 + i p x_2} + A_3 e^{i p x_1 + i k x_2},		&\text{in region 3 ($x_1< x_2 <0$);}\\
	B_2 e^{i k x_1 + i p x_2} + A_2 e^{i p x_1 + i k x_2},		&\text{in region 2 ($x_1<0, x_2>0$);}\\
	B_1 e^{i k x_1 + i p x_2} + A_1 e^{i p x_1 + i k x_2},		&\text{in region 1 ($x_2>x_1>0$).}
	\end{cases}
\end{equation}The wavefunction in other regions is defined by boson symmetry. The goal of the computations, based upon the Bethe ansatz, is then to \emph{check} that such a form indeed satisfies the appropriate equations of motion, and in the process of checking, to determine all constraints relating the $A$'s and $B$'s coefficients. Since by construction, $g(x_1, x_2)$ already satisfies the equations of motion in regions 1, 2, and 3:
\begin{equation}
\left(-i \frac{\partial}{\partial x_1}  -i \frac{\partial}{\partial x_2} - E\right) g(x_1, x_2) =0
\end{equation} for $x_1 \neq 0$ and $x_2\neq 0$, all we need is to use the boundary conditions [Eq.~(\ref{E:BC1}) and (\ref{E:BC2})] and the self-consistency condition [Eq.~(\ref{E:YangBaxter})] to determine the constraints on $A$'s and $B$'s.

At the boundary between quadrant II and III, since in region 3  ($x_1<x_2 <0$), $g(x_1, x_2)= B_3 e^{i k x_1 + i p x_2} + A_3 e^{i p x_1 + i k x_2}$, one then has
\begin{equation}
g(x_1, 0^-) = B_3 e^{i k x_1} + A_3 e^{i p x_1}.
\end{equation}
Therefore, using Eq.~(\ref{E:ODESolution2}), we have, for $x < 0$,
\begin{equation}\label{E:RegionIe}
e(x) =  \sqrt{2} V \left(\frac{B_3 e^{i k x}}{p-\Omega + i \Gamma/2}+\frac{A_3 e^{i p x}}{k-\Omega  + i \Gamma/2}\right).
\end{equation}

Plugging $e(x)$ to Eq.~(\ref{E:Eq1}), we obtain
\begin{align}
g(x_1, 0^+) &= g(x_1, 0^-) - i \frac{V}{\sqrt{2}} e(x_1)\notag\\
& = B_3 e^{i k x_1} + A_3 e^{i p x_1}  - i \frac{V}{\sqrt{2}} (\sqrt{2} V) \left(\frac{B_3 e^{i k x_1}}{p-\Omega + i \Gamma/2}+\frac{A_3 e^{i p x_1}}{k-\Omega  + i \Gamma/2}\right)\notag\\
&= B_3 e^{i k x_1}\frac{p-\Omega - i \Gamma/2}{p-\Omega + i \Gamma/2} + A_3 e^{i p x_1}\frac{k-\Omega - i \Gamma/2}{k-\Omega + i \Gamma/2}\notag\\
&= t_p B_3 e^{i k x_1} + t_k A_3 e^{i p x_1},
\end{align}
and therefore, in the whole quadrant II ($x_1<0$, $x_2>0$), using the Bethe ansatz form of $g(x_1, x_2)$ [Eq.~(\ref{E:g})], we have
\begin{equation}
B_2 = t_p B_3; \quad A_2 = t_k A_3.
\end{equation}
One can understand this expression by realizing that when going from quadrant III to quadrant II, $x_1$ is unchanged, while $x_2: 0^- \rightarrow 0^+$. Consequently the part of the wave function $B_3 e^{i k x_1 +  i p x_2}$ acquires a transmission coefficient $t_p$, and the part of the wave function $A_3 e^{i p x_1 +  i k x_2}$ acquires a transmission coefficient $t_k$.

In addition, from the expression of $e(x <0 )$ [Eq.~(\ref{E:RegionIe})], we have 
\begin{equation}\label{E:eZeroM}
e(0^-) =  \sqrt{2} V \left(\frac{B_3}{p-\Omega + i \Gamma/2}+\frac{A_3}{k-\Omega  + i \Gamma/2}\right).
\end{equation}

We apply the same procedures to the next boundary. The boundary conditions on the boundary of quadrants II and I ($x_2 > 0$) are (reproduced here from Eq.~(\ref{E:BC2})):
\begin{align*}
-i \left(g(0^+, x_2) - g(0^-, x_2)\right)  &+\frac{V}{\sqrt{2}} e(x_2)=0,\notag\\
\left(-i\frac{\partial}{\partial x_2}-(E-\Omega)\right) e(x_2) &+  \frac{V}{\sqrt{2}}(g(0^+, x_2) + g(0^-, x_2))=0.
\end{align*}
As previously, from the first equation, we have 
\begin{equation}
g(0^+, x_2) = g(0^-, x_2) - i \frac{V}{\sqrt{2}} e(x_2).
\end{equation}
Substitute into the second equation, we obtain
\begin{equation}
-i \frac{\partial}{\partial x_2} e(x_2) = \left((E-\Omega) + i \frac{\Gamma}{2}\right) e(x_2)- \sqrt{2} V g(0^-, x_2).
\end{equation}
Since
\begin{equation}
g(0^-, x_2) = t_p B_3 e^{i p x_2} + t_k A_3 e^{i k x_1}, 
\end{equation}
we have, for $x >0$
\begin{equation}\label{E:exLarge0}
e(x)=  \sqrt{2} V \left(\frac{t_p B_3 e^{i p x}}{k-\Omega + i \Gamma/2}+\frac{t_k A_3 e^{i k x}}{p-\Omega  + i \Gamma/2}\right),
\end{equation}
and
\begin{align}
g(0^{+}, x_2) &= g(0^-, x_2) - i \frac{V}{\sqrt{2}} e(x_2)\notag\\
&=\left(t_p B_3 e^{i p x_2} + t_k A_3 e^{i k x_1}\right) + (-i \Gamma)\left(\frac{t_p B_3 e^{i p x_2}}{k-\Omega + i \Gamma/2}+\frac{t_k A_3 e^{i k x_2}}{p-\Omega  + i \Gamma/2}\right)\notag\\
&=\left(1-\frac{i \Gamma}{k-\Omega + i\Gamma/2}\right) t_p B_3 e^{i p x_2} + \left(1-\frac{i \Gamma}{p-\Omega + i\Gamma/2}\right) t_k B_3 e^{i k x_2}\notag\\
&= t_p t_k \left(B_3 e^{i p x_2} + A_3 e^{ i k x_2}\right).
\end{align}
Therefore, in region I ($x_2 > x_1$ region of quadrant I), using the Bethe ansatz again,
\begin{equation}
B_1 = t_p t_k B_3;\quad A_1 =t_p t_k A_3.
\end{equation}
One can understand this expression by realizing that when going from quadrant II to quadrant I, $x_2$ is unchanged, while $x_1: 0^- \rightarrow 0^+$. Consequently the part of the wave function $t_p B_3 e^{i k x_1 +  i p x_2}$ acquires a transmission coefficient $t_k$, and the part of the wave function $t_k A_3 e^{i p x_1 +  i k x_2}$ acquires a transmission coefficient $t_p$. 

Also, from the expression of $e(x>0)$ [Eq.~(\ref{E:exLarge0})], one has
\begin{equation}\label{E:eZeroP}
e(0^+) =  \sqrt{2} V \left(\frac{t_p B_3}{k-\Omega + i \Gamma/2}+\frac{t_k A_3}{p-\Omega  + i \Gamma/2}\right).
\end{equation}
Combining Eqs.~(\ref{E:eZeroM}) and (\ref{E:eZeroP}), together with the self-consistency condition $e(0^-) = e(0^+)$ [Eq.~(\ref{E:YangBaxter})], we can determine the ratio of $B_3/A_3$ from
\begin{equation}\label{E:NotS}
\left(\frac{B_3}{p-\Omega + i \Gamma/2}+\frac{A_3}{k-\Omega  + i \Gamma/2}\right) =  \left(\frac{t_p B_3}{k-\Omega + i \Gamma/2}+\frac{t_k A_3}{p-\Omega  + i \Gamma/2}\right),
\end{equation}
which simplifies to
\begin{equation}\label{E:ABRatio}
\frac{B_3}{A_3} = \frac{k-p - i \Gamma}{k -p + i \Gamma}.
\end{equation}
As can be seen from Eq.~(\ref{E:TwoPhotonGeneral}), the two-photon wavefunction $g(x_1, x_2)$ and the amplitude $e(x)$ completely determine the interacting eigenstate of $H_e$. Fig.~\ref{Fi:Extended} summarizes the two-photon wavefunction $g(x_1, x_2)$ in the entire $x_1$-$x_2$ plane, as well as $e(x)$ for all $x$.


We now need to extract the information of the in- and out-states from the interacting eigenstate $|i^+\rangle$. As shown in Appendix~\ref{A:ReadOff}, the in-state $|i\rangle$ and the out-state $|f_i\rangle$ are
\begin{align}\label{E:InStateAsWiegmann}
|i\rangle & \equiv \int dx_1 dx_2\, g_i(x_1, x_2) \frac{1}{\sqrt{2}}c_{e}^{\dagger}(x_1)c_{e}^{\dagger}(x_2)|\emptyset, -\rangle,\notag\\
|f_i\rangle & \equiv \int dx_1 dx_2\, g_f(x_1, x_2) \frac{1}{\sqrt{2}} c_{e}^{\dagger}(x_1)c_{e}^{\dagger}(x_2)|\emptyset, -\rangle,
\end{align}where, for \emph{all} $x_1$ and $x_2$,
\begin{align}\label{E:InAsWiegmann}
g_i(x_1, x_2) & = g(x_1<0, x_2<0) \quad\mbox{[$g(x_1, x_2)$ in quadrant III]},\notag\\
g_f(x_1, x_2) & = g(x_1>0, x_2>0) \quad\mbox{[$g(x_1, x_2)$ in quadrant I]}.
\end{align}Note this result is consistent with the intuitive notion that the in-state is the ``incoming'' part, \emph{i.e.}, the $x_1, x_2<0$ region, of the full interacting state; while the out-state is the ``outgoing'' part,  \emph{i.e.}, the $x_1, x_2>0$ region, of the full interacting state. 

The in- and out-states, when explicitly spelled out in real-space, have non-trivial structures. $g(x_1, x_2)$ in the \emph{full} quadrant III ($x_1, x_2 <0$) is,
\begin{align}
g(x_1, x_2) &=  \left(A_3 e^{i k x_1 + i p x_2} + B_3 e^{i p x_1 + i k x_2}\right) \theta(x_1 - x_2) + \left(B_3 e^{i k x_1 + i p x_2} + A_3 e^{i p x_1 + i k x_2}\right) \theta(x_2 - x_1)\notag\\
&\propto \left[(k-p+i \Gamma)e^{i k x_1 + i p x_2} + (k -p - i \Gamma)e^{i p x_1 + i k x_2} \right] \theta(x_1 - x_2)\notag\\ &+ \left[(k-p-i \Gamma)e^{i k x_1 + i p x_2} + (k -p + i \Gamma)e^{i p x_1 + i k x_2} \right] \theta(x_2 - x_1)\notag\\
&=(k-p)\left(e^{i k x_1+i p x_2} +e^{i k x_2+i p x_1}\right) + i \Gamma \left(e^{i k x_1+i p x_2} -e^{i k x_2+i p x_1}\right) \mbox{sgn}(x_1-x_2)\notag\\
&\propto (k-p)\, S_{k, p}(x_1, x_2) + i \Gamma A_{k, p}(x_1, x_2), 
\end{align}where $S_{k, p}(x_1, x_2)$ and $A_{k, p}(x_1, x_2)$ are defined in Eq.~(\ref{E:DefSkp}) and (\ref{E:DefAkp}), respectively. Therefore, in the \emph{entire} $x_1$-$x_2$ plane, the in-state photon wavefunction $g_i(x_1, x_2)$ is
\begin{equation}\label{E:ExtendedEntire}
g_i(x_1, x_2) \propto (k-p)\,S_{k, p}(x_1, x_2) + i \Gamma\, A_{k, p}(x_1, x_2). 
\end{equation}
Similarly, in the \emph{entire} $x_1$-$x_2$ plane, the out-state photon wavefunction $g_f(x_1, x_2)$ is
\begin{equation}\label{E:FinalExtendedEntire}
g_f(x_1, x_2) \propto t_k t_p \left[(k-p)\, S_{k, p}(x_1, x_2)+ i \Gamma\, A_{k, p}(x_1, x_2)\right].
\end{equation}
Note that $g(x_1, x_2)$ is equal to zero when $k = p$ in the entire $x_1$-$x_2$ plane.

\subsection{In- and out-states from the Wiegmann-Andrei state}\label{P:WiegmannState}

Following the discussions of the previous section, we therefore define
\begin{equation}\label{E:WDef}
|\tilde{W}_{k,p}\rangle_{ee}\equiv (k-p)|S_{k,p}\rangle_{ee} + i \Gamma |A_{k,p}\rangle_{ee},
\end{equation}and discuss some of the general properties of $|\tilde{W}_{k, p}\rangle_{ee}$. These states are obviously important for the scattering problems, since they are the eigenstates of the S-matrix, $\mathbf{S}_{ee}$, with eigenvalues $t_k t_p$, as can be seen from Eqs.~(\ref{E:ExtendedEntire}) and (\ref{E:FinalExtendedEntire}). Each state therefore is directly analogous to a so-called ``scattering channel'' in the partial wave expansion~\cite{Taylor:1972, Sakurai:1994}. Below we will normalize these states and show that they are orthogonal to each other (as expected, since they are, after all, eigenstates of the S-matrix with different eigenvalues). Most importantly, and perhaps surprisingly, even though they directly arise from the standard Bethe ansatz approach, they are in fact \emph{incomplete} and thereby can not span the free two-photon Hilbert space. 

From the definition, Eq.~(\ref{E:WDef}), it is clear that 
\begin{subequations}
\begin{equation}\label{E:WProp1}
|\tilde{W}_{k, p}\rangle_{ee} =0, \qquad \mbox{when $k=p$},
\end{equation}
\begin{equation}\label{E:WProp2}
|\tilde{W}_{p, k}\rangle_{ee} = -|\tilde{W}_{k, p}\rangle_{ee}, \qquad \mbox{for any $k$, $p$}.
\end{equation}
\end{subequations}
The normalization and the check for orthogonality is straightforward:
\begin{align}
&\phantom{}_{ee}\langle\tilde{W}_{k_1, p_1}|\tilde{W}_{k_2, p_2}\rangle_{ee}\notag\\
= \, &(k_1 - p_1)(k_2-p_2)\, \phantom{}_{ee}\langle S_{k_1, p_1}|S_{k_2, p_2}\rangle_{ee}+ \Gamma^2 \cdot \phantom{}_{ee}\langle A_{k_1, p_1}|A_{k_2, p_2}\rangle_{ee}\notag\\
&+ (k_1 - p_1) (i \Gamma) \cdot \phantom{}_{ee}\langle S_{k_1, p_1}|A_{k_2, p_2}\rangle_{ee} + (k_2 - p_2) (-i \Gamma) \cdot \phantom{}_{ee}\langle A_{k_1, p_1}|S_{k_2, p_2}\rangle_{ee}\notag\\
= \,& (k_1 - p_1) (k_2 - p_2) \left[\delta(k_1 - k_2) \delta(p_1 - p_2) +\delta(k_1 - p_2) \delta(p_1 - k_2)\right]\notag\\
&+ \Gamma^2 \left[\delta(k_1 - k_2) \delta(p_1 - p_2) -\delta(k_1 - p_2) \delta(p_1 - k_2)\right]\notag\\
=\,& \left[(k_1 - p_1) (k_2 - p_2)+\Gamma^2\right]\delta(k_1 - k_2) \delta(p_1 - p_2)\notag\\
&+ \left[(k_1 - p_1) (k_2 - p_2)-\Gamma^2\right]\delta(k_1 - p_2) \delta(p_1 - k_2),
\end{align}where we have used the overlap between various $|S\rangle_{ee}$ and $|A\rangle_{ee}$ states, as provided in Appendix~\ref{A:Overlap}. One thus is led to the definition of $|W_{k,p}\rangle_{ee}$:
\begin{align}\label{E:WiegmannState}
|W_{k,p}\rangle_{ee}&\equiv \frac{1}{\sqrt{(k-p)^2 +\Gamma^2}}|\tilde{W}_{k,p}\rangle_{ee}=\frac{1}{\sqrt{4\Delta^2 + \Gamma^2}}|\tilde{W}_{k,p}\rangle_{ee}\notag\\
&=\frac{1}{\sqrt{4\Delta^2 + \Gamma^2}}\left(2\Delta |S_{k, p}\rangle_{ee} + i\Gamma|A_{k,p}\rangle_{ee}\right),
\end{align}with $|W_{k,p}\rangle_{ee}$ being normalized to
\begin{equation}\label{E:ExtendedState}
\phantom{}_{ee}\langle W_{k_1, p_1}|W_{k_2, p_2} \rangle_{ee} = \delta(k_1 - k_2) \delta(p_1 - p_2),
\end{equation}when $k_{1} < p_{1}$ and $k_{2} < p_{2}$, or $k_{1} > p_{1}$ and $k_{2} > p_{2}$. For other cases, one has
\begin{equation}\label{E:ExtendedState2}
\phantom{}_{ee}\langle W_{k_1, p_1}|W_{k_2, p_2} \rangle_{ee} = -\delta(k_1 - k_2) \delta(p_1 - p_2),
\end{equation}which arises from Eq.~(\ref{E:WProp2}).

Using the normalized $|W_{k,p}\rangle$, the in-state and out-state thus are 
\begin{align}
|i\rangle &=  |W_{k,p}\rangle_{ee} \equiv \int dx_1 dx_2\,  W_{k,p}(x_1, x_2) \frac{1}{\sqrt{2}} c_e^{\dagger}(x_1)c_e^{\dagger}(x_2)|\emptyset, -\rangle,\notag\\
|f_i\rangle & = t_k t_p |W_{k,p}\rangle_{ee} = \int dx_1 dx_2\, t_k t_p W_{k,p}(x_1, x_2) \frac{1}{\sqrt{2}} c_e^{\dagger}(x_1)c_e^{\dagger}(x_2)|\emptyset, -\rangle,
\end{align}with
\begin{align}
W_{k, p}(x_1, x_2) &= \phantom{}_{ee}\langle x_1, x_2|W_{k, p}\rangle_{ee}\notag\\
&\equiv \frac{1}{\sqrt{(k-p)^2 +\Gamma^2}}\left[(k-p)\,S_{k, p}(x_1, x_2) + i \Gamma\, A_{k, p}(x_1, x_2)\right]\notag\\
&= \frac{\sqrt{2}}{2\pi} e^{i E x_c}\left[2\Delta\cos(\Delta x)-\Gamma\mbox{sgn}(x) \sin(\Delta x) \right]. 
\end{align}

In describing the scattering process, we will need to find all the eigenvalues of the S-matrix. The set of these eigenstates then span the free two-photon Hilbert space. To check whether $\{|W_{k',p'}\rangle_{ee}:  k' \leq p'\}$ is complete, one could start with an arbitrary state, for example, $|S_{k,p}\rangle_{ee}$, project out all $|W_{k',p'}\rangle$ components and calculate
\begin{equation}
|\delta_{k, p}\rangle \equiv |S_{k,p}\rangle_{ee} -\sum_{k' \leq p'} \phantom{}_{ee}\langle W_{k', p'}|S_{k,p}\rangle_{ee} \,|W_{k', p'}\rangle_{ee}.
\end{equation}If the set $\{|W_{k',p'}\rangle_{ee}:  k' \leq p'\}$ were complete, such a computation should yield $|\delta_{k, p}\rangle =0$ for arbitrary $k$ and $p$. This computation is performed in Appendix~\ref{A:Completeness}. Surprisingly, independent of the choice of $|S_{k,p}\rangle_{ee}$, the computation results in $|\delta_{k, p}\rangle \propto |B_{E}\rangle_{ee}$, where
\begin{equation}\label{E:BoundDef}
|B_E\rangle_{ee} \equiv \int dx_1 dx_2\, B_E (x_1, x_2) \frac{1}{\sqrt{2}}c_{e}^{\dagger}(x_1)c_{e}^{\dagger}(x_2)|\emptyset, -\rangle,
\end{equation}with
\begin{equation}\label{E:BoundStateWavefunction}
\phantom{}_{ee}\langle x_1, x_2|B_E\rangle_{ee}=B_E (x_1, x_2) \equiv \frac{\sqrt{\Gamma}}{\sqrt{4\pi}} e^{i E x_c - \frac{\Gamma}{2}|x|}, 
\end{equation}and normalized as
\begin{equation}\label{E:BoundState}
\phantom{}_{ee}\langle B_{E'}|B_{E}\rangle_{ee} = \delta(E-E'),
\end{equation}where $x_c \equiv (x_1 + x_2)/2$, and $x\equiv x_1 - x_2$. The defining feature of $B_E (x_c, x)$ is that, when $x \rightarrow \pm\infty$, $|B_E (x_c, x)| \rightarrow 0$, and therefore $|B_{E}\rangle$ is a \emph{two-photon bound state}.

The set $\{|W_{k,p}\rangle: \forall \,k \leq p\}\bigcup\, \{|B_{E}\rangle\}$ together forms a complete basis of states, and any symmetric functions of $x_1$ and $x_2$ can be expanded using $\{W_{k,p}(x_1, x_2), B_{E}(x_1, x_2)\}$. The completeness of this basis is crucial for discussing the transport properties of scattering problems.

\subsection{Two-Photon bound state is an eigenstate of the S-matrix}\label{P:BoundState}

We now show  that the two-photon bound state is an eigenstate of the S-matrix, with eigenvalue 
\begin{equation}
t_E = \frac{E-2\Omega-2i\Gamma}{E-2\Omega+2i\Gamma}.
\end{equation} This therefore concludes the calculations of the S-matrix. 

Suppose that in region 3 ($x_1 < x_2  <0$), $g(x_1, x_2)$ takes the following form
\begin{equation}
g(x_1, x_2) = e^{i E x_c + \frac{\Gamma}{2} x} = e^{i(E-i\Gamma)x_1/2} e^{i(E+i\Gamma)x_2/2}.
\end{equation} 
We then apply the same procedures as previously to obtain $g(x_1, x_2)$ in any other regions. One first has
\begin{equation}
g(x_1<0, 0^-) = e^{i E x_1/2 + \Gamma x_1/2}.
\end{equation}
With the same boundary conditions between quadrant III and quadrant II, we have
\begin{align}
e(x_1<0) &=  i  e^{+ i (E-\Omega+ i \Gamma/2)x_1}\int_{-\infty}^{x_1} e^{- i (E-\Omega+ i \Gamma/2)  x'} (-\sqrt{2} V) e^{i E x'/2 + \Gamma x'/2} dx'\notag\\
&= \frac{2 \sqrt{2}V}{E-2\Omega+ 2 i\Gamma} e^{i E x_1/2 + \Gamma x_1/2},
\end{align}
where we have used $V^2 =\Gamma$. Note that the resonance occurs at $E = 2\Omega$. This is in contrast to the single particle excitation in $\{|\tilde{W}_{k,p}\rangle_{ee}\}$ where the resonances occur at $k = \Omega$ or $p = \Omega$.

Proceed as before, 
\begin{align}
g(x_1, 0^+) &= -i\frac{V}{\sqrt{2}} e(x_1) + g(x_1, 0^-)\notag\\
&=  -i\frac{V}{\sqrt{2}} \frac{2 \sqrt{2}V}{E-2\Omega+ 2 i\Gamma} e^{i E x_1/2 + \Gamma x_1/2} + e^{i E x_1/2 + \Gamma x_1/2}\notag\\
&= \frac{E-2\Omega}{E-2\Omega+ 2 i \Gamma}e^{i E x_1/2 + \Gamma x_1/2}.
\end{align} 

Therefore, in quadrant II ($x_1 < 0 < x_2$), in accord with the Bethe ansatz, we postulate
\begin{equation}
g(x_1<0, x_2>0) = \frac{E-2\Omega}{E-2\Omega+2 i \Gamma} e^{i E x_c +\Gamma x/2}.
\end{equation}

To extend to quadrant I, we first obtain $g(0^-, x_2>0)$:
\begin{equation}
g(0^-, x_2>0) = \frac{E-2\Omega}{E-2\Omega+2 i \Gamma} e^{i E x_2/2 -\Gamma x_2/2},
\end{equation}thus
\begin{align}
& e(x_2>0)\notag\\
 =&\,  i  e^{+ i (E-\Omega+ i \Gamma/2)x_2}\int_{-\infty}^{x_2} e^{- i (E-\Omega+ i \Gamma/2)  x'} (-\sqrt{2}V)\frac{E-2\Omega}{E-2\Omega+ 2 i\Gamma} e^{i E x'/2 - \Gamma x'/2} dx'\notag\\
= &\,\frac{2\sqrt{2}V}{E-2\Omega+2i\Gamma}\,e^{i E x_2/2 -\Gamma/2 x_2}.
\end{align}

\begin{figure}[thb]
\scalebox{0.6}{\includegraphics{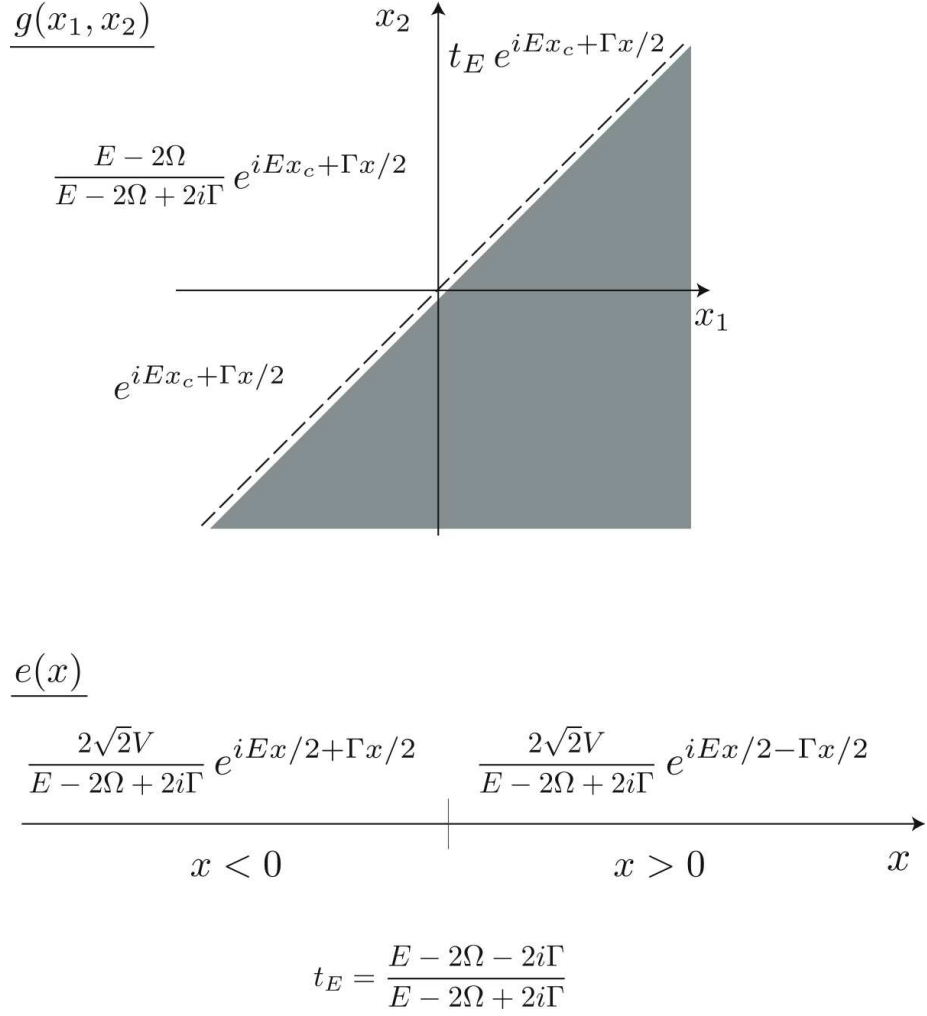}}
\caption{The wavefunction of the interacting eigenstate $|i^+\rangle$ of $H_e$ cosisting of two-photon bound state. The shaded region is obtained by symmetry, \emph{i.e.}, $g(x_2, x_1) = + g(x_1, x_2)$. Also shown is $e(x)$ for all $x$.}\label{Fi:Bound}
\end{figure}

From this, we obtain
\begin{align}
g(0^+, x_2>0) &= -i\frac{V}{\sqrt{2}} e(x_2) + g(0^-, x_2)\notag\\
&= \left( -\frac{2i\Gamma}{E-2\Omega + 2 i\Gamma}+\frac{E-2\Omega}{E-2\Omega+2 i\Gamma}\right)e^{i E x_2/2 -\Gamma x_2/2}\notag\\
&= \frac{E-2\Omega-2i\Gamma}{E-2\Omega+2 i \Gamma} e^{i E x_2/2 -\Gamma x_2/2},
\end{align}and thus, in region I ($x_2 > x_1>0$), applying the Bethe ansatz again,
\begin{equation}
g(x_1, x_2) = \frac{E-2\Omega-2i\Gamma}{E-2\Omega+2 i \Gamma} e^{i E x_c +\Gamma x/2}.
\end{equation}
Therefore, in the \emph{full} quadrant I,
\begin{align}
g(x_1>0, x_2>0) &=  \frac{E-2\Omega-2i\Gamma}{E-2\Omega+2 i \Gamma} e^{i E x_c +\Gamma x/2}\theta(-x) +  \frac{E-2\Omega-2i\Gamma}{E-2\Omega+2 i \Gamma} e^{i E x_c -\Gamma x/2}\theta(x)\notag\\
&=\frac{E-2\Omega-2i\Gamma}{E-2\Omega+2 i \Gamma} e^{i E x_c -\Gamma |x|/2}\notag\\
& \equiv t_E e^{i E x_c -\Gamma |x|/2}.
\end{align}While in the \emph{full} quadrant III,
\begin{align}
g(x_1>0, x_2>0) &=   e^{i E x_c +\Gamma x/2}\theta(-x) +  e^{i E x_c -\Gamma x/2}\theta(x)\notag\\
& = e^{i E x_c -\Gamma |x|/2}.
\end{align}
Finally, note that for the two-photon bound state, the self-consistency condition $e(0^-) = e(0^+)$ is automatically satisfied. This proves that $\mathbf{S}_{ee}|B_E\rangle_{ee} = t_E |B_E\rangle_{ee}$, and therefore $|B_E\rangle_{ee}$ is an eigenstate of $\mathbf{S}_{ee}$. Fig.~\ref{Fi:Bound} summarizes $g(x_1, x_2)$ in the entire $x_1$-$x_2$ plane, as well as $e(x)$ for all $x$, for the two-photon bound state $|B_{E}\rangle$.


\subsection{The S-Matrix for $\mathbf{S}_{ee}$}\label{P:S}

From the definition of the S-matrix, $\mathbf{S}_{ee} = \sum_{|\mbox{\scriptsize in}\rangle} |\mbox{out}\rangle \langle \mbox{in}|$, the two-photon one-mode S-matrix therefore is 
\begin{equation}\label{E:SMatrix}
\mathbf{S}_{ee}\equiv\sum_{k \leqslant p}  t_k t_p |W_{k,p}\rangle_{ee}\, \phantom{}_{ee}\langle W_{k, p}| + \sum_{E} t_E |B_E\rangle_{ee} \,\phantom{}_{ee}\langle B_E|.
\end{equation} For $|\mbox{in}\rangle = |W_{k,p}\rangle_{ee}$, or $|B_E\rangle_{ee}$, the out-state  $|\mbox{out}\rangle = t_k t_p |W_{k,p}\rangle_{ee}$, and $ t_E |B_E\rangle_{ee}$, respectively. 

It should be explicitly pointed out that the S-matrix defined above, Eq.~(\ref{E:SMatrix}), describes the physical scattering process that the photon in-state is mapped to the out-state via $|\mbox{out}\rangle = \mathbf{S}_{ee} |\mbox{in}\rangle$. This definition of the S-matrix is exactly the same as that in the usual scattering theory. In the literatures on Bethe ansatz, unfortunately, sometimes a different definition is adopted~\cite{Wiegmann:1983a, Hewson:1997}. There, the S-matrix is defined to be Eq.~(\ref{E:ABRatio}), the relative phase of the two plane waves of the wavefunction in region 3. 

Below we summarize several computations that are needed for two-mode calculations later. The details for these computations are provided in Appendix~\ref{A:SDetails}. We first mention the results of $\phantom{}_{ee}\langle S_{k_2, p_2}|\mathbf{S}_{ee}|S_{k_1, p_1}\rangle_{ee}$, the momentum distribution of of the out-state $\phantom{}_{ee}\langle S_{k_2, p_2} |\mbox{out}\rangle$ for in-state $|S_{k_1, p_1}\rangle_{ee}$ in $ee$ subspace:
\begin{equation}\label{E:SMatrixElement}
\phantom{}_{ee}\langle S_{k_2, p_2}|\mathbf{S}_{ee}|S_{k_1, p_1}\rangle_{ee} = t_{k_1}t_{p_1}\delta(k_1 -k_2)\delta(p_1 - p_2) + t_{k_1} t_{p_1}\delta(k_1 - p_2)\delta(k_2 - p_1)+ B\delta(E_1 -E_2),
\end{equation}where the first two terms of product of delta functions indicate the uncorrelated part of the S-matrix, which are simply the direct and exchange terms of each individual incident momentum, and can also be written as $ t_{k_1}t_{p_1}\delta(\Delta_1 -\Delta_2)\delta(E_1 - E_2) + t_{k_1} t_{p_1}\delta(\Delta_1 +\Delta_2)\delta(E_1 - E_2)$. The third term 
\begin{equation}
B =
\frac{16 i \Gamma^2}{\pi}\frac{E_1-2\Omega + i\Gamma}{\left[4\Delta_1^2 -(E_1 - 2\Omega + i\Gamma)^2\right] \left[4\Delta_2^2 -(E_1 - 2\Omega + i\Gamma)^2\right]},
\end{equation} in contrast, indicates the strong correlations between the two photons, and manifests as the background fluorescence due to the scattering. Note that this term does not conserve individual energy of each photon, but only the total energy.
When $\Delta_1 \neq \Delta_2$, $|B(E_1, \Delta_1, \Delta_2)|^2$ is the probability density for the outgoing photon pair in $(E_1, \Delta_2)$ state, when the incoming photon pair is in $(E_1, \Delta_1)$ state. 

The uncorrelated part in Eq.~(\ref{E:SMatrixElement}) comes entirely from the first term in Eq.~(\ref{E:SMatrix}), $\sum_{k \leqslant p}  t_k t_p |W_{k,p}\rangle_{ee}\, \phantom{}_{ee}\langle W_{k, p}|$; while the correlated part in Eq.~(\ref{E:SMatrixElement}), $B\delta(E_1 -E_2)$, has contributions from both $|W_{k,p}\rangle$ and $|B_{E}\rangle$ in Eq.~(\ref{E:SMatrix}).

For the same in-state $|\mbox{in}\rangle = |S_{k_1, p_1}\rangle_{ee} = |S_{E_1, \Delta_1}\rangle_{ee}$, one could also write down the real-space representation of the out-state: 
\begin{align}\label{E:Out}
\phantom{}_{ee}\langle x_c, x |\mbox{out}\rangle_{ee} &= \phantom{}_{ee}\langle x_c, x|\mathbf{S}_{ee}|\mbox{in}\rangle\notag\\
& =\sum_{E_2, \Delta_2\leq 0} S_{E_2, \Delta_2}(x_c, x) \phantom{}_{ee}\langle S_{E_2, \Delta_2}|\mathbf{S}_{ee}|S_{E_1, \Delta_1}\rangle_{ee}\notag\\
&=\sum_{E_2, \Delta_2\leq 0} S_{E_2, \Delta_2}(x_c, x) \left(t_{k_1} t_{p_1}  \,\phantom{}_{ee}\langle S_{E_2, \Delta_2}|S_{E_1, \Delta_1}\rangle_{ee} + B \delta(E_2 - E_1)\right)\notag\\
&= t_{k_1} t_{p_1}  S_{E_1, \Delta_1}(x_c, x)+ \sum_{\Delta_2 \leq 0} B S_{E_1, \Delta_2}(x_c, x)\notag\\
&=e^{i E_1 x_c}\frac{\sqrt{2}}{2\pi}\left(t_{k_1} t_{p_1} \cos\left(\Delta_1 x\right)-\frac{4\Gamma^2}{4\Delta_1^2 -(E_1-2\Omega +  i\Gamma)^2} e^{i (E_1-2\Omega) |x|/2 -\Gamma |x|/2}\right)
\end{align}
which takes the form $e^{i E_1 x_c} \langle x|\phi\rangle$, where $\langle x |\phi\rangle$ is the wavefunction in the relative coordinate $x$.  
The deviation of the out-state wavefunctions from that of interaction-free case is large when $\Delta_1 \simeq \pm (E_1/2 -\Omega)$, i.e., when at least one of the incident photons is close to resonance. 

\section{Two-Photon case II : Two-Mode model}\label{P:TwoPhotonII}

We now compute the two-mode two-photon scattering properties. To analyze a two-photon scattering experiment, one first projects the wave packets describing the two photons to each $|S_{k, p}\rangle$, and applies the previous discussions to each component. Specifically, consider an in-state
\begin{equation}\label{E:InRR}
|\mbox{in}\rangle \equiv |S_{k_1,p_1}\rangle_{RR} =\int dx_1 dx_2 \frac{1}{2\pi\sqrt{2}} \left(e^{i k_1 x_1 + i p_1 x_2} + e^{i k_1 x_2 + i p_1 x_1}\right) \frac{1}{\sqrt{2}}c_R^{\dagger}(x_1) c_R^{\dagger}(x_2) |\emptyset, -\rangle,
\end{equation}which describes two incident photons of plane waves from the left with momenta $k$ and $p$ respectively. To apply the decomposition relation, Eq.~(\ref{E:Decomposition}), we first decompose the in-state $|S_{k_1, p_1}\rangle$ to the components in $ee$, $oo$, and $eo$ subspaces, followed by computing the scattering states in each subspace, and finally transform the results back to $RR$, $LL$, and $RL$ spaces. The two-mode out-state thus obtained is (please refer to Appendix~\ref{A:S2Out} for details)
\begin{align}\label{E:RealSpaceOutState}
|\mbox{out}\rangle &= \mathbf{S} |\mbox{in}\rangle\notag\\
& = \int dx_1 dx_2 \,t_2(x_1, x_2) \frac{1}{\sqrt{2}}c_R^{\dagger}(x_1) c_R^{\dagger}(x_2)|\emptyset, -\rangle \notag\\
& + \int dx_1 dx_2 \,r_2(x_1, x_2) \frac{1}{\sqrt{2}}c_L^{\dagger}(x_1) c_L^{\dagger}(x_2)|\emptyset, -\rangle \notag\\
& + \int dx_1 dx_2 \,rt(x_1, x_2) c_R^{\dagger}(x_1) c_L^{\dagger}(x_2)|\emptyset, -\rangle
\end{align}where
\begin{align}\label{E:t2Wavefunction}
t_2(x_1, x_2) 
&=e^{i E_1 x_c}\frac{\sqrt{2}}{2\pi}\left(\bar{t}_{k_1} \bar{t}_{p_1} \cos\left(\Delta_1 x\right)-\frac{\Gamma^2}{4\Delta_1^2 -(E_1-2\Omega +  i\Gamma)^2} e^{i (E_1-2\Omega) |x|/2 -\Gamma |x|/2}\right),
\end{align}
\begin{align}\label{E:r2Wavefunction}
r_2(x_1, x_2) 
&=e^{-i E_1 x_c}\frac{\sqrt{2}}{2\pi}\left(\bar{r}_{k_1} \bar{r}_{p_1} \cos\left(\Delta_1 x\right)-\frac{\Gamma^2}{4\Delta_1^2 -(E_1-2\Omega +  i\Gamma)^2} e^{i (E_1-2\Omega) |x|/2 -\Gamma |x|/2}\right),
\end{align}and 
\begin{align}\label{E:rtWavefunction}
& rt(x_1, x_2)\notag\\
& = \frac{1}{2\pi} e^{i \frac{E_1}{2}x}\left(\bar{t}_{k_1} \bar{r}_{p_1} e^{2 i \Delta_1 x_c} + \bar{r}_{k_1} \bar{t}_{p_1} e^{-2 i \Delta_1 x_c}-\frac{2\Gamma^2}{4\Delta_1^2 -(E_1-2\Omega +  i\Gamma)^2} e^{i (E_1-2\Omega) |x_c| -\Gamma |x_c|}\right)
\end{align}
where $\bar{t}_{k_1}$, $\bar{t}_{p_1}$ are the two-mode single photon transmission amplitudes, and $\bar{r}_{k_1}$, $\bar{r}_{p_1}$ the two-mode single photon reflection amplitudes [Eq.~(\ref{E:TwoModetr})]. Note the locations of $x$ and $x_c$ in $rt(x_1, x_2)$ compared with $t_2(x_1, x_2)$ and $r_2(x_1, x_2)$. 

\begin{figure}[thb]
\scalebox{1}{\includegraphics[width=\columnwidth]{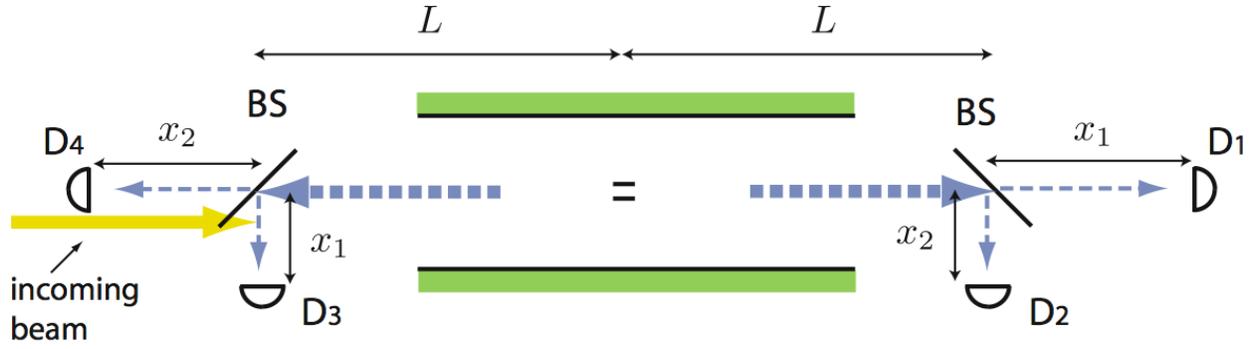}}
\caption{(Color online). Schematic experimental setups for concurrence measurements of   (a) $|t_2(x_1, x_2)|^2$, (b) $|r_2(x_1, x_2)|^2$, and (c) $|rt(x_1, x_2)|^2$. $D_1$, $D_2$ are photo-detectors with adjustable positions. BS, beam splitter. The ``='' symbol inside the one-dimensional waveguide denotes the two-level quantum impurity.}\label{Fi:Measurement}
\end{figure}

$t_2(x_1, x_2)$, $r_2(x_1, x_2)$, and $rt(x_1, x_2)$ represent two-photon wavefunctions in parts of the out-state, in which either both photons are transmitted or reflected, or one photon is transmitted while the other reflected. Experimentally, at least in principle, the magnitude of these wavefunctions can be measured in the setup shown in Fig.~\ref{Fi:Measurement}, where a beam splitter with a single-photon counter on each of the output arm, is placed at the entrance and the exit of the one-dimensional waveguide. In the forward (backward) direction, these photon counters are labeled $D_1$, $D_2$ ($D_3$, $D_4$), and are placed at a distance $x_1$, $x_2$ from the beam splitter, respectively. The experiments can be carried out by injecting a weak classical beam such that the average numbers photons per pulse is far smaller than 2, and such that the pulse repetition rate is much smaller than the inverse of the spontaneous emission lifetime. It can also be carried out with two-photon sources. $|t_2(x_1, x_2)|^2$ corresponds to those events where both $D_1$ and $D_2$ click simultaneously. The dependency on $x_1$, $x_2$ can be measured by varying the distance of the photo-detectors from the beam splitters, since $|t_2(x_1, x_2)|^2$ depends only upon $x_1-x_2$. Similar coincidence detection can be used in the backward direction to detect $|r_2(x_1, x_2)|^2$. For $|rt(x_1, x_2)|^2$, one could measure the coincidence rate for $D_1$ in the forward direction and $D_4$ in the backward directions. Alternatively, one could employ the Hanbury Brown and Twiss arrangement wherein the two photo-detectors on each side are kept at the same distance from the beam splitter. In this setup one measures the delay time $\tau$, which is proportional to $x_1-x_2$, between two consecutive clicks on the two detectors. 

\begin{figure}[thb]
\scalebox{1}{\includegraphics[width=\columnwidth]{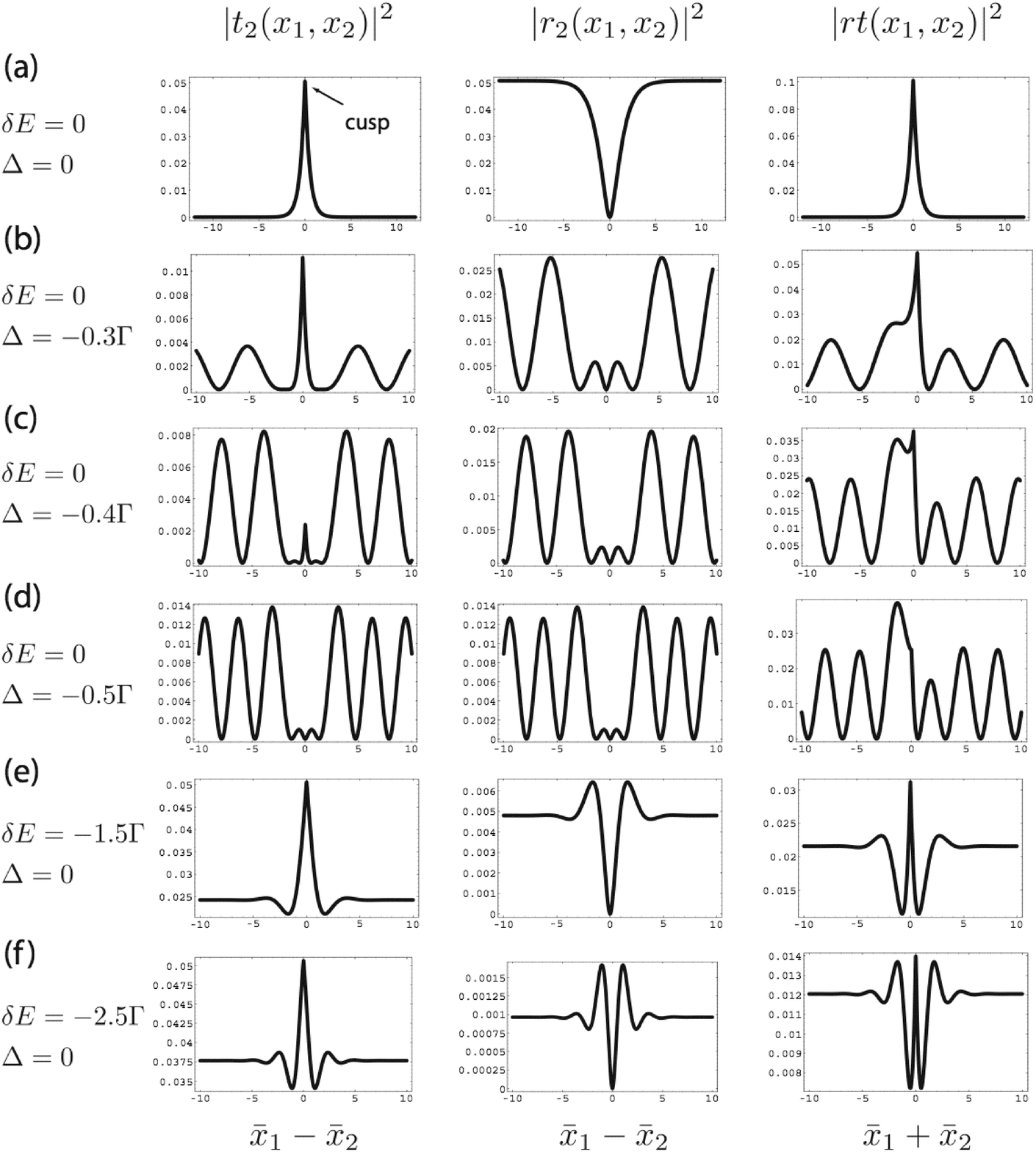}}
\caption{$|t_2(x_1, x_2)|^2$, $|r_2(x_1, x_2)|^2$, and $|rt(x_1, x_2)|^2$ for various  photon-pair energy detuning $\delta E\equiv E-2\Omega$, and energy difference $\Delta$. (a) $\delta E=0$, $\Delta=0$,  (b) $\delta E=0$, $\Delta=-0.3\Gamma$, (c) $\delta E=0$, $\Delta=-0.4\Gamma$, (d) $\delta E=0$, $\Delta=-0.5\Gamma$, and (e) $\delta E=-1.5\Gamma$, $\Delta=0$, (f) $\delta E=-2.5\Gamma$, $\Delta=0$. $\bar{x}\equiv \Gamma x/2$.}\label{Fi:AllWavefunctions}
\end{figure}

In Fig.~\ref{Fi:AllWavefunctions}, we plot $|t_2(x_1, x_2)|^2$, $|r_2(x_1, x_2)|^2$, and $|rt(x_1, x_2)|^2$ for various total energy detuning $\delta E \equiv E-2\Omega$, and energy difference $\Delta \equiv (k-p)/2$. Before going into details, we mention some general properties of $|t_2(x_1, x_2)|^2$, $|r_2(x_1, x_2)|^2$, and $|rt(x_1, x_2)|^2$, from the analytic expressions [Eqs.~(\ref{E:t2Wavefunction}), (\ref{E:r2Wavefunction}), (\ref{E:rtWavefunction})]. First of all, all $|t_2(x_1, x_2)|^2$, $|r_2(x_1, x_2)|^2$, and $|rt(x_1, x_2)|^2$ are even functions of $E-2\Omega$ and of $\Delta$, thus it suffices to investigate only, say, the range where $E-2\Omega \leq 0$, and $\Delta \leq 0$. Also, when $x_2 = x_1$ (\emph{i.e.,} $x=0$), $r_2(x_1, x_2)$ is always zero for all $E$ and $\Delta$, \emph{i.e.}, the two photons are always \emph{anti-bunching} in the backward direction. Finally, when $\Delta_1=0$, we always have $|t_2(x_1, x_2=x_1)| = \sqrt{2}/2\pi$, regardless of the photon pair energy, $E$. 

We now discuss the effects of varying both $\delta E$ and $\Delta$. When the two incident photons are degenerate and on resonance with the atom, \emph{i.e.,} $\delta E = \Delta =0$, the out-wavefucntions are
\begin{align}
t_2(x_1, x_2) &= \frac{\sqrt{2}}{2\pi} e^{+i 2 \Omega x_c} \left(- e^{-\Gamma |x|/2}\right),\notag\\
r_2(x_1, x_2) &= \frac{\sqrt{2}}{2\pi} e^{-i 2 \Omega x_c} \left(1- e^{-\Gamma |x|/2}\right),\notag\\
rt(x_1, x_2) & = \frac{1}{2\pi} e^{i 2 \Omega (x_1-x_2)/2} e^{-\Gamma |x_1 + x_2|/2}\times (-2),
\end{align}as plotted in Fig.~\ref{Fi:AllWavefunctions}(a). 
$|t_2(x_1, x_2)|^2$ decays exponentially as $|x|\equiv |x_1-x_2|$ becomes large, and thus the two transmitted photons are in a \emph{bound state}. Moreover, when $|x|$ is small, $|t_2|^2 \propto 1-\Gamma |x|$ shows a cusp at $x=0$, while $|r_2|^2 \propto x^2$ does not. This should manifest in the measurement of the $g^{(2)}(\tau)$ function in each case.

When the photon-pair energy is kept on resonance with the quantum impurity ($\delta E =0)$ while the energy difference between the two photons, $|\Delta|$, is gradually increased from zero to $\Gamma/2$, as shown in Fig.~\ref{Fi:AllWavefunctions} (a) -- (d), the peak at $x=0$ in $|t_2(x_1, x_2)|^2$ reduces from its maximum to zero. The transmitted photons thus change from \emph{bunching} to \emph{anti-bunching}.  Hence the quantum impurity can induce either an effective repulsion or attraction between two photons. $|r_2(x_1, x_2)|^2$ is always zero when $x_1=x_2$, as previously mentioned. Both $|t_2(x_1, x_2)|^2$ and $|r_2(x_1, x_2)|^2$ are even functions of $x_1-x_2$. On the other hand, $|rt(x_1, x_2)|^2$ can show asymmetry as a function of $x_1 + x_2$ when $\Delta \neq 0$.  A symmetric peak at $x_1+x_2=0$ occurs when $\delta E =\Delta =0$ (Fig.~\ref{Fi:AllWavefunctions}(a)), and becomes asymmetric when $|\Delta|$ increases (Fig.~\ref{Fi:AllWavefunctions}(b) -- (d)). When $\Delta \neq 0$,  the maximum of $|rt(x_1, x_2)|^2$ always occurs at $x_1+x_2 <0$, which indicates the reflected photon leaves the impurity earlier than the transmitted photon. In addition, at $\delta E=0$, $\Delta \neq 0$,  all the two-photon out-wavefunctions show oscillations for large $x_1-x_2$ or $x_1+x_2$. On the other hand, when $\delta E \neq 0$, but $\Delta=0$, the oscillations at large $x_1-x_2$ or $x_1+x_2$ disappear, as shown in Fig.~\ref{Fi:AllWavefunctions} (e) and (f).  

The anti-bunching in $r_2(x_1, x_2)$ at $x=0$ for all $\delta E$ and $\Delta$ in fact has similar physical origin as the anti-bunching experimentally observed in resonance fluorescence from a single trapped ion~\cite{Hoffges:1997}. Since $r_2(x_1, x_2)$ arises entirely from the emission of the atom with not contribution from the incident light, $r_2(x=0)=0$ simply indicates that two photons can not be simultaneously emitted by a single atom. As a further validation of this argument, as well as a somewhat indirect experimental support of our theory, we note that our calculated $|r_2(x_1, x_2)|^2$, as shown in Fig.~\ref{Fi:AllWavefunctions}(a), (e), and (f), in fact agrees excellently, after normalization, with the experimentally measured $g^{(2)}(\tau)$ for a single trapped ion subject to a weak beam~\cite{EnergyDetuningBy2}. On the other hand, the predictions here for $t_2(x_1, x_2)$ and $rt(x_1, x_2)$ involves interference between the incident and emitted photons and therefore represent new physical effects.


The momentum distributions in each case can also be computed directly. In the forward direction, the momentum distribution is (again, please refer to Appendix~\ref{A:S2Out} for details)
\begin{equation}\label{E:t2Momenta}
\phantom{}_{RR}\langle S_{k_2, p_2}|\mathbf{S}|S_{k_1, p_1}\rangle_{RR} = \bar{t}_{k_1} \bar{t}_{p_1}\left[\delta(k_1 - k_2)\delta(p_1 - p_2)+ \delta(k_1 - p_2)\delta(p_1 - k_2)\right] + \frac{1}{4} B\delta(E_1 - E_2),
\end{equation}and the momentum distribution in the backward direction:
\begin{equation}\label{E:r2Momenta}
\phantom{}_{LL}\langle S_{k_2, p_2}|\mathbf{S}|S_{k_1, p_1}\rangle_{RR} = \bar{r}_{k_1} \bar{r}_{p_1}\left[\delta(k_1 + k_2)\delta(p_1 + p_2)+ \delta(k_1 + p_2)\delta(p_1 + k_2)\right] + \frac{1}{4} B\delta(E_1 - E_2).
\end{equation}
Define
\begin{equation}
|k_2^{R}, p_2^{L}\rangle_{RL} \equiv \int dx_1 dx_2 \frac{1}{2\pi}e^{i k_2 x_1 + i p_2 x_2}c^{\dagger}_{R}(x_1) c^{\dagger}_{L}(x_2)|\emptyset, -\rangle,
\end{equation}the momentum distribution in the $RL$ subspace is
\begin{equation}\label{E:rtMomenta}
\phantom{}_{RL}\langle k_2^{R}, p_2^{L}|\mathbf{S}|S_{k_1, p_1}\rangle_{RR} = \bar{t}_{k_1} \bar{r}_{p_1}\delta(k_2-k_1)\delta(p_2 + p_1)+  \bar{r}_{k_1} \bar{t}_{p_1}\delta(k_2 - p_1)\delta(p_2 + k_1) + \frac{1}{4} B\delta(E_1 - E_2),
\end{equation}where
\begin{equation}
B =
\frac{16 i \Gamma^2}{\pi}\frac{E_1-2\Omega + i\Gamma}{\left[4\Delta_1^2 -(E_1 - 2\Omega + i\Gamma)^2\right] \left[4\Delta_2^2 -(E_1 - 2\Omega + i\Gamma)^2\right]}.
\end{equation}

\begin{figure}[thb]
\scalebox{1}{\includegraphics{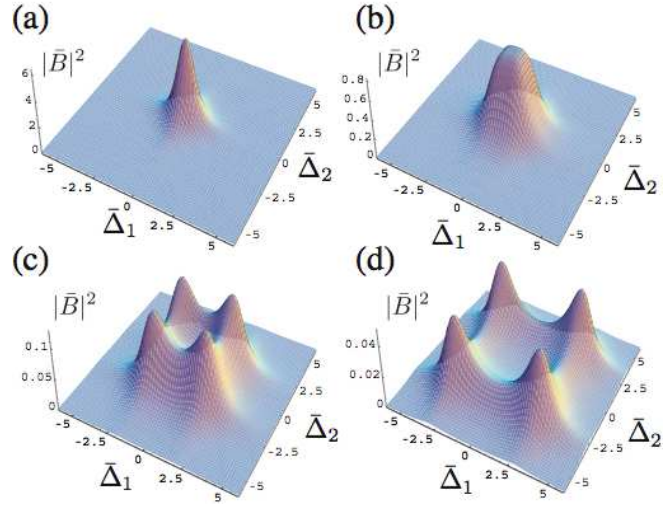}}
\caption{(Color online) Background fluorescence as a function of $\bar{\Delta}_1$ and $\bar{\Delta}_2$ at various energy. (a) $\bar{E}=0$. (b) $\bar{E}=2$. (c) $\bar{E}=4$. (d) $\bar{E}=6$. $\bar{B}\equiv (\Gamma/2) B$, $\bar{E}\equiv (E-2\Omega)/(\Gamma/2)$, and $\bar{\Delta}\equiv\Delta/(\Gamma/2)$. For any given $E$, the in- and out-states can be completely specified by one quadrant in the $\Delta_1$-$\Delta_2$ plane.}\label{Fi:Background_3D}
\end{figure}

In each momentum distribution of Eq.~(\ref{E:t2Momenta}), (\ref{E:r2Momenta}), and (\ref{E:rtMomenta}), the delta function terms correspond to the uncorrelated part of the two-photon transport. The $\frac{1}{4} B\delta(E_1 - E_2)$ term, however, is the signature of the strong correlation between the two photons and represents the background fluorescence. Specifically, $B$ is the momentum distribution of the two photons scattered out of the original values $k_1$ and $p_1$. This term originates from the $ee$ subspace, and gives the same contributions in the $RR$, $LL$, and $RL$ subspaces. 
Fig.~\ref{Fi:Background_3D} plots normalized $|B(E,\Delta_1, \Delta_2)|^2$ as a function of $\Delta_1$ and $\Delta_2$ for various photon-pair energy $E$.  Since the locations of the poles in $B$ are at $k_{1,2}=p_{1,2} =\Omega - i\Gamma/2$, which correspond approximately to either one of the photons having an energy at $\Omega$, one can picture the background fluorescence as one photon inelastically scattering off a composite transient object formed by the atom absorbing the other photon.  

The out-state wavefunctions, $t_2(x_1, x_2)$, $r_2(x_1, x_2)$, and $rt(x_1, x_2)$ [Eqs.~(\ref{E:t2Wavefunction}), (\ref{E:r2Wavefunction}), and (\ref{E:rtWavefunction})], together with the corresponding momentum distributions [Eqs.~(\ref{E:t2Momenta}), (\ref{E:r2Momenta}), (\ref{E:rtMomenta})] provide a complete full quantum-mechanical description for the two-photon in-state $|S_{k_1, p_1}\rangle_{RR}$ scattering off a two-level system. In a classic paper, B. R. Mollow investigated the power spectrum of light scattered by two-level systems in a three-dimensional system, using a semiclassical treatment, wherein the two-level atom is driven near resonance by a monochromatic classical electric field~\cite{Mollow:1969}. We note that in Mollow's paper, the power spectrum of the scattered field, in the limit of very low incident field intensity, has exactly the same lineshape as the momentum distribution $\phantom{}_{LL}\langle S_{k_2, p_2}|\mathbf{S}|S_{k_1, p_1}\rangle_{RR}$ [Eq.~(\ref{E:rtMomenta})] in the present work~\cite{MollowNote}. In particular, the inelastic part of the power spectrum in Mollow's paper corresponds directly to the background fluorescence, $|B|^2$. In his case, however, the strength of inelastic scattering vanishes in the weak field limit, while in our case, strong inelastic scattering occurs even with only two incident photons. Therefore, the strong interference in one-dimension greatly enhances the inelastic components. Also, the full quantum-mechanical treatment gives the correct $g^{(2)}(\tau)$ correlation function, and points out the connection between $g^{(2)}(\tau)$ correlation function and the background fluorescence, which could not be obtained in the semiclassical treatment.

\section{Three Photon Case}\label{P:ThreePhoton}

The above procedures can be generalized to multi-photon case. For example, when there are three photons and one two-level system in the one-dimensional waveguide, the self-consistency condition becomes
\begin{equation}\label{E:YangBaxter3}
e(0^-, 0^-) = e(0^-, 0^+) = e(0^+, 0^+),
\end{equation} which is the generalization of Eq.~(\ref{E:YangBaxter}). Here $e(x_1, x_2)$ is the two-photon amplitude when one photon is absorbed and the two-level system is in the excited state. By equating the components, Eq.~(\ref{E:YangBaxter3}) gives a set of six self-consistent equations. These equations are exactly the Yang-Baxter equations~\cite{Yang:1967,Wiegmann:1983a}, and are connected with the integrability of the Hamiltonian. The details will be presented elsewhere~\cite{Shen:2007c}.

\section{Conclusion and Outlook}

In this paper, we present an exact and complete solution of the transport properties of two-photons interacting with a single two-level system, when the photons are confined to a one-dimensional waveguide. Because the two-level system, at a given time, can only absorb one photon, the solution exhibits rich features, including, for example, the effects of background fluorescence and two-photon bound states. These results could be of relevance for many on-going quantum optics experiments.

Also, from a formalism point of view, here we outline a general approach, based upon the Bethe ansatz, to solve for the transport properties of multi-particle states in a class of quantum-impurity problem in one-dimension. In particular, we introduce a rigorous program to extract the ``in-'' and ``out-''states from the eigenstates of the interacting Hamiltonian, as well as a systematic approach to construct the complete scattering matrix of the system based upon these in- and out-states. This approach should be of general importance for a wide range of theoretical problems both in quantum optics and in condensed matter physics.

A key observation from our solution is that the in- and out-states, as obtained from the standard Bethe ansatz solution, is in fact not complete, at least for the photon-Hamitonian (as well as the Anderson Hamiltonian in the infinite-$U$ limit). The completeness of the Bethe ansatz solution for the interacting Hamiltonian was a subject of debate~\cite{Lai:1981, Schulz:1982, Hewson:1982} in the first few years since the publication of the pioneering papers by Wiegmann \emph{et al.}~\cite{Wiegmann:1980, Wiegmann:1981, Wiegmann:1983a} and by Andrei  \emph{et al.}~\cite{Andrei:1980} Here we note, that the proof of completeness by Schulz~\cite{Schulz:1982}, as cited by a  comprehensive review article in this area~\cite{Tsvelick:1983a}, in fact only proves that completeness of the $S_{k,p}(x_1, x_2)$ and $A_{k,p}(x_1, x_2)$ states. Since the completeness of in- and out-states is particularly important when constructing the full scattering matrix, one needs to carefully re-examine the recent works of applying Bethe ansatz to the interacting resonance level system for open systems~\cite{Mehta:2006}, where the crucial property of completeness of the solution is not explicitly checked.

\begin{acknowledgments}
S. Fan acknowledges financial support by the David and Lucile Packard Foundation.
\end{acknowledgments}

\appendix
\section{Computing the in-state $|i\rangle$ and the out-state $|f_i\rangle$ from the interacting eigenstate $|i^+\rangle$}\label{A:ReadOff}

In this appendix, we detail the derivations of the in-state and out-state from the eigenstates of the interacting Hamiltonian for both the one-photon and the two-photon cases, which are mentioned previously in Sec.~\ref{P:OnePhoton}, and Sec.~\ref{P:HeEigenstates}, respectively. Since the discussions are in ``$e$'' and ``$ee$'' subspaces, in this appendix, we suppress the label ``$e$'' and ``$ee$'' when there is no confusion.

\subsection{One-Photon Case}\label{A:ReadOffOnePhoton}
Here, we seek to prove the forms of one-photon in-state and out-state, Eq.~(\ref{E:OnePhotonInOut}), starting from the eigenstate $|k^+\rangle_{e}$ in Eq.~(\ref{E:kPlus}). To do so, we first note that the real-space representation of the advanced Green's function for $H_{0}^{e} = \int dx\,c_{e}^{\dagger}(x)\left(-i \frac{\partial}{\partial x}\right) c_e(x)$ and $E \equiv k$, is
\begin{equation}
\langle x, -|G^{A}_0|x', -\rangle=\langle x,-| \frac{1}{k-H_{0}^{e} - i\epsilon}|x', -\rangle = \theta(x'-x) (+i) e^{i k (x -x')},
\end{equation}where
\begin{equation}
|x, -\rangle \equiv c^{\dagger}(x)|\emptyset, -\rangle
\end{equation}
Let the out-state $|f_k\rangle$ be
\begin{equation}
|f_k\rangle \equiv \int dx \,\phi_f(x) c^{\dagger}(x)|\emptyset, -\rangle,
\end{equation}where $\phi_f(x) = \langle x,-|f_k\rangle$ is the one-photon wavefunction. Projecting the Lippmann-Schwinger equation, Eq.~(\ref{E:LippmannA}) to $\langle x, -|$, we have
\begin{equation}
\phi_f(x) = \phi(x) - \langle x, -|G^{A}_0 H_{\mbox{\scriptsize int}} |k^+\rangle.
\end{equation}

Inserting complete sets before and after $H_{\mbox{\scriptsize int}}$, since $G^{A}_0$ does not excite the impurity, and $H_{\mbox{\scriptsize int}}$ connects $\langle x', -|$ to the state $|\emptyset, +\rangle$, and does not vanish only when $x'=0$, we have
\begin{align}
\phi_f(x) &= \phi(x) - \int dx'\,\langle  x,-|G^{A}_0|x', -\rangle \langle x', -|H_{\mbox{\scriptsize int}}|\emptyset, +\rangle \langle +, \emptyset |k^+\rangle\notag\\
&= \phi(x) - \langle  x,-|G^{A}_0|0, -\rangle V e_k\notag\\
&= \phi(x) - \left(\theta(-x)(+i) e^{i k x} V\right)\left( \frac{1}{\sqrt{2\pi}}\frac{V}{k-\Omega+i \Gamma/2}\right)\notag\\
&= \left(\theta(-x)\frac{e^{i k x}}{\sqrt{2\pi}}+\theta(x) t_k \frac{e^{i k x}}{\sqrt{2\pi}}\right)- \theta(-x)  \frac{e^{i k x}}{\sqrt{2\pi}}  \frac{i\Gamma}{k-\Omega+i \Gamma/2}\notag\\
&= \left(\theta(-x)+\theta(x)\right)t_k \frac{e^{i k x}}{\sqrt{2\pi}}=t_k \frac{e^{i k x}}{\sqrt{2\pi}}.
\end{align}Thus the out-state one-photon wavefunction $\phi_f(x) = t_k \frac{e^{i k x}}{\sqrt{2\pi}} = t_k \langle x|k\rangle $ for all $x$. In the above derivations, we have used 
\begin{align}
\langle x',-|H_{\mbox{\scriptsize int}}|\emptyset, +\rangle & = \langle x',-| \int dx V\delta(x)\left(c^{\dagger}(x)\sigma + c(x)\sigma_{+}\right)|\emptyset, +\rangle\notag\\
&=\int dx V\delta(x) \langle \emptyset,-|c(x')c^{\dagger}(x)\sigma|\emptyset, +\rangle\notag\\
&= \int dx V\delta(x) \delta(x'-x)\notag\\
&= V\delta(x'),
\end{align}and
\begin{align}
\langle \emptyset,+ |k^+\rangle &=  \langle \emptyset,+|\left(\int dx \,\phi(x) c^{\dagger}(x) +  e_k \sigma_{+}\right) |\emptyset, -\rangle\notag\\
&= e_k.
\end{align}
Similarly, by using Eq.~(\ref{E:LippmannR}) and the retarded Green's function
\begin{equation}
\langle  x,-|G^{A}_0|x', -\rangle=\langle  x,-| \frac{1}{k-H_{0}^{e} + i\epsilon}|x', -\rangle = \theta(x-x') (-i) e^{i k (x -x')},
\end{equation}
the in-state can be shown to be
\begin{equation}
|k\rangle \equiv \int dx\, \phi_i(x) c^{\dagger}(x)|\emptyset,-\rangle,
\end{equation}with $\phi_i(x)=\langle x|k\rangle = \frac{e^{i k x}}{\sqrt{2\pi}}$ for all $x$.

\subsection{Two-Photon Case}

\begin{figure}[thb]
\scalebox{0.6}{\includegraphics{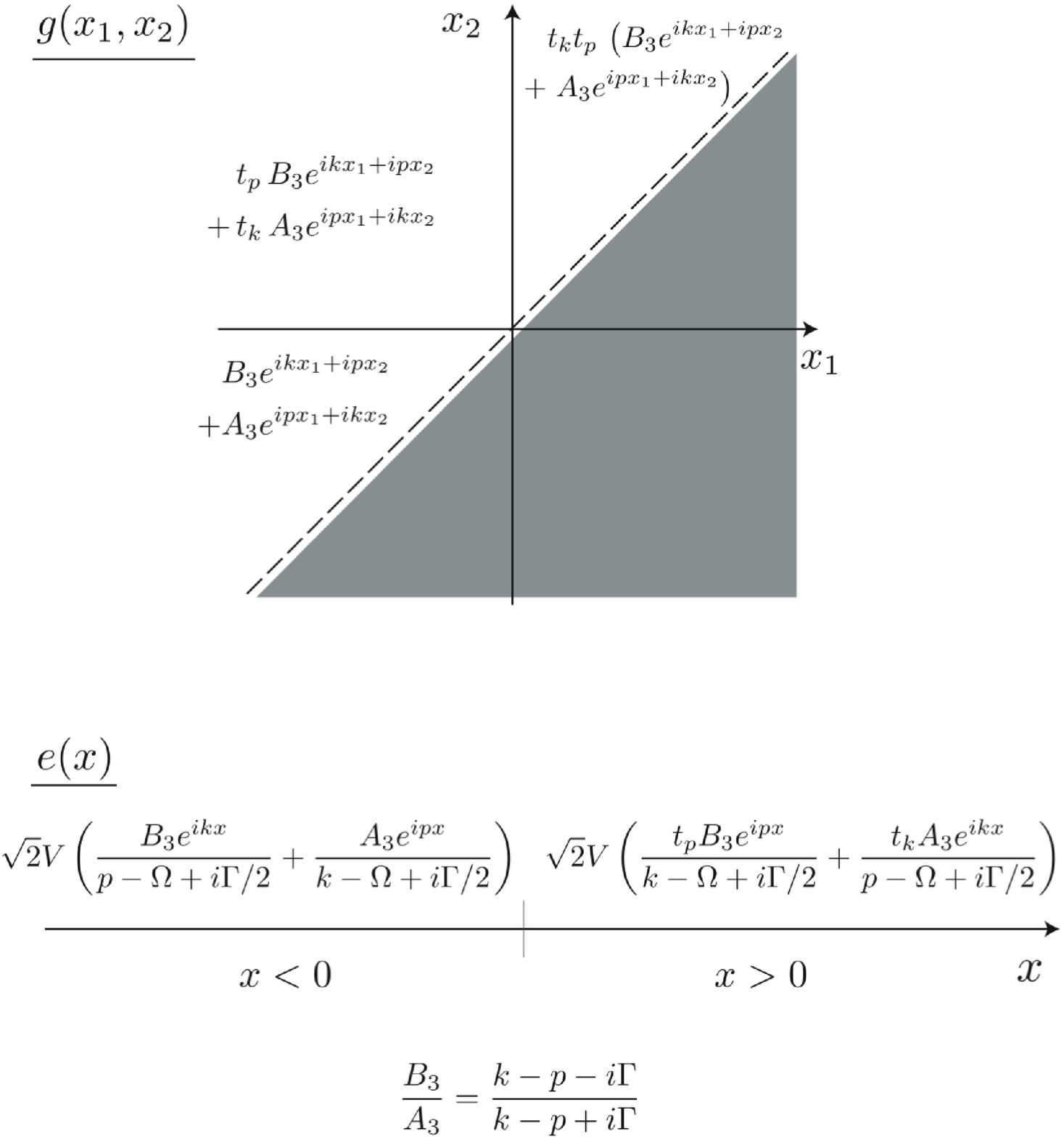}}
\caption{The wavefunction of the interacting eigenstate $|i^+\rangle$ of $H_e$ as constructed from the standard Bethe ansatz approach. The shaded region is obtained by symmetry, \emph{i.e.}, $g(x_2, x_1) = + g(x_1, x_2)$. Also shown is $e(x)$ for all $x$.}\label{Fi:Extended}
\end{figure}

The one-mode two-photon eigenstate $|i^+\rangle$ for the Hmiltonian $H_{e}$ is computed in Sec.~\ref{P:HeEigenstates} and \ref{P:BoundState}, and has the following form:
\begin{equation}\label{E:iPlus}
|i^+\rangle = \int dx_1 dx_2 \,g(x_1, x_2) \frac{1}{\sqrt{2}}c^{\dagger}(x_1) c^{\dagger}(x_2)|\emptyset, -\rangle + \int dx\,e(x) c^{\dagger}(x)|\emptyset, +\rangle,
\end{equation}with $g(x_1, x_2)$ and $e(x)$ being from either the extended Wiegmann-Andrei state in Sec.~\ref{P:HeEigenstates}, as summarized in Fig.~(\ref{Fi:Extended}),  or the bound state in Sec.~\ref{P:BoundState}, as summarized Fig.~(\ref{Fi:Bound}). The aim here is to prove the forms of the in- and out-state, \emph{i.e.}, Eq.~(\ref{E:InStateAsWiegmann}) from Eq.~(\ref{E:iPlus}).

Let the out-state be $|f\rangle$ with the following form:
\begin{equation}
|f\rangle \equiv \int dx_1 dx_2 \,g_f(x_1, x_2) \frac{1}{\sqrt{2}}c^{\dagger}(x_1) c^{\dagger}(x_2)|\emptyset, -\rangle
\end{equation} 

It is easy to see that
\begin{equation}\label{E:Identity}
\sum_{\xi = \pm}\int dx_1 dx_2 |x_1, x_2, \xi\rangle \langle \xi, x_1, x_2|,
\end{equation}where $\xi$ labels the atomic state, is the identity operator in the two-photon subspace.

Projecting $\langle  x_1, x_2, -|$ from the left to Eq.~(\ref{E:LippmannA}), and inserting the identity operator in the form of  Eq.~(\ref{E:Identity}) between $G^{0}_{A}$ and $H_{\mbox{\scriptsize int}}$, we have
\begin{align}\label{E:Out2}
g_{f}(x_1, x_2) &=  g(x_1, x_2)\notag\\
 &- \int dx_1' dx_2' dx'' \langle x_1, x_2,-|G^{0}_{A}|x_1', x_2', -\rangle \langle x_1', x_2', -|H_{\mbox{\scriptsize int}}|x'', +\rangle\langle x'', +|i^+\rangle,
\end{align}where we have used the fact that $G^{0}_{A}$ connects $\langle -|$ only to $|-\rangle$. Only the matrix element $\langle x_1', x_2', -|H_{\mbox{\scriptsize int}}|x'', +\rangle$ appears because of the form of $H_{\mbox{\scriptsize int}}$.

We compute each matrix element in the integral. First we have
\begin{align}
\langle x'', +|i^+\rangle &= \int dx \,e(x) \langle\emptyset, +|c(x'')c^{\dagger}(x)|\emptyset, +\rangle\notag\\
&= e(x''),
\end{align}and
\begin{align}
&\langle x_1', x_2', -|H_{\mbox{\scriptsize int}}|x'', +\rangle\notag\\
= & \frac{1}{\sqrt{2}} \langle \emptyset,-|c(x_2') c(x_1')\int dx V\delta(x) \left(c(x)\sigma_{+}+c^{\dagger}(x)\sigma_{-}\right) c^{\dagger}(x'')|\emptyset, +\rangle\notag\\
=&\frac{V}{\sqrt{2}}\int dx\, \delta(x)\langle \emptyset,-|c(x_2') c(x_1')c^{\dagger}(x)c^{\dagger}(x'')|\emptyset, -\rangle\notag\\
=& \frac{V}{\sqrt{2}}\int dx\, \delta(x)\left(\delta(x_1'-x)\delta(x_2'-x'')+\delta(x_1'-x'')\delta(x_2'-x)\right)\notag\\
=& \frac{V}{\sqrt{2}}\left(\delta(x_1')\delta(x_2'-x'')+\delta(x_1'-x'')\delta(x_2')\right).
\end{align}
Putting back to Eq.~(\ref{E:Out2}), we have
\begin{align}\label{E:Out3}
g_{f}(x_1, x_2) &= g(x_1, x_2)\notag\\
&-\frac{V}{\sqrt{2}}\int dx_1' dx_2' dx'' \langle  x_1, x_2, -|G^{0}_{A}|x_1', x_2', -\rangle \left(\delta(x_1')\delta(x_2'-x'')+\delta(x_1'-x'')\delta(x_2')\right) e(x'')\notag\\
&= g(x_1, x_2) -\frac{V}{\sqrt{2}}\int dx'' \left(\langle x_1, x_2, -|G^{0}_{A}|0, x'', -\rangle + \langle x_1, x_2, -|G^{0}_{A}|x'', 0, -\rangle\right) e(x'')\notag\\
&= g(x_1, x_2)-\sqrt{2}V \int dx'' \langle x_1, x_2, -|G^{0}_{A}|0, x'', -\rangle e(x'').
\end{align}

Let $\langle x_1, x_2,-|G^{0}_{A}|x_1', x_2', -\rangle \equiv G^{0}_{A}(x_1, x_2; x_1', x_2')$. Since $G^{0}_{A}$ satisfies
\begin{equation}
\left(E_0-H_{0}^{e}-i\epsilon\right) G^{0}_{A} = \mathbf{1},
\end{equation}we have
\begin{equation}
\left(E_0+ i\frac{\partial}{\partial x_1}+ i\frac{\partial}{\partial x_2}-i\epsilon\right)G^{0}_{A}(x_1, x_2; x_1', x_2') = \langle x_1, x_2|x_1', x_2' \rangle.
\end{equation} 

Inserting the identity, 
\begin{equation}
\mathbf{1} = \frac{1}{2}\iint_{-\infty}^{+\infty}dk dp\, |S_{k, p}\rangle\langle S_{k, p}|, 
\end{equation}
we have
\begin{equation}\label{E:GTwoPhotonsReal}
\left(E_0+ i\frac{\partial}{\partial x_1}+ i\frac{\partial}{\partial x_2}-i\epsilon\right)G^{0}_{A}(x_1, x_2; x_1', x_2') = \frac{1}{2}\iint dk dp\, \langle x_1, x_2|S_{k,p}\rangle\langle S_{k,p}|x_1', x_2' \rangle,
\end{equation}which can be solved by Fourier expanding $G^{0}_{A}(x_1, x_2; x_1', x_2')$ using $S_{k,p}(x_1, x_2)$:
\begin{align}\label{E:GTwoPhotonsMomentum}
G^{0}_{A}(x_1, x_2; x_1', x_2') &= \langle x_1, x_2,-|G^{0}_{A}|x_1', x_2', -\rangle\notag\\
&= \iint \langle x_1, x_2|S_{k, p}\rangle\langle S_{k, p}|G^{0}_{A}|x_1', x_2'\rangle \,dk dp\notag\\
&\equiv \iint \langle x_1, x_2|S_{k, p}\rangle G_{k,p}(x_1', x_2') \,dk dp.
\end{align}We have suppressed the idle atomic degree of freedom ``$-$''. Inserting Eq.~(\ref{E:GTwoPhotonsMomentum}) into Eq.~(\ref{E:GTwoPhotonsReal}), we then have
\begin{align}
&\left(E_0-(k+p)-i\epsilon\right)G_{k, p}(x_1', x_2') = \frac{1}{2} \langle S_{k,p}|x_1', x_2' \rangle\\
\Rightarrow & \, G_{k, p}(x_1', x_2') = \frac{1}{2}\frac{1}{E_0-(k+p)-i\epsilon}\langle S_{k,p}|x_1', x_2' \rangle\\
\Rightarrow & \, G^{0}_{A}(x_1, x_2; x_1', x_2') =\iint \frac{1}{2}\frac{1}{E_0-(k+p)-i\epsilon}\langle S_{k,p}|x_1', x_2' \rangle \langle x_1, x_2|S_{k,p}\rangle dk dp.
\end{align}
Therefore,
\begin{align}
G^{0}_{A}(x_1, x_2; 0, x'') &= \iint \frac{1}{2}\frac{1}{E_0-(k+p)-i\epsilon}\langle S_{k,p}|0, x'' \rangle \langle x_1, x_2|S_{k,p}\rangle dk dp\notag\\
&= \iint dk dp\,\frac{1}{2}\frac{1}{E_0-(k+p)-i\epsilon}\left(\frac{1}{2\pi\sqrt{2}}\right)^2\left(e^{-i p x''}+ e^{- i k x''}\right)\left(e^{i k x_1 + i p x_2}+e^{i p x_1+ i k x_2}\right)\notag\\
&=\iint dk dp\,\frac{1}{2}\frac{1}{E_0-(k+p)-i\epsilon} \left(\frac{\sqrt{2}}{2\pi}\right)^2 e^{-i E x''/2}\cos \Delta x'' e^{i E x_c}\cos \Delta x\notag\\
&= -\frac{1}{(2\pi)^2}\left(\int_{-\infty}^{\infty}d\Delta \cos\Delta x'' \cos \Delta x\right)\left(\int_{-\infty}^{\infty}dE \frac{1}{E-E_0 +i\epsilon}e^{i E(x_c -x''/2)}\right)\notag\\
&= -\frac{1}{(2\pi)^2}\left[\pi\left(\delta(x+x'')+\delta(x-x'')\right)\right]\left[\theta(x''/2-x_c)(-2\pi i)e^{i E_0 (x_c - x''/2)}\right]\notag\\
&= \frac{i}{2}\theta(x''/2-x_c)e^{i E_0 (x_c - x''/2)}\left[\delta(x+x'')+\delta(x-x'')\right],
\end{align}where again $x \equiv x_1 - x_2$, $x_c \equiv (x_1 + x_2)/2$. In this calculation, we have used
\begin{align}
&\int_{-\infty}^{\infty}d\Delta \cos\Delta x'' \cos \Delta x\notag\\
 &= 2\int_{0}^{\infty}d\Delta \cos\Delta x'' \cos \Delta x\notag\\
&=\int_{0}^{\infty}d\Delta\left(\cos \Delta(x+x'')+\cos \Delta(x-x'')\right)\notag\\
&=\pi\left(\delta(x+x'')+\delta(x-x'')\right).
\end{align}

Putting back to Eq.~(\ref{E:Out3}), we have
\begin{align}\label{E:Out4}
g_{f}(x_1, x_2) &= g(x_1, x_2) -\frac{i}{\sqrt{2}}V\left[\theta(-x/2 - x_c) e^{i E_0 (x_c + x/2)} e(-x)+ \theta(x/2 - x_c) e^{i E_0 (x_c - x/2)} e(x)\right]\notag\\
&= g(x_1, x_2) -\frac{i}{\sqrt{2}}V\left[\theta(-x_1) e^{i E_0 x_1} e(x_2-x_1)+ \theta(-x_2) e^{i E_0 x_2} e(x_1-x_2)\right].
\end{align}

Using Eq.~(\ref{E:Out4}), one can compute the out-state directly. Here we perform the explicit check for the case when $|i^+\rangle$ is the extended Wiegmann-Andrei state in Sec.~\ref{P:WiegmannState}. The photon wavefunction $g(x_1, x_2)$,  and $e(x)$, are  shown in Fig.~\ref{Fi:Extended}. The calculations are done separately in the four quadrants of the $x_1$-$x_2$ plane. 

\begin{enumerate}
\item In $x_1>0$ and $x_2>0$ region, due to the step functions, $\theta(-x_1)$ and $\theta(-x_2)$ in Eq.~(\ref{E:Out4}), $g_{f}(x_1, x_2) = g(x_1, x_2) = t_k t_p (B_3 e^{i k x_1 + i p x_2}+ A_3 e^{i p x_1 + i k x_2})$.
\item In $x_1<0$ and $x_2>0$ region, since $x_2 - x_1>0$, we have
\begin{align}
g_{f}(x_1, x_2) &= g(x_1, x_2) -\frac{i}{\sqrt{2}}V e^{i E_0 x_1} e(x_2-x_1)\notag\\
&=\left(t_p B_3 e^{i k x_1+ i p x_2}+ t_k A_3 e^{i p x_1 + i k x_2}\right)-i\Gamma e^{i E_0 x_1}\left(\frac{t_p B_3 e^{i p(x_2-x_1)}}{k -\Omega + i\Gamma/2}+ \frac{t_k A_3 e^{i k (x_2 - x_1)}}{p -\Omega + i\Gamma/2}\right)\notag\\
&=\left(t_p B_3 e^{i k x_1+ i p x_2}+ t_k A_3 e^{i p x_1 + i k x_2}\right)-i\Gamma \left(\frac{t_p B_3 e^{i k x_1 + i p x_2}}{k -\Omega + i\Gamma/2}+ \frac{t_k A_3 e^{i p x_1 + i k x_2}}{p -\Omega + i\Gamma/2}\right)\notag\\
&= t_p B_3 e^{i k x_1 + i p x_2}\left(1-\frac{i\Gamma}{k-\Omega + i\Gamma/2}\right)+ t_k A_3 e^{i p x_1 + i k x_2}\left(1-\frac{i\Gamma}{p-\Omega + i\Gamma/2}\right)\notag\\
&=  t_k t_p \left(B_3 e^{i k x_1 + i p x_2}+ A_3 e^{i p x_1 + i k x_2}\right),
\end{align}where we have used $e^{i E_0 x_1}e^{i p (x_2-x_1)} = e^{i (k+p) x_1}e^{i p(x_2-x_1)} = e^{i k x_1 + i p x_2}$, and $e^{i E_0 x_1}e^{i k (x_2-x_1)} = e^{i (k+p) x_1}e^{i k(x_2-x_1)} = e^{i p x_1 + i k x_2}$.
\item In $x_1 < x_2<0$ region, we have
\begin{align}
&g_{f}(x_1, x_2)\notag\\
&= g(x_1, x_2) -\frac{i}{\sqrt{2}}V \left[e^{i E_0 x_1} e(x_2-x_1)+ e^{i E_0 x_2}e(x_1-x_2)\right]\notag\\
&= \left(B_3 e^{i k x_1 + i p x_2}+ A_3 e^{i p x_1 + i k x_2}\right)\notag\\
& - i\Gamma\left[e^{i E_0 x_1}\left(\frac{t_p B_3 e^{i p(x_2-x_1)}}{k -\Omega + i\Gamma/2}+ \frac{t_k A_3 e^{i k (x_2 - x_1)}}{p -\Omega + i\Gamma/2}\right)+ e^{i E_0 x_2}\left(\frac{B_3 e^{i k(x_1-x_2)}}{p -\Omega + i\Gamma/2}+ \frac{A_3 e^{i p (x_1 - x_2)}}{k -\Omega + i\Gamma/2}\right)\right]\notag\\
&=\left(B_3 e^{i k x_1 + i p x_2}+ A_3 e^{i p x_1 + i k x_2}\right)\notag\\
& - i\Gamma\left[\frac{t_p B_3 e^{i k x_1 + i p x_2}}{k -\Omega + i\Gamma/2}+ \frac{t_k A_3 e^{i p x_1 + i k x_2}}{p -\Omega + i\Gamma/2}+ \frac{B_3 e^{i kx_1 + i p x_2)}}{p -\Omega + i\Gamma/2}+ \frac{A_3 e^{i p x_1 + i k x_2)}}{k -\Omega + i\Gamma/2}\right]\notag\\
&=B_3 e^{i k x_1 + i p x_2}\left(1-\frac{i\Gamma t_p}{k -\Omega +i\Gamma/2}-\frac{i\Gamma}{p-\Omega+i\Gamma/2}\right)\notag\\
&+A_3 e^{i p x_1 + i k x_2}\left(1-\frac{i\Gamma t_k}{p -\Omega +i\Gamma/2}-\frac{i\Gamma}{k-\Omega+i\Gamma/2}\right)\notag\\
&=t_k t_p (B_3 e^{i k x_1 + i p x_2}+ A_3 e^{i p x_1 + i k x_2}).
\end{align}
\end{enumerate}
The $x_1> x_2$ region is obtained by $g_{f}(x_2, x_1) = + g_{f}(x_1, x_2)$. 

Thus, we explicitly demonstrate that for the interacting eigenstate of the extended Wiegmann-Andrei form in Sec.~\ref{P:WiegmannState}, the out-state photon wavefunction $g_f(x_1, x_2)$ is $t_k t_p (B_3 e^{i k x_1 + i p x_2}+ A_3 e^{i p x_1 + i k x_2})$, in the \emph{entire} $x_1$-$x_2$ plane. This is consistent with the usual ``read-off'' of the out-state by taking the $x_1, x_2 >0$ region of the interacting eigenstate and extend to the entire $x_1$-$x_2$ plane.

The in-state can be computed in exactly the same fashion by starting from Eq.~(\ref{E:LippmannR}) and shown to be $g_{i}(x_1, x_2) = B_3 e^{i k x_1 + i p x_2}+ A_3 e^{i p x_1 + i k x_2}$, in the \emph{entire} $x_1$-$x_2$ plane. Again, it is consistent with the usual read-off of the in-state by taking the $x_1, x_2<0$ region of the interacting eigenstate and extend to the entire  $x_1$-$x_2$ plane. Similar computations have been done for the two-photon bound state $|B_{E}\rangle$, and the same conclusion has been reached.

\section{Overlaps of Various States}\label{A:Overlap}
In this appendix, we summarize the properties of the two complete sets $\{|S_{k,p}\rangle_{ee}: k\leq p\}$ and $\{|A_{k,p}\rangle_{ee}: k\leq p\}$ defined in Sec.~\ref{P:TwoPhotonI}. These properties are used to normalized the Wiegmann-Andrei state in Sec.~\ref{P:WiegmannState} as well as in the completeness check in Appendix~\ref{A:Completeness}. In this section, we suppress the $ee$ label since there is no confusion. 

We first mention the following identities:
\begin{subequations}
\begin{equation}
\int_{0}^{\infty}\cos k x \,dk = \pi \delta(x)
\end{equation}
\begin{equation}
\int_{0}^{\infty}\sin k x \,dk = \mathcal{P}\frac{1}{x}
\end{equation}
\end{subequations}
where $\mathcal{P}$ denotes the Cauchy principal value. Recall that
\begin{align}
\langle x_1, x_2|S_{k, p}\rangle &\equiv \frac{1}{2\pi}\frac{1}{\sqrt{2}}\left(e^{i k x_1} e^{i p x_2} +e^{i k x_2} e^{i p x_1}\right)= \frac{\sqrt{2}}{2\pi} e^{i E x_c}\cos\left(\Delta x\right)\notag\\
\langle x_1, x_2|A_{k,p}\rangle &\equiv \frac{1}{2\pi}\frac{1}{\sqrt{2}}\,\mbox{sgn}(x)\left(e^{i k x_1} e^{i p x_2} -e^{i k x_2} e^{i p x_1}\right)
= \frac{\sqrt{2}i}{2\pi}\,\mbox{sgn}(x) \, e^{i E x_c}\sin\left(\Delta x\right)
\end{align}
where $\Delta \equiv (k-p)/2 = k -E/2$.

The overlap between $|S_{k_1,p_1}\rangle$ and $|A_{k_2,p_2}\rangle$ is
\begin{align}
&\langle S_{k_1,p_1}|A_{k_2, p_2}\rangle\notag\\
&= \left(\frac{1}{2\pi}\right)^2 \sqrt{2} (\sqrt{2} i) \int_{-\infty}^{\infty} dx_c\, e^{(E_2 - E_1)x_c} \int_{-\infty}^{\infty} dx\, \mbox{sgn}(x) \cos\Delta_1 x \sin\Delta_2 x\notag\\
&=\frac{i}{2\pi^2}\times \left(2\pi\delta(E_2 - E_1)\right)\times 2 \int_{0}^{\infty} dx \cos\Delta_1 x \sin\Delta_2 x\notag\\
&=\frac{i}{\pi}\times\delta(E_2 - E_1)\times 2 \int_{0}^{\infty} dx \left[\frac{1}{2}\sin[(\Delta_2 + \Delta_1)x] + \frac{1}{2}\sin[(\Delta_2 - \Delta_1)x]\right]\notag\\
&=\frac{i}{\pi}\delta(E_2 - E_1)\times\left[\mathcal{P}\frac{1}{\Delta_2+\Delta_1}+\mathcal{P}\frac{1}{\Delta_2-\Delta_1}\right]\notag\\
&=\frac{i}{\pi}\delta(E_2 - E_1)\times \left(2\Delta_2\right) \,\mathcal{P}\frac{1}{\Delta_2^2-\Delta_1^2}.
\end{align}

The overlap between $|S_{k_1,p_1}\rangle$ and $|S_{k_2,p_2}\rangle$ is
\begin{align}\label{E:SSOverlap}
&\langle S_{k_1,p_1}|S_{k_2, p_2}\rangle\notag\\
&= (\frac{1}{2\pi})^2 (\sqrt{2})^2 \int_{-\infty}^{\infty} dx_c\, e^{(E_2 - E_1)x_c} \int_{-\infty}^{\infty} dx \cos\Delta_1 x \cos\Delta_2 x\notag\\
&=\frac{1}{2\pi^2} \times \left(2\pi\delta(E_2 - E_1)\right)\times  2\int_{0}^{\infty} dx \left[\frac{1}{2}\cos[(\Delta_1 - \Delta_2)x] + \frac{1}{2}\cos[(\Delta_1 + \Delta_2)x]\right]\notag\\
&=\frac{1}{\pi}\delta(E_2 - E_1)\times \pi\left(\delta(\Delta_1- \Delta_2) + \delta(\Delta_1+\Delta_2)\right)\notag\\
&=\delta(E_2 - E_1)\left[\delta(\Delta_1- \Delta_2) + \delta(\Delta_1+\Delta_2)\right]\notag\\
&=\delta(k_1-k_2)\delta(p_1 - p_2)+\delta(k_1-p_2)\delta(k_2 - p_1)\\
&=\mbox{direct term}+\mbox{exchange term}.\notag
\end{align}

The overlap between $|A_{k_1,p_1}\rangle$ and $|A_{k_2,p_2}\rangle$ is
\begin{align}
&\langle A_{k_1,p_1}|A_{k_2, p_2}\rangle\notag\\
&= (\frac{1}{2\pi})^2 (-\sqrt{2} i)(\sqrt{2} i) \int_{-\infty}^{\infty} dx_c\, e^{(E_2 - E_1)x_c} \int_{-\infty}^{\infty} dx \sin\Delta_1 x \sin\Delta_2 x\notag\\
&=\frac{1}{2\pi^2}\times \left(2\pi\delta(E_2 - E_1)\right)\times  2\int_{0}^{\infty} dx \left[\frac{1}{2}\cos[(\Delta_1 - \Delta_2)x] - \frac{1}{2}\cos[(\Delta_1 + \Delta_2)x]\right]\notag\\
&=\frac{1}{\pi}\delta(E_2 - E_1)\times \pi\left(\delta(\Delta_1- \Delta_2) - \delta(\Delta_1+\Delta_2)\right)\notag\\
&=\delta(E_2 - E_1)\left[\delta(\Delta_1- \Delta_2) - \delta(\Delta_1+\Delta_2)\right]\notag\\
&=\delta(k_1-k_2)\delta(p_1 - p_2)-\delta(k_1-p_2)\delta(k_2 - p_1)\\
&=\mbox{direct term}-\mbox{exchange term}.\notag
\end{align}

Various calculations in this paper involve evaluation of overlap with the state [Eq.~(\ref{E:WiegmannState})]
\begin{equation}
|W_{k,p}\rangle =\frac{1}{\sqrt{4\Delta^2 + \Gamma^2}}\left(2\Delta |S_{k, p}\rangle + i\Gamma|A_{k,p}\rangle\right).
\end{equation} For example,
\begin{align}\label{E:WSOverlap}
\langle W_{k, p}|S_{k_1, p_1}\rangle &= \frac{1}{\sqrt{4\Delta^2 + \Gamma^2}}\left(2\Delta \langle S_{k, p}|S_{k_1, p_1}\rangle -i\Gamma \langle A_{k, p}|S_{k_1, p_1}\rangle\right)\notag\\
&= \frac{2\Delta}{\sqrt{4\Delta^2 + \Gamma^2}}\left(\delta(\Delta-\Delta_1) + \delta(\Delta+\Delta_1) -\frac{\Gamma}{\pi}\mathcal{P}\frac{1}{\Delta^2 -\Delta_1^2}\right)\delta(E-E_1),
\end{align}where $\mathcal{P}$ denotes Cauchy principal value. 

In both the completeness check, as well as in the evaluation of the S-matrix, one needs to calculate the product $\langle S_{k_2, p_2}|W_{k, p}\rangle \langle W_{k, p}|S_{k_1, p_1}\rangle$. Using Eq.~(\ref{E:WSOverlap}), we have
\begin{align}\label{E:SWWSRaw}
\langle S_{k_2, p_2}|W_{k, p}\rangle \langle W_{k, p}|S_{k_1, p_1}\rangle &=\frac{4\Delta^2}{4\Delta^2 + \Gamma^2}\delta(E-E_1)\delta(E-E_2)\times\left\{\phantom{\left(-\frac{\Gamma}{\pi}\right)}\right.\notag\\
&\left[\delta(\Delta-\Delta_1) + \delta(\Delta+\Delta_1)\right]\left[\delta(\Delta-\Delta_2) + \delta(\Delta+\Delta_2)\right] &  &\text{(term 1)}\notag\\
+&\left(-\frac{\Gamma}{\pi}\right)\left[\delta(\Delta-\Delta_1) + \delta(\Delta+\Delta_1)\right]\mathcal{P}\frac{1}{\Delta^2-\Delta_2^2} & &\text{(term 2)}\notag\\
+&\left(-\frac{\Gamma}{\pi}\right)\left[\delta(\Delta-\Delta_2) + \delta(\Delta+\Delta_2)\right]\mathcal{P}\frac{1}{\Delta^2-\Delta_1^2} & &\text{(term 3)}\notag\\
+&\left.\left(\frac{\Gamma}{\pi}\right)^2\mathcal{P}\frac{1}{\Delta^2-\Delta_1^2}\mathcal{P}\frac{1}{\Delta^2-\Delta_2^2}\right\}. & &\text{(term 4)}
\end{align}
We now evaluate these terms.\\
\textit{term 2}: including the prefactor $\frac{4\Delta^2}{4\Delta^2 + \Gamma^2}$, term 2 can be simplified as
\begin{equation}
\left(-\frac{\Gamma}{\pi}\right)\frac{4\Delta_1^2}{4\Delta_1^2 + \Gamma^2}\mathcal{P}\frac{1}{\Delta_1^2-\Delta_2^2}\delta(E-E_1)\delta(E-E_2)\left[\delta(\Delta-\Delta_1)+\delta(\Delta+\Delta_1)\right]. 
\end{equation}
\textit{term 3}: including the prefactor $\frac{4\Delta^2}{4\Delta^2 + \Gamma^2}$, term 3 can be simplified as
\begin{equation}
\left(-\frac{\Gamma}{\pi}\right)\frac{4\Delta_2^2}{4\Delta_2^2 + \Gamma^2}\mathcal{P}\frac{1}{\Delta_2^2-\Delta_1^2}\delta(E-E_1)\delta(E-E_2)\left[\delta(\Delta-\Delta_2)+ \delta(\Delta+\Delta_2)\right]. 
\end{equation}
\textit{term 4}: in evaluating term 4, we first note the Poincar\'{e} - Bertrand formula~\cite{PoincareCite}:
\begin{align}\label{E:Poincare}
& \mathcal{P}\frac{1}{X-Y}\mathcal{P}\frac{1}{X-Z} \notag\\
=\, &\mathcal{P}\frac{1}{Y-Z}\left(\mathcal{P}\frac{1}{X-Y}-\mathcal{P}\frac{1}{X-Z} \right) +\pi^2\delta(X-Y)\delta(X-Z),
\end{align}for three arbitrary variables, $X$, $Y$, and $Z$.
Hence,
\begin{align}\label{E:PoincareProduct}
&\mathcal{P}\frac{1}{\Delta^2-\Delta_1^2}\mathcal{P}\frac{1}{\Delta^2-\Delta_2^2}\notag\\
&=\mathcal{P}\frac{1}{\Delta_1^2-\Delta_2^2}\left(\mathcal{P}\frac{1}{\Delta^2-\Delta_1^2}-\mathcal{P}\frac{1}{\Delta^2-\Delta_2^2} \right) +\pi^2\delta(\Delta^2-\Delta_1^2)\delta(\Delta^2-\Delta_2^2)\notag\\
&= \mathcal{P}\frac{1}{\Delta_1^2-\Delta_2^2}\left(\mathcal{P}\frac{1}{\Delta^2-\Delta_1^2}-\mathcal{P}\frac{1}{\Delta^2-\Delta_2^2} \right)\notag\\
& +\frac{\pi^2}{4|\Delta_1| |\Delta_2|}\left[\delta(\Delta-\Delta_1)+\delta(\Delta+\Delta_1)\right]\left[\delta(\Delta-\Delta_2)+\delta(\Delta+\Delta_2)\right]
\end{align}
The terms with $\delta$-fucntions in Eq.~(\ref{E:PoincareProduct}), including all prefactors in Eq.~(\ref{E:SWWSRaw}), yield
\begin{align}
&\frac{4\Delta^2}{4\Delta^2 + \Gamma^2}\left(\frac{\Gamma}{\pi}\right)^2 \left(\frac{\pi^2}{4|\Delta_1| |\Delta_2|}\left[\delta(\Delta-\Delta_1)+\delta(\Delta+\Delta_1)\right]\left[\delta(\Delta-\Delta_2)+\delta(\Delta+\Delta_2)\right]\right)\notag\\
&=\frac{4\Gamma^2}{4\Delta^2 + \Gamma^2}\left[\delta(\Delta-\Delta_1)+\delta(\Delta+\Delta_1)\right]\left[\delta(\Delta-\Delta_2)+\delta(\Delta+\Delta_2)\right],
\end{align}which can be combined together with term 1 in Eq.~(\ref{E:SWWSRaw}) to yield 
\begin{equation}
\delta(E-E_1)\delta(E-E_2)\times\left\{\left[\delta(\Delta-\Delta_1)+\delta(\Delta+\Delta_1)\right]\left[\delta(\Delta-\Delta_2)+\delta(\Delta+\Delta_2)\right]\right\}.
\end{equation}

Therefore, the end result is
\begin{align}\label{E:SWWS}
&\langle S_{k_2, p_2}|W_{k, p}\rangle \langle W_{k, p}|S_{k_1, p_1}\rangle\notag\\
&=\delta(E-E_1)\delta(E-E_2)\times\left\{\phantom{\mathcal{P}\frac{1}{\Delta^2}}\right.\notag\\
&\left[\delta(\Delta-\Delta_1)+\delta(\Delta+\Delta_1)\right]\left[\delta(\Delta-\Delta_2)+\delta(\Delta+\Delta_2)\right]\notag\\
&+\left(-\frac{\Gamma}{\pi}\right)\frac{4\Delta_1^2}{4\Delta_1^2 + \Gamma^2}\mathcal{P}\frac{1}{\Delta_1^2-\Delta_2^2}\left[\delta(\Delta-\Delta_1)+\delta(\Delta+\Delta_1)\right]\notag\\
&+\left(-\frac{\Gamma}{\pi}\right)\frac{4\Delta_2^2}{4\Delta_2^2 + \Gamma^2}\mathcal{P}\frac{1}{\Delta_2^2-\Delta_1^2}\left[\delta(\Delta-\Delta_2)+ \delta(\Delta+\Delta_2)\right]\notag\\
&+ \left.\left(\frac{\Gamma}{\pi}\right)^2 \frac{4\Delta^2}{4\Delta^2 + \Gamma^2}\mathcal{P}\frac{1}{\Delta_1^2-\Delta_2^2}\left(\mathcal{P}\frac{1}{\Delta^2-\Delta_1^2}-\mathcal{P}\frac{1}{\Delta^2-\Delta_2^2} \right)\right\}
\end{align}

\section{Completeness check}\label{A:Completeness}

In this appendix, we carry out the explicit check of the completeness of the eigenstates $\{|W_{k, p}\rangle, |B_{E}\rangle\}$ in Sec.~\ref{P:WiegmannState}. Again, since the discussions below are in the $ee$ subspace, we omit the subscript when there is no confusion.

As noted in Sec.~\ref{P:WiegmannState}, to check whether $\{|W_{k, p}\rangle:  k  \leq p\}$ is complete, one could start with an arbitrary state, for example, $|S_{k_1,p_1}\rangle$, project out all $|W_{k,p}\rangle$ components and calculate
\begin{equation}\label{E:Delta}
|\delta_{k_1,p_1}\rangle \equiv |S_{k_1,p_1}\rangle -\sum_{k \leq p}\langle W_{k, p}|S_{k_1,p_1}\rangle |W_{k, p}\rangle.
\end{equation}If the set $\{|W_{k,p}\rangle:  k \leq p\}$ were complete, such a computation should yield $|\delta_{k_1,p_1}\rangle =0$ for arbitrary $k_1$ and $p_1$. To calculate $|\delta_{k_1,p_1}\rangle$, we first project $|\delta_{k_1,p_1}\rangle$ to $\langle S_{k_2, p_2}|$:
\begin{equation}
\langle S_{k_2, p_2}|\delta_{k_1,p_1}\rangle = \langle S_{k_2, p_2}|S_{k_1,p_1}\rangle -\sum_{k \leq p}\langle W_{k, p}|S_{k_1,p_1}\rangle \langle S_{k_2, p_2}|W_{k, p}\rangle.
\end{equation}

The first term in the right hand side is [from Eq.~(\ref{E:SSOverlap})]
\begin{equation}
\langle S_{k_2, p_2}|S_{k_1,p_1}\rangle = \delta(k_1-k_2)\delta(p_1-p_2)+\delta(k_1-p_2)\delta(p_1-k_2),
\end{equation}while in the second term, the restriction $k\leq p$ can be dropped, using the symmetry property of $|W_{k, p}\rangle$, Eq.~(\ref{E:WProp2}):
\begin{equation}
\sum_{k, p, k\leq p} \langle S_{k_2, p_2}|W_{k, p}\rangle \langle W_{k, p}|S_{k_1, p_1}\rangle = \frac{1}{2} \sum_{k, p} \langle S_{k_2, p_2}|W_{k, p}\rangle \langle W_{k, p}|S_{k_1, p_1}\rangle.
\end{equation}
Using Eq.~(\ref{E:SWWS}) for $\langle S_{k_2, p_2}|W_{k, p}\rangle \langle W_{k, p}|S_{k_1, p_1}\rangle$, the second term becomes
\begin{align}\label{E:IntSWWS}
& \frac{1}{2}\sum_{k, p}\langle S_{k_2, p_2}|W_{k,p}\rangle \langle W_{k,p}| S_{k_1, p_1}\rangle \notag\\
=&\, \frac{1}{2}\int_{-\infty}^{+\infty}dE \int_{-\infty}^{+\infty}d\Delta\,\delta(E-E_1)\delta(E-E_2)\times\left\{\phantom{\mathcal{P}\frac{1}{\Delta^2}}\right.\notag\\
&\left[\delta(\Delta-\Delta_1)+\delta(\Delta+\Delta_1)\right]\left[\delta(\Delta-\Delta_2)+\delta(\Delta+\Delta_2)\right]\notag\\
&+\left(-\frac{\Gamma}{\pi}\right)\frac{4\Delta_1^2}{4\Delta_1^2 + \Gamma^2}\mathcal{P}\frac{1}{\Delta_1^2-\Delta_2^2}\left[\delta(\Delta-\Delta_1)+\delta(\Delta+\Delta_1)\right]\notag\\
&+\left(-\frac{\Gamma}{\pi}\right)\frac{4\Delta_2^2}{4\Delta_2^2 + \Gamma^2}\mathcal{P}\frac{1}{\Delta_2^2-\Delta_1^2}\left[\delta(\Delta-\Delta_2)+ \delta(\Delta+\Delta_2)\right]\notag\\
&+ \left.\left(\frac{\Gamma}{\pi}\right)^2 \frac{4\Delta^2}{4\Delta^2 + \Gamma^2}\mathcal{P}\frac{1}{\Delta_1^2-\Delta_2^2}\left(\mathcal{P}\frac{1}{\Delta^2-\Delta_1^2}-\mathcal{P}\frac{1}{\Delta^2-\Delta_2^2} \right)\right\}.
\end{align}
The first term in Eq.~(\ref{E:IntSWWS}) yields
\begin{align}
&\delta(E_1-E_2)\left[\delta(\Delta_1-\Delta_2)+\delta(\Delta_1+\Delta_2)\right]\notag\\
=&\, \delta(k_1-k_2)\delta(p_1-p_2)+ \delta(k_1 - p_2)\delta(k_2-p_1)\notag\\
=&\, \langle S_{k_2, p_2}|S_{k_1, p_1}\rangle.
\end{align}
The integrations of the second and third terms are straightforward, and the sum of both terms give
\begin{equation}
-\frac{4\Gamma^3}{\pi}\frac{1}{4\Delta_1^2+\Gamma^2}\frac{1}{4\Delta_2^2+\Gamma^2}\delta(E_1-E_2).
\end{equation}
The last term can be calculated using a contour integral. The only non-vanishing contribution comes from the pole at $\Delta = + i\Gamma/2$, when the integration contour is chosen to be completed in the upper half plane. Hence the integration yields
\begin{align}
&\left[\frac{1}{2}2\pi i \left(\frac{\Gamma}{\pi}\right)^2 \left(-\frac{\Gamma^2}{4 i\Gamma}\right)\mathcal{P}\frac{1}{\Delta_1^2 -\Delta_2^2}\left(\frac{1}{(i\Gamma/2)^2 -\Delta_2^2}-\frac{1}{(i\Gamma/2)^2 -\Delta_1^2}\right)\right]\delta(E_1-E_2)\notag\\
=&\, -\frac{4\Gamma^3}{\pi}\frac{1}{4\Delta_1^2+\Gamma^2}\frac{1}{4\Delta_2^2+\Gamma^2}\delta(E_1-E_2).
\end{align}

The final result therefore is
\begin{align}
&\sum_{k, p, k\leq p} \langle S_{k_2, p_2}|W_{k, p}\rangle \langle W_{k, p}|S_{k_1, p_1}\rangle\notag\\
=&\, \delta(k_1-k_2)\delta(p_1-p_2)+ \delta(k_1 - p_2)\delta(k_2-p_1)\notag\\
&+ \left(-\frac{8\Gamma^3}{\pi}\right) \frac{1}{4\Delta_1^2+\Gamma^2}\frac{1}{4\Delta_2^2+\Gamma^2}\delta(E_1-E_2).
\end{align}Thus
\begin{align}\label{E:DeltaProjection}
\langle S_{k_2, p_2}|\delta_{k_1,p_1}\rangle &= \left(+\frac{8\Gamma^3}{\pi}\right) \frac{1}{4\Delta_1^2+\Gamma^2}  \frac{1}{4\Delta_2^2+\Gamma^2}\delta(E_1-E_2)\notag\\
&= \left(+\frac{8\Gamma^3}{\pi}\right) \frac{1}{(k_1-p_1)^2+\Gamma^2}  \frac{1}{(k_2-p_2)^2+\Gamma^2}\delta(E_1-E_2).
\end{align}Since $\langle S_{k_2, p_2}|\delta_{k_1,p_1}\rangle \neq 0$, this directly proves that the set $\{|W_{k, p}\rangle: k \leq p\}$ is incomplete.

A very important observation regarding Eq.~(\ref{E:DeltaProjection}) is that, independent of the choice of $k_1$, $p_1$, the resulting state $\{|\delta_{k_1, p_1}\rangle\}$ calculated in Eq.~(\ref{E:DeltaProjection}) is always proportional to the same state
\begin{align}
|\delta_{k_1, p_1}\rangle &= \frac{1}{2}\int_{-\infty}^{+\infty}dE_2\int_{-\infty}^{+\infty}d\Delta_2 \langle S_{k_2, p_2}|\delta_{k_1, p_1}\rangle |S_{k_2, p_2}\rangle\notag\\
&\propto \int_{-\infty}^{+\infty}dE_2\int_{-\infty}^{+\infty} d\Delta_2 \frac{1}{4\Delta_2^2 +\Gamma^2} |S_{k_2, p_2}\rangle\delta(E_1-E_2)\notag\\
&\equiv |\tilde{B}_{E_1}\rangle,
\end{align}Therefore, the set $\{|\delta_{k_1, p_1}\rangle\}$ in fact forms a one-dimensional Hilbert space. Thus, only one extra state $|\tilde{B}_{E}\rangle$ is needed in order to span the two-photon Hilbert space. Since 
\begin{align}
\langle \tilde{B}_{E}|\tilde{B}_{E'}\rangle &= 2\int_{-\infty}^{+\infty}d\Delta \left(\frac{1}{4\Delta^2+\Gamma^2}\right)^2 \delta(E-E')\notag\\
&=\frac{\pi}{2\Gamma^3}\delta(E-E'),
\end{align}
this extra state, when normalized, is
\begin{align}\label{E:BStateDef1}
|B_{E}\rangle &= \sqrt{\frac{2\Gamma^3}{\pi}}\int_{-\infty}^{\infty}d\Delta\,\frac{1}{4\Delta^2 +\Gamma^2}|S_{k, p}\rangle.
\end{align}
This concludes the proof that $\left\{|W_{k, p}\rangle, |B_{E}\rangle\right\}$ forms a complete basis of the two-photon Hilbert space.

To see the physical meaning of $|B_{E}\rangle$, we rewrite Eq.~(\ref{E:BStateDef1}) as 
\begin{equation}\label{E:BStateDef}
|B_{E}\rangle = \int dx_c dx\, B_E(x_1,x_2) \frac{1}{\sqrt{2}}c^{\dagger}(x_1) c^{\dagger}(x_1)|\emptyset, -\rangle,
\end{equation}with
\begin{equation}\label{E:BDef}
B_E(x_1,x_2) \equiv B_E(x_c,x) = e^{i E x_c}\frac{\sqrt{\Gamma}}{\sqrt{4\pi}} e^{-\frac{\Gamma}{2}|x|}.
\end{equation}In the above derivation, we have used
\begin{equation}
\int_{-\infty}^{\infty}d\Delta \frac{1}{4\Delta^2 + \Gamma^2}\cos(\Delta x) = \frac{\pi}{2\Gamma} e^{-\frac{\Gamma}{2} |x|}.
\end{equation}
Thus, the state $|B_{E}\rangle$ in fact defines a two-photon bound state.

\section{Derivations of S-Matrix}\label{A:SDetails}

In this appendix, we provide the detailed calculations of Eq.~(\ref{E:SMatrixElement}), the matrix element of the S-matrix in the $ee$ subspace, $\mathbf{S}_{ee}$. Again, since the discussions below are in the $ee$ subspace, we omit the subscript when there is no confusion.

The S-matrix  in the $ee$ subspace, $\mathbf{S}_{ee}$ is defined in Eq.~(\ref{E:SMatrix}):
\begin{equation}\label{E:See}
\mathbf{S}_{ee}\equiv\sum_{k \leq p}  t_k t_p |W_{k,p}\rangle \langle W_{k, p}| + \sum_{E} t_E |B_E\rangle \langle B_E|.
\end{equation}Our goal is to compute the matrix element, Eq.~(\ref{E:SMatrixElement}),
\begin{equation}\label{E:See2}
\langle S_{k_2, p_2}|\mathbf{S}_{ee}|S_{k_1, p_1}\rangle = \langle S_{k_2, p_2}|\left(\sum_{k\leq p}  t_k t_p |W_{k,p}\rangle \langle W_{k, p}|\right)|S_{k_1, p_1}\rangle + \langle S_{k_2, p_2}|\left(\sum_{E} t_E |B_E\rangle \langle B_E|\right)|S_{k_1, p_1}\rangle.
\end{equation}We will compute the two terms on the right hand side separately in the following.

\subsection*{D1. First term of Eq.~(\ref{E:See2})}

Using Eq.~(\ref{E:SWWS}) for $\langle S_{k_2, p_2} |W_{k,p}\rangle \langle W_{k, p}|S_{k_1, p_1}\rangle$, the first term in Eq.~(\ref{E:See2}) becomes
\begin{align}\label{E:InttktpSWWS}
&\sum_{k\leq p}t_k t_p\langle S_{k_2, p_2}|W_{k,p}\rangle \langle W_{k,p}| S_{k_1, p_1}\rangle \notag\\
=&\, \frac{1}{2}\sum_{k, p}t_k t_p\langle S_{k_2, p_2}|W_{k,p}\rangle \langle W_{k,p}| S_{k_1, p_1}\rangle \notag\\
=&\, \frac{1}{2}\int_{-\infty}^{+\infty}dE \int_{-\infty}^{+\infty}d\Delta\,\delta(E-E_1)\delta(E-E_2)\times t_k t_p\left\{\phantom{\mathcal{P}\frac{1}{\Delta^2}}\right.\notag\\
&\left[\delta(\Delta-\Delta_1)+\delta(\Delta+\Delta_1)\right]\left[\delta(\Delta-\Delta_2)+\delta(\Delta+\Delta_2)\right]\notag\\
&+\left(-\frac{\Gamma}{\pi}\right)\frac{4\Delta_1^2}{4\Delta_1^2 + \Gamma^2}\mathcal{P}\frac{1}{\Delta_1^2-\Delta_2^2}\left[\delta(\Delta-\Delta_1)+\delta(\Delta+\Delta_1)\right]\notag\\
&+\left(-\frac{\Gamma}{\pi}\right)\frac{4\Delta_2^2}{4\Delta_2^2 + \Gamma^2}\mathcal{P}\frac{1}{\Delta_2^2-\Delta_1^2}\left[\delta(\Delta-\Delta_2)+ \delta(\Delta+\Delta_2)\right]\notag\\
&+ \left.\left(\frac{\Gamma}{\pi}\right)^2 \frac{4\Delta^2}{4\Delta^2 + \Gamma^2}\mathcal{P}\frac{1}{\Delta_1^2-\Delta_2^2}\left(\mathcal{P}\frac{1}{\Delta^2-\Delta_1^2}-\mathcal{P}\frac{1}{\Delta^2-\Delta_2^2} \right)\right\}.
\end{align}
Note that
\begin{align}
t_k t_p &= t_{\Delta + E/2}  t_{-\Delta + E/2} \notag\\
&=  \left(\frac{\Delta+(E-2\Omega)/2-i\Gamma/2}{\Delta+(E-2\Omega)/2+i\Gamma/2}\right)\times \left(\frac{-\Delta+(E-2\Omega)/2-i\Gamma/2}{-\Delta+(E-2\Omega)/2+i\Gamma/2}\right) \notag\\
&=  \left(\frac{\Delta+(E-2\Omega)/2-i\Gamma/2}{\Delta+(E-2\Omega)/2+i\Gamma/2}\right)\times\left(\frac{\Delta-(E-2\Omega)/2+i\Gamma/2}{\Delta-(E-2\Omega)/2-i\Gamma/2}\right),
\end{align}
the first term in Eq.~(\ref{E:InttktpSWWS}) can be evaluated as
\begin{align}\label{E:First}
& t_{k_1} t_{p_1} \delta(E_1-E_2)\left[\delta(\Delta_1-\Delta_2)+\delta(\Delta_1-\Delta_2)\right]\notag\\
=&\, t_{k_1} t_{p_1}\delta(k_1-k_2)\delta(p_1-p_2)+ t_{k_1} t_{p_1}\delta(k_1-p_2)\delta(p_1-k_2).
\end{align}
The second and the third term in Eq.(\ref{E:InttktpSWWS}) can be combined to give 
\begin{equation}\label{E:SecondThird}
\frac{\Gamma}{\pi}\left(t_{k_2} t_{p_2} \frac{4\Delta_2^2}{4\Delta_2^2 + \Gamma^2}\mathcal{P}\frac{1}{\Delta_1^2 -\Delta_2^2} + t_{k_1}t_{p_1} \frac{4\Delta_1^2}{4\Delta_1^2 + \Gamma^2}\mathcal{P}\frac{1}{\Delta_2^2 -\Delta_1^2}\right)\delta(E_1 - E_2).
\end{equation}
In evaluating the fourth term in Eq.~(\ref{E:InttktpSWWS}), we use a contour integral, by completing the contour in the upper half plane. The only poles that give non-zero contributions are located at $\Delta = (E-2\Omega +i\Gamma)/2$, and $\Delta = i\Gamma/2$. The result is
\begin{align}\label{E:Fourth}
&\delta(E_1-E_2)\left(-\frac{4\Gamma^3}{\pi}\frac{E_1 - 2\Omega - 2 i\Gamma}{E_1 - 2\Omega + 2 i\Gamma}\frac{1}{4\Delta_1^2 + \Gamma^2}\frac{1}{4\Delta_2^2 + \Gamma^2}\right.\notag\\
&\left.-\frac{16\Gamma^3}{\pi}\frac{E_1 - 2\Omega + i\Gamma}{E_1 - 2\Omega + 2 i\Gamma}\frac{1}{4\Delta_1^2 - (E_1 - 2\Omega +i\Gamma)^2}\frac{1}{4\Delta_2^2 - (E_1 - 2\Omega +i\Gamma)^2}\right).
\end{align}
Summing  Eqs.~(\ref{E:First}), (\ref{E:SecondThird}), and (\ref{E:Fourth}) together, we thus have
\begin{align}\label{E:WContribution}
&\langle S_{k_2, p_2}|\left(\sum_{k\leq p}  t_k t_p |W_{k,p}\rangle\langle W_{k, p}|\right)|S_{k_1, p_1}\rangle\notag\\
=&\, t_{k_1} t_{p_1}\delta(k_1-k_2)\delta(p_1-p_2) + t_{k_1} t_{p_1}\delta(k_1-p_2)\delta(p_1-k_2)\notag\\
&+ \frac{\Gamma}{\pi}\delta(E_1 - E_2)\left(t_{k_2} t_{p_2} \frac{4\Delta_2^2}{4\Delta_2^2 + \Gamma^2}\mathcal{P}\frac{1}{\Delta_1^2 -\Delta_2^2} + t_{k_1}t_{p_1} \frac{4\Delta_1^2}{4\Delta_1^2 + \Gamma^2}\mathcal{P}\frac{1}{\Delta_2^2 -\Delta_1^2}\right)\notag\\
&+\delta(E_1-E_2)\left(-\frac{4\Gamma^3}{\pi}\frac{E_1 - 2\Omega - 2 i\Gamma}{E_1 - 2\Omega + 2 i\Gamma}\frac{1}{4\Delta_1^2 + \Gamma^2}\frac{1}{4\Delta_2^2 + \Gamma^2}\right.\notag\\
&\left.-\frac{16\Gamma^3}{\pi}\frac{E_1 - 2\Omega + i\Gamma}{E_1 - 2\Omega + 2 i\Gamma}\frac{1}{4\Delta_1^2 - (E_1 - 2\Omega +i\Gamma)^2}\frac{1}{4\Delta_2^2 - (E_1 - 2\Omega +i\Gamma)^2}\right).
\end{align}

\subsection*{D2. Second term of Eq.~(\ref{E:See2})}

We first evaluate the overlap between $|B_{E}\rangle$ and $|S_{k_1, p_1}\rangle$:
\begin{align}
&\langle B_{E}|S_{k_1, p_1}\rangle\notag\\
=& \,\frac{\sqrt{\Gamma}}{\sqrt{4\pi}} \int_{-\infty}^{\infty}dx_c \int_{-\infty}^{\infty}dx \left(e^{-i Ex_c -\Gamma/2 |x|}\right) \times\left[\frac{1}{2\pi}\frac{1}{\sqrt{2}}e^{i E_1 x_c}\left(e^{i\Delta_1 x} +e^{-i\Delta_1 x} \right)\right]\notag\\
=&  \,\frac{\sqrt{\Gamma}}{\sqrt{4\pi}}\frac{1}{2\pi}\frac{1}{\sqrt{2}}\times2\pi\delta(E_1-E_2)\times 2 \int_{0}^{\infty}dx (e^{(i\Delta_1 -\Gamma/2) x}+e^{(-i\Delta_1 -\Gamma/2) x})\notag\\
=&  \,\frac{\sqrt{\Gamma}}{\sqrt{2\pi}}\delta(E_1-E_2) \left(\frac{-1}{i\Delta_1 -\Gamma/2}+\frac{-1}{-i\Delta_1 -\Gamma/2}\right)\notag\\
=& \,\frac{\sqrt{\Gamma}}{\sqrt{2\pi}}\frac{4\Gamma}{4\Delta_1^2 + \Gamma^2}\delta(E_1-E_2),
\end{align}
Thus, the second term of Eq.~(\ref{E:See2}), the bound state contribution, is
\begin{align}\label{E:BContribution}
&\langle S_{k_2, p_2}|\left(\sum_{E} t_{E}|B_{E}\rangle\langle B_{E}|\right)|S_{k_1, p_1}\rangle\notag\\
=& \frac{\Gamma}{2\pi} \int_{-\infty}^{\infty}\left(\frac{E-2\Omega-2 i \Gamma}{E-2\Omega+2 i\Gamma}\right)\frac{4\Gamma}{4\Delta_1^2+\Gamma^2}\frac{4\Gamma}{4\Delta_2^2+\Gamma^2}\delta(E-E_1)\delta(E-E_2)\notag\\
=& \,\frac{8\Gamma^3}{\pi}\frac{E_1-2\Omega-2 i\Gamma}{E_1 -2\Omega+ 2 i\Gamma}\frac{1}{4\Delta_1^2+\Gamma^2}\frac{1}{4\Delta_2^2+\Gamma^2}\delta(E_1 -E_2)
\end{align}

Summing Eqs.~(\ref{E:WContribution}) and (\ref{E:BContribution}), we obtain the S-matrix $\langle S_{k_2, p_2}|\mathbf{S}_{ee}|S_{k_1, p_1}\rangle$, Eq.~(\ref{E:SMatrixElement}):
\begin{equation}
\langle S_{k_2, p_2}|\mathbf{S}_{ee}|S_{k_1, p_1}\rangle = t_{k_1}t_{p_1}\delta(k_1 -k_2)\delta(p_1 - p_2) + t_{k_1} t_{p_1}\delta(k_1 - p_2)\delta(k_2 - p_1)+ B\delta(E_1 -E_2),
\end{equation}with
\begin{align}
B &=
\frac{16 i \Gamma^2}{\pi}\frac{E_1-2\Omega + i\Gamma}{\left[4\Delta_1^2 -(E_1 - 2\Omega + i\Gamma)^2\right] \left[4\Delta_2^2 -(E_1 - 2\Omega + i\Gamma)^2\right]}.
\end{align}

\section{Derivations of two-mode out-state}\label{A:S2Out}

In this appendix, we present the details of the derivations of the two-mode out-state two-photon wavefunciton, $t_2(x_1, x_2)$ [Eq.~(\ref{E:t2Wavefunction})], $r_2(x_1, x_2)$ [Eq.~(\ref{E:r2Wavefunction})], and $rt(x_1, x_2)$ [Eq.~(\ref{E:rtWavefunction})].

The in-state is a state of two right-going photons, $|S_{k_1, p_1}\rangle_{RR}$. We first decompose the in-state into $ee$, $oo$, and $eo$ subspaces: 
\begin{align}
|\mbox{in}\rangle &\equiv |S_{k_1, p_1}\rangle_{RR}\notag\\
& = \int dx_1 dx_2 \frac{1}{2\pi\sqrt{2}} \left(e^{i k x_1 + i p x_2} + e^{i k x_2 + i p x_1}\right) \frac{1}{\sqrt{2}}c_R^{\dagger}(x_1) c_R^{\dagger}(x_2) |\emptyset,-\rangle\notag\\
&=\int dx_1 dx_2 \frac{1}{2\pi\sqrt{2}} \left(e^{i k x_1 + i p x_2} + e^{i k x_2 + i p x_1}\right)\frac{1}{\sqrt{2}}\frac{1}{2}\left(c_{e}^{\dagger}(x_1) + c_{o}^{\dagger}(x_1)\right)\left(c_{e}^{\dagger}(x_2) + c_{o}^{\dagger}(x_2)\right)|\emptyset,-\rangle\notag\\
&=\frac{1}{2}\int dx_1 dx_2 \frac{1}{2\pi\sqrt{2}} \left(e^{i k x_1 + i p x_2} + e^{i k x_2 + i p x_1}\right) \frac{1}{\sqrt{2}}c_e^{\dagger}(x_1) c_e^{\dagger}(x_2) |\emptyset,-\rangle\notag\\
&+\frac{1}{2}\int dx_1 dx_2 \frac{1}{2\pi\sqrt{2}} \left(e^{i k x_1 + i p x_2} + e^{i k x_2 + i p x_1}\right) \frac{1}{\sqrt{2}}c_o^{\dagger}(x_1) c_o^{\dagger}(x_2) |\emptyset,-\rangle\notag\\
&+\frac{1}{2}\frac{1}{\sqrt{2}}\int dx_1 dx_2 \frac{1}{2\pi\sqrt{2}} \left(e^{i k x_1 + i p x_2} + e^{i k x_2 + i p x_1}\right) \left(c_e^{\dagger}(x_1) c_o^{\dagger}(x_2)+c_o^{\dagger}(x_1) c_e^{\dagger}(x_2)\right) |\emptyset,-\rangle\notag\\
&=\frac{1}{2}|S_{k_1, p_1}\rangle_{ee}+\frac{1}{2}|S_{k_1, p_1}\rangle_{oo}+ \frac{1}{2}\frac{1}{\sqrt{2}}\int dx_1 dx_2 \frac{2}{2\pi\sqrt{2}} \left(e^{i k x_1 + i p x_2} + e^{i k x_2 + i p x_1}\right) c_e^{\dagger}(x_1) c_o^{\dagger}(x_2)|\emptyset,-\rangle\notag\\
&\equiv |\mbox{in}\rangle_{ee} + |\mbox{in}\rangle_{oo}+|\mbox{in}\rangle_{eo}
\end{align}

Employing the decomposition relation, Eq.~(\ref{E:Decomposition}), we carry out the calculations in each individual subspace:
\begin{align}
\mathbf{S} |\mbox{in}\rangle_{ee}&=\mathbf{S}_{ee}|\mbox{in}\rangle_{ee}=\mathbf{S}_{ee}\frac{1}{2}|S_{k_1, p_1}\rangle_{ee}\notag\\
&=\frac{1}{2}\sum_{E_2, \Delta_2\leq 0}|S_{E_2, \Delta_2}\rangle_{ee}\,\phantom{}_{ee}\langle S_{E_2, \Delta_2}|\mathbf{S}_{ee}|S_{k_1, p_1}\rangle_{ee}\notag\\
&= \frac{1}{2}\left(t_{k_1} t_{p_1}|S_{k_1, p_1}\rangle_{ee}+\sum_{\Delta_2\leq 0} B |S_{E_1, \Delta_2}\rangle_{ee}\right)\notag\\ 
&=\frac{1}{2}\int dx_1 dx_2\, \phantom{}_{ee}\langle x_c, x|\mbox{out}\rangle_{ee} \frac{1}{\sqrt{2}}c_e^{\dagger}(x_1) c_e^{\dagger}(x_2)|\emptyset,-\rangle\notag\\
&\equiv\frac{1}{2}\int dx_1 dx_2\, \phi_{ee}(x_1, x_2) \frac{1}{\sqrt{2}}c_e^{\dagger}(x_1) c_e^{\dagger}(x_2)|\emptyset,-\rangle,
\end{align}$\phi_{ee}(x_1, x_2)$ is the out-state wavefunction in $ee$ subspace, Eq.~(\ref{E:Out}).
\begin{align}
\mathbf{S} |\mbox{in}\rangle_{oo}&=\mathbf{S}_{oo}|\mbox{in}\rangle_{oo}=\mathbf{S}_{oo}\frac{1}{2}|S_{k_1, p_1}\rangle_{oo}\notag\\
&= \frac{1}{2}|S_{k_1, p_1}\rangle_{oo}\notag\\
&=\frac{1}{2}\int dx_1 dx_2\, \phantom{}_{oo}\langle x_1, x_2|S_{k_1, p_1}\rangle_{oo} \frac{1}{\sqrt{2}}c_o^{\dagger}(x_1)c_o^{\dagger}(x_2)|\emptyset,-\rangle\notag\\
&\equiv\frac{1}{2}\int dx_1 dx_2\,S_{k_1, p_1} \frac{1}{\sqrt{2}}c_o^{\dagger}(x_1)c_o^{\dagger}(x_2)|\emptyset,-\rangle
\end{align}and
\begin{align}
&\mathbf{S} |\mbox{in}\rangle_{eo}=\mathbf{S}_{eo}|\mbox{in}\rangle_{eo}\notag\\
&=\mathbf{S}_{eo}\left[\frac{1}{2}\frac{1}{\sqrt{2}}\int dx_1 dx_2 \frac{1}{2\pi\sqrt{2}} \left(e^{i k x_1 + i p x_2} + e^{i k x_2 + i p x_1}\right) \left(c_e^{\dagger}(x_1) c_o^{\dagger}(x_2)+c_o^{\dagger}(x_1) c_e^{\dagger}(x_2)\right) |0\rangle\right]\notag\\
&=\frac{1}{\sqrt{2}}\int dx_1 dx_2 \frac{1}{2\pi\sqrt{2}}\left(t_k e^{i k x_1 + i p x_2}+t_p e^{i k x_2 + i p x_1}\right)c_e^{\dagger}(x_1)c_o^{\dagger}(x_2)|\emptyset,-\rangle.
\end{align}

Using the transformation formula, Eq.~(\ref{E:Transformation}), and collect terms according to the operators in Eq.~(\ref{E:RealSpaceOutState}), we obtain $t_2(x_1, x_2)$, $r_2(x_1, x_2)$, and $rt(x_1, x_2)$:
\begin{align}
t_2(x_1, x_2) 
&= \frac{1}{4}\left[\phi_{ee}(x_1, x_2) + S_{k_1, p_1}(x_1, x_2)+  (t_k + t_p) S_{k_1, p_1}(x_1, x_2)\right]\notag\\
&= \frac{1}{4}\left[t_{k_1} t_{p_1}   S_{k_1, p_1}(x_1, x_2) + \sum_{\Delta_2 \leq E_1/2} B S_{E_1, \Delta_2}(x_c, x) +S_{k_1, p_1}(x_1, x_2) +  (t_k + t_p)S_{k_1, p_1}(x_1, x_2)\right]\notag\\
&= \frac{1}{4}\left[(1+t_{k_1})(1+t_{p_1})S_{k_1, p_1}(x_1, x_2) +  \sum_{\Delta_2 \leq E_1/2} B S_{E_1, \Delta_2}(x_c, x)\right]\\
&=\bar{t}_{k_1}\bar{t}_{p_1} S_{k_1, p_1}(x_1, x_2)+ \frac{1}{4}\sum_{\Delta_2 \leq E_1/2} B  S_{E_1, \Delta_2}(x_c, x)\notag\\
&=e^{i E_1 x_c}\frac{\sqrt{2}}{2\pi}\left(\bar{t}_{k_1} \bar{t}_{p_1} \cos\left(\Delta_1 x\right)-\frac{\Gamma^2}{4\Delta_1^2 -(E_1-2\Omega +  i\Gamma)^2} e^{i (E_1-2\Omega) |x|/2 -\Gamma |x|/2}\right),
\end{align}
\begin{align}
r_2(x_1, x_2) 
&= \frac{1}{4}\left[\phi_{ee}(-x_1, -x_2) + S_{k_1, p_1}(-x_1, -x_2) -  (t_k + t_p)S_{k_1, p_1}(-x_1, -x_2)\right]\notag\\
&= \frac{1}{4}\left[(1-t_{k_1})(1-t_{p_1})S_{k_1, p_1}(-x_1, -x_2) +  \sum_{\Delta_2 \leq E_1/2} B S_{E_1, \Delta_2}(-x_c, -x)\right]\notag\\
&=\bar{r}_{k_1} \bar{r}_{p_1} S_{k_1, p_1}(-x_1, -x_2)+ \frac{1}{4}\sum_{\Delta_2 \leq E_1/2} B S_{E_1, \Delta_2}(-x_c, -x)\notag\\
&=e^{-i E_1 x_c}\frac{\sqrt{2}}{2\pi}\left(\bar{r}_{k_1} \bar{r}_{p_1} \cos\left(\Delta_1 x\right)-\frac{\Gamma^2}{4\Delta_1^2 -(E_1-2\Omega +  i\Gamma)^2} e^{i (E_1-2\Omega) |x|/2 -\Gamma |x|/2}\right),
\end{align}and 
\begin{align}
& rt(x_1, x_2)\notag\\
&=\frac{1}{4\sqrt{2}}\left[\phi_{ee}(x_1, -x_2) - S_{k_1, p_1}(x_1, -x_2)+  (t_p - t_k)\frac{1}{2\pi\sqrt{2}}\left(e^{i k x_1 - i p x_2} - e^{-i k x_2 + i p x_1}\right)\right]\notag\\
& = \frac{1}{2\pi} e^{i \frac{E_1}{2}x}\left(\bar{t}_{k_1} \bar{r}_{p_1} e^{2 i \Delta_1 x_c} + \bar{r}_{k_1} \bar{t}_{p_1} e^{-2 i \Delta_1 x_c}-\frac{2\Gamma^2}{4\Delta_1^2 -(E_1-2\Omega +  i\Gamma)^2} e^{i (E_1-2\Omega) |x_c| -\Gamma |x_c|}\right).
\end{align}

The momentum distributions can be computed directly. In the forward direction,
\begin{align}
&\phantom{}_{RR}\langle S_{k_2, p_2}|\mathbf{S}|S_{k_1, p_1}\rangle_{RR}\notag\\
=&\,\int dx_1 dx_2 \,S^*_{k_2, p_2}(x_1, x_2) t_2(x_1, x_2)\notag\\
=&\,\frac{1}{4}\left(\phantom{}_{RR}\langle S_{k_2, p_2}|\mbox{out}\rangle_{RR} + \phantom{}_{RR}\langle S_{k_2, p_2}|S_{k_1, p_1}\rangle_{RR}+t_{k_1} t_{p_1} \phantom{}_{RR}\langle S_{k_2, p_2}|S_{k_1, p_1}\rangle_{RR} \right)\notag\\
=&\,\frac{1}{4}\left[(t_{k_1}+1)(t_{p_1}+1)\left(\delta(k_1 - k_2)\delta(p_1-p_2)+\delta(k_1-p_2)\delta(p_1-k_2)\right)+B\delta(E_1 - E_2)\right]\notag\\
=&\,\bar{t}_{k_1}\bar{t}_{p_1}\left(\delta(k_1 - k_2)\delta(p_1-p_2)+\delta(k_1-p_2)\delta(p_1-k_2)\right)+\frac{1}{4}B\delta(E_1 - E_2).
\end{align} In the backward direction:
\begin{align}
&\phantom{}_{LL}\langle S_{k_2, p_2}|\mathbf{S}|S_{k_1, p_1}\rangle_{RR}\notag\\
=&\,\int dx_1 dx_2 \,S^*_{k_2, p_2}(x_1, x_2) r_2(x_1, x_2)\notag\\
=&\,\frac{1}{4}\left[(1-t_{k_1})(1-t_{p_1})\left(\delta(k_1 + k_2)\delta(p_1+p_2)+\delta(k_1+p_2)\delta(p_1+k_2)\right)+B\delta(E_1 - E_2)\right]\notag\\
=&\,\bar{r}_{k_1}\bar{r}_{p_1}\left(\delta(k_1 + k_2)\delta(p_1+p_2)+\delta(k_1+p_2)\delta(p_1+k_2)\right)+\frac{1}{4}B\delta(E_1 - E_2),
\end{align}while in the $RL$ subspace:
\begin{align}
&\phantom{}_{RL}\langle k_2^{R}, p_2^{L}|\mathbf{S}|S_{k_1, p_1}\rangle_{RR}\notag\\
=&\,\int dx_1 dx_2 \,\left(\frac{1}{2\pi}e^{i k_2 x_1+i p_2 x_2}\right)^* rt(x_1, x_2)\notag\\
=&\,\bar{t}_{k_1} \bar{r}_{p_1}\delta(k_2-k_1)\delta(p_2 + p_1)+  \bar{r}_{k_1} \bar{t}_{p_1}\delta(k_2 - p_1)\delta(p_2 + k_1) + \frac{1}{4} B\delta(E_1 - E_2).
\end{align}
In the above calculations, we have adopted the following sign convention for the left-moving photons:
\begin{align}
|k\rangle_{L} &\equiv \int dx\, \frac{e^{i k x}}{\sqrt{2\pi}} c^{\dagger}_{L}(x)|\emptyset, -\rangle, \quad \mbox{with $k <0$.}\notag\\
|S_{k, p}\rangle_{LL} &\equiv \iint dx_1 dx_2\, \frac{1}{2\pi\sqrt{2}}\left(e^{i k x_1+i p x_2}+ e^{i k x_2 + i p x_1}\right)\frac{1}{\sqrt{2}}c^{\dagger}_{L}(x_1)c^{\dagger}_{L}(x_2)|\emptyset, -\rangle, \quad \mbox{with $k, p <0$.}\notag\\
|k^{R}, p^{L}\rangle_{RL} &\equiv \iint dx_1 dx_2 \frac{1}{2\pi} e^{i k x_1 + i p x_2}c^{\dagger}_{R}(x_1)c^{\dagger}_{L}(x_2)|\emptyset, -\rangle, \quad \mbox{with $k>0$, $p <0$.}
\end{align}


\pagebreak
\newpage
\begin{figure}[thb]
\scalebox{0.5}{\includegraphics{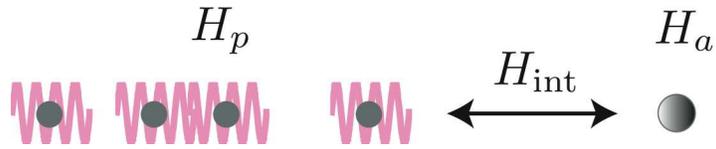}}
\caption{(Color online) Pictorial representation of the Hamiltonian of Eq.~(\ref{E:TotalHamiltonian}). The dynamics of the free quantum many-particles are described by $H_p$, which can have more than one incident particle. The dynamics of the impurity (the ``atom'') is described by $H_{a}$. The interactions between the quantum particles and the impurity is described by $H_{\mbox{\scriptsize int}}$.}\label{Fi:Hamiltonian}
\end{figure}

\pagebreak
\newpage

\pagebreak
\newpage

\pagebreak
\newpage

\pagebreak
\newpage

\pagebreak
\newpage

\pagebreak
\newpage

\pagebreak
\newpage

\pagebreak
\newpage

\pagebreak
\newpage

\pagebreak
\newpage

\end{document}